\documentclass[sigconf]{acmart}
\AtBeginDocument{%
  }

\setcopyright{acmlicensed}
\copyrightyear{2026}
\acmYear{2026}
\acmDOI{XXXXXXX.XXXXXXX}
\acmConference[SIGMOD]{Proceedings of the ACM on Management of Data}{May 31 -- June 05,
  2026}{Bengaluru, India}

\usepackage{algorithm}
\usepackage{multirow}
\usepackage{makecell}
\usepackage[noend]{algpseudocode}
\newcommand\ef{\texttt{ef}}
\usepackage{tikz}

\begin{document}
\title{Distribution-Aware Exploration for Adaptive HNSW Search}

\author{Chao Zhang}
\affiliation{%
  \institution{University of Waterloo}
  \country{Waterloo, Canada}
}
\email{chao.zhang@uwaterloo.ca}

\author{Renée J. Miller}
\affiliation{%
  \institution{University of Waterloo}
  \country{Waterloo, Canada}
}
\email{rjmiller@uwaterloo.ca}

\begin{abstract}
Hierarchical Navigable Small World (HNSW) is widely adopted for approximate nearest neighbor search (ANNS) for its ability to deliver high recall with low latency  on large-scale, high-dimensional embeddings.
The exploration factor, commonly referred to as \ef, is a key parameter in HNSW-based vector search that balances accuracy and efficiency.
However, existing systems typically rely on manually and statically configured \ef\ values that are uniformly applied across all queries.
This results in a distribution-agnostic configuration that fails to account for the non-uniform and skewed nature of real-world embedding data and query workloads.
As a consequence, HNSW-based systems suffer from two key practical issues:
(i) the absence of recall guarantees, and 
(ii) inefficient ANNS performance due to over- or under-searching.
In this paper, we propose Adaptive-\ef\ (\texttt{Ada-\ef}), a data-driven, update-friendly, query-adaptive approach that dynamically configures \ef\ for each query at runtime to approximately meet a declarative target recall with minimal computation.
The core of our approach is a theoretically grounded statistical model that captures the similarity distribution between each query and the database vectors.
Based on this foundation, we design a query scoring mechanism that distinguishes between queries requiring only small \ef\ and those that need larger \ef\ to  meet a target recall, and accordingly assigns an appropriate \ef\ to each query.
Experimental results on real-world embeddings produced by state-of-the-art Transformer models from OpenAI and Cohere show that, compared with state-of-the-art learning-based adaptive approaches, our method achieves the target recall while avoiding both over- and under-searching, reducing online query latency by up to 4$\times$, offline computation time by 50$\times$, and offline memory usage by 100$\times$.
\end{abstract}

\maketitle
\begin{tikzpicture}[remember picture, overlay]
  \node[anchor=south west, align=left, fill=white, inner sep=5pt] 
    at (current page.south west) 
    [xshift=2cm, yshift=2cm] 
    {
      \textbf{Accepted for publication in SIGMOD 2026}
    };
\end{tikzpicture}

\section{Introduction}\label{sec:intro}
The rapid advancement of embedding models~\cite{NIPS2013_9aa42b31}, particularly Transformer-based architectures~\cite{NIPS2017_3f5ee243,devlin-etal-2019-bert,reimers-gurevych-2019-sentence,dosovitskiy2021an}, has revolutionized the representation and processing of data.
These models encode diverse types of information, such as text~\cite{karpukhin-etal-2020-dense} and images~\cite{Radford2021LearningTV}, into high-dimensional semantic embeddings that capture contextual meaning in vector space.
To compare these embeddings effectively, inner product and cosine similarity
\cite{reimers-gurevych-2019-sentence, neelakantan2022textcodeembeddingscontrastive, you2025semanticsanglecosinesimilarity,Radford2021LearningTV}
have become standard, due to their ability to reflect semantic alignment between modern semantic embeddings.
Building on this foundation, vector indexing techniques are employed to efficiently retrieve the most relevant embeddings from large collections, making them critical components in real-time semantic search and retrieval systems.

Various types of vector indexes have been developed, including tree-based~\cite{4587638,bernhardsson2017annoy,arora18hdindex, 10.14778/3397230.3397240,10.1145/3539618.3591651}, hashing-based~\cite{10.1145/276698.276876,10.14778/2850469.2850470, 10.14778/3397230.3397243, 10.14778/3579075.3579084}, quantization-based~\cite{5432202,PAN202059,10.14778/2856318.2856324}, and graph-based~\cite{dong2011kgraph,jin2014IEH,MALKOV2014NSW,iwasaki2015ngt,Harwood2016FANNG,fu2016efanna,chen2018sptag,li2019dpg,Malkov2020HNSW,Fu2019NSG,Vargas2019HCNNG,Subramanya2019Vamana,Fu2022NSSG,Azizi2023ELPIS,Azizi2023ELPIS} methods. Among these, graph-based indexes have emerged as the most effective for high-dimensional similarity search, leveraging proximity graphs to efficiently structure vector data. 
One of the most widely adopted graph-based solutions for vector search is Hierarchical Navigable Small World (HNSW)~\cite{Malkov2020HNSW}.
It serves as a core indexing technique in vector search libraries like Faiss~\cite{douze2024faiss}, and powers search engines such as Elasticsearch~\cite{elastic_hnsw} and Lucene~\cite{lucene_hnswgraph}. Additionally, it is the backbone of specialized vector databases, including Milvus~\cite{10.1145/3448016.3457550}, Pinecone~\cite{pinecone_website}, and Weaviate~\cite{weaviate_website}, and has also been integrated into relational databases such as PostgreSQL~\cite{10.1145/3318464.3386131,pgvector_github}, SingleStore~\cite{10.14778/3685800.3685805}, and DuckDB~\cite{gabrielsson2024vectorsimilarity,10.1145/3299869.3320212}, and graph databases like TigerGraph~\cite{TigerVector} and Kùzu~\cite{feng2023kuzu}.

Given a set of data vectors $\mathbf{V}$, HNSW constructs a hierarchical proximity graph to serve as a vector index for approximate nearest neighbor search (ANNS) over $\mathbf{V}$. 
The number of nodes decreases logarithmically from the bottom layer to the top level, with the bottom level containing all $|\mathbf{V}|$ nodes, each representing a data vector. Edges in the HNSW graph encode the similarity relationship between vectors.
During query time, the search traverses the hierarchical graph from top to bottom to identify an entry point at the base layer. 
Starting from this point, the search proceeds using a best-first strategy guided by a fixed-size priority queue, whose size is a user-defined parameter, commonly referred to as \ef\ or \texttt{efSearch}, short for \textit{exploration factor}. The search terminates when the furthest element currently in the queue is closer to the query than the closest candidate node to be visited, indicating that no better candidate is likely to be found.

\begin{figure*}
    \centering
    \includegraphics[width=0.24\linewidth]{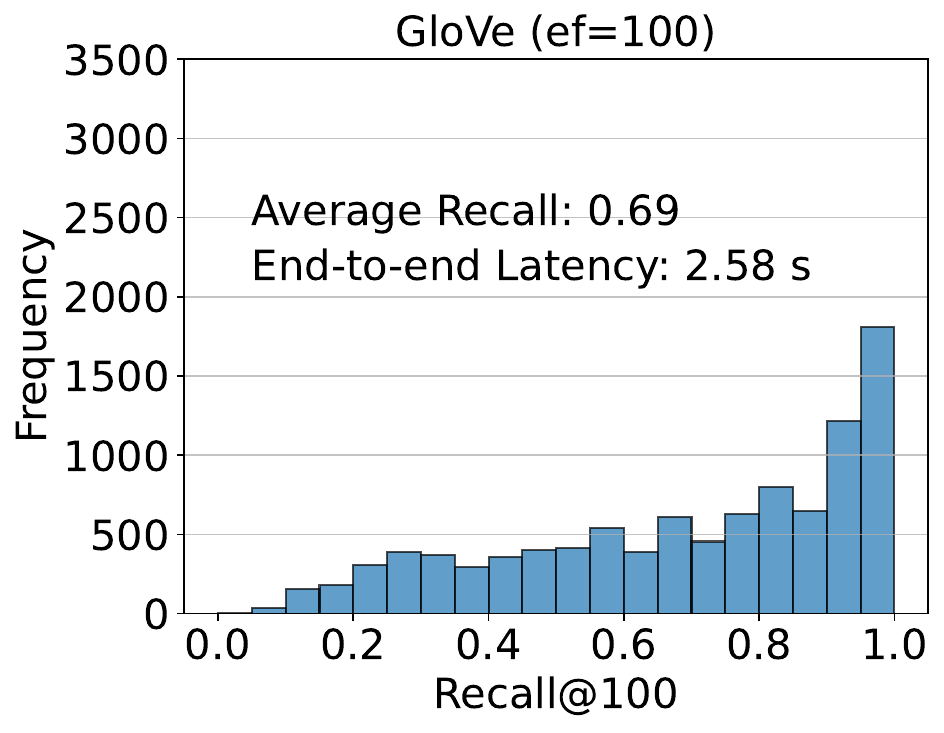}
    \includegraphics[width=0.24\linewidth]{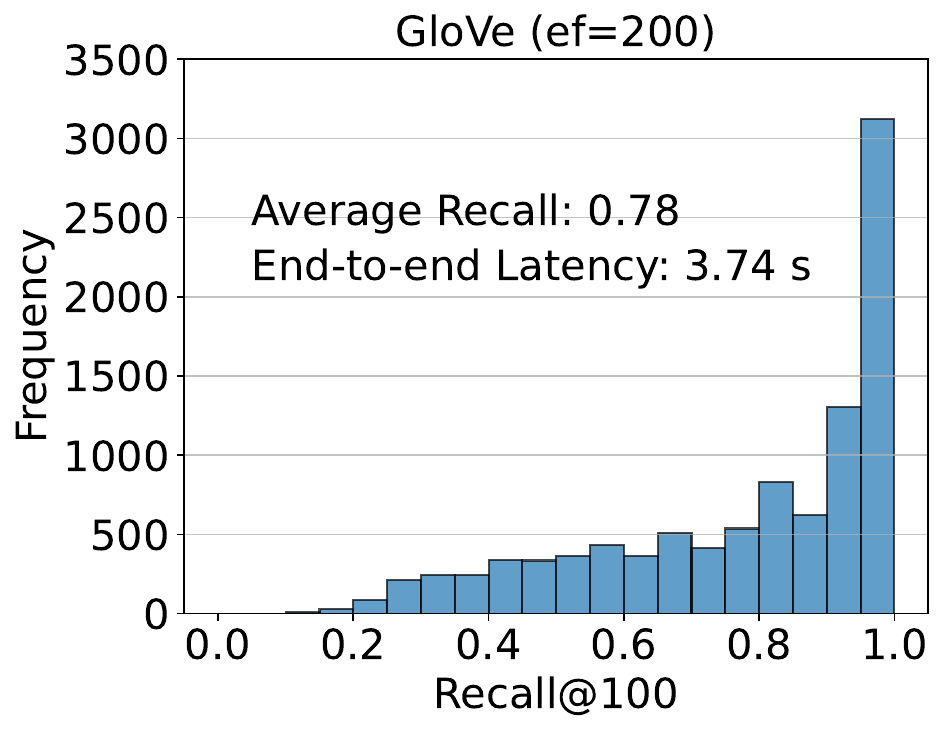}
    \hfill
    \hfill
    \includegraphics[width=0.24\linewidth]{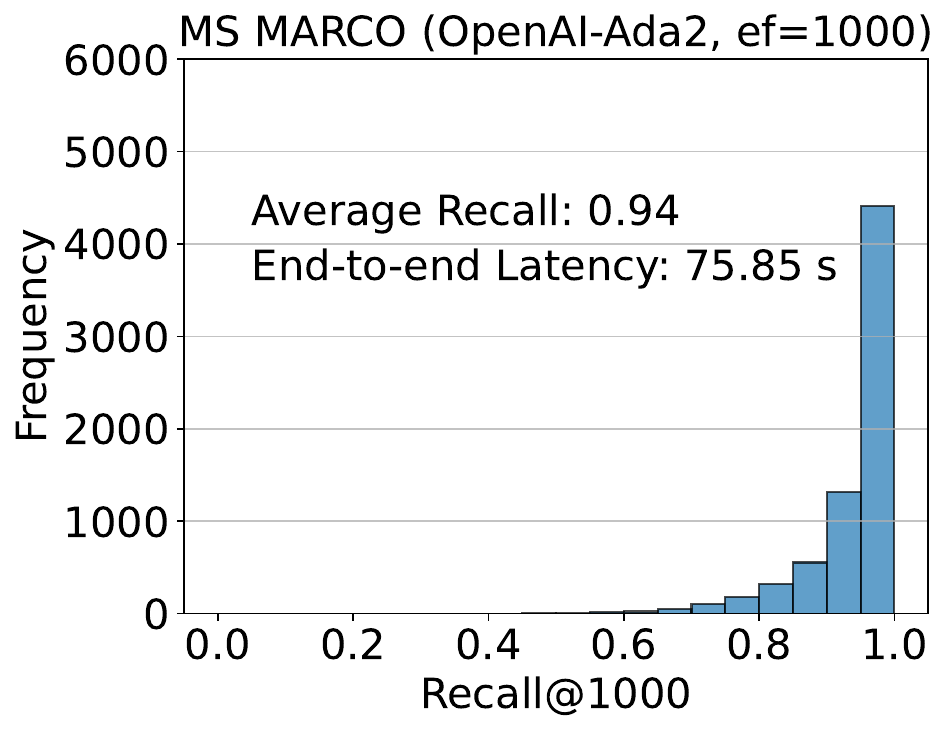}
    \includegraphics[width=0.24\linewidth]{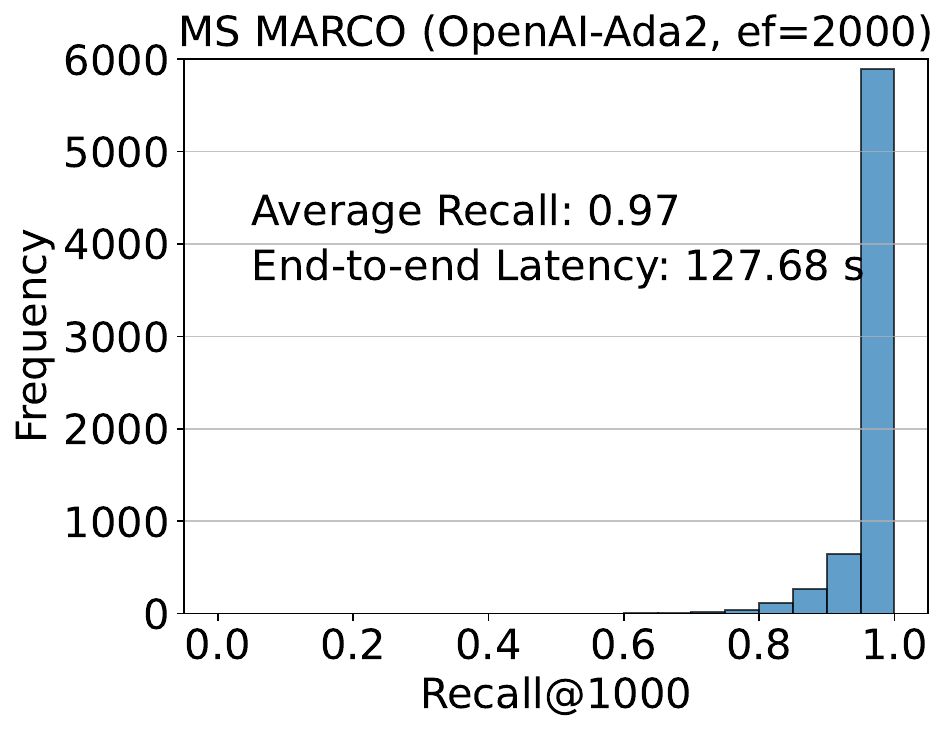}
    \caption{
    Recall distribution of HNSW search. GloVe~\cite{pennington-etal-2014-glove}: 1.8M vectors (100D), 10K queries, Top-100 ANNS with \ef\ = 100 and 200. MS MARCO (OpenAI Ada-002 embeddings)~\cite{msmarco_openai}: 8.8M vectors (1536D),  6.9K queries, Top-1000 ANNS with \ef\ = 1000 and 2000.
    }
    \label{fig:recall_distribution}
\end{figure*}

The parameter \ef\ controls the trade-off between query latency and result accuracy. 
A larger \ef\ typically yields a higher recall at the cost of increased query time, whereas a smaller \ef\ results in faster queries but may compromise accuracy. 
The default setting of \ef\ in existing systems~\cite{practical_ef_recall} is either a static  \ef\ value, \textit{e.g.}, \ef=40 in PGVector~\cite{pgvector_github} and \ef=16 in Faiss~\cite{douze2024faiss}, or setting \ef\ to $k$ for Top-$k$ ANNS, where the goal is to retrieve the $k$ nearest vectors to a given query, \textit{e.g.}, OpenSearch (Lucene)~\cite{opensearch_lucene}.

Once configured, the \ef\ value remains fixed and is applied uniformly across the entire query workload, under the expectation that it will consistently provide high recall.  However, this is not guaranteed so to compensate, users typically set \ef\ to a sufficiently large value, often based on empirical heuristics or subjective estimation.
However, this static \ef\ configuration introduces two key issues in production environments: 
(i) the absence of recall guarantees; and (ii) 
inappropriate \ef\ values for individual queries, leading to either over-searching or under-searching.

More fundamentally, applying a fixed \ef\ uniformly across queries is structurally misaligned with the inherently non-uniform and skewed nature of semantic embedding spaces. 
Semantic embeddings, including word, sentence, and image embeddings, are widely known to exhibit highly skewed, non-uniform distributions. 
In word embeddings, this skew is often tied to frequency bias: frequent words dominate the embedding space with larger norms and concentrated directional variance, a phenomenon shown to degrade similarity search due to the emergence of hub-points that appear as nearest neighbors to many unrelated vectors~\cite{mu2018allbutthetop,yokoi2024zipfian}. 
Similar skew is observed in sentence embeddings from transformer-based models like BERT and RoBERTa, which tend to produce anisotropic vector spaces where embeddings cluster along dominant directions, limiting their discriminative power~\cite{ethayarajh-2019-contextual,li-etal-2020-sentence,pmlr-v233-nielsen24a}. Cross-modal embeddings, such as CLIP's joint image–text vectors, also suffer from skew and hubness: normalized vectors 
lie in narrow cones, with retrieval dominated by a small set of hub images or texts~\cite{9878857,DBLP:journals/information/TyshchukKSPRP23}.

\textbf{Motivating example}.
To illustrate these issues, we built HNSW indexes using \texttt{HNSWlib}~\cite{hnswlib} for two datasets (see Table~\ref{table:datasets} for dataset statistics) and conduct Top-$k$ ANNS with static \ef\ values. 
The results in Figure~\ref{fig:recall_distribution} highlight several important observations.
First, different datasets require different \ef\ to achieve high average recall. 
For example, setting \ef\ to $k$ yields relatively high recall in MS MARCO (average recall is 94\%), but performs less effectively in GloVe (average recall is only 69\%).
Second, a significant portion of the queries exhibit recall substantially higher or lower than the average, indicating that average recall alone does not fully capture actual search performance.
While increasing \ef\ from $k$ to $2k$ can improve average recall, it also leads to over-searching for queries that already achieve high recall. In the GloVe dataset, approximately 1800 queries attain recall close to 1.0 with \ef=$k$; increasing \ef\ to $2k$ offers little improvement for these queries but adds unnecessary computation overhead. A similar trend is observed in the MS MARCO dataset, where over 4000 queries already perform well with \ef = $k$, making further increases redundant.
Additionally, even though the average recall in MS MARCO reaches 0.97 with \ef = $2k$, approximately 1000 queries still have recall below 0.9. In the GloVe dataset, a significant number of queries fall well below the average.

In real-world production environments, selecting an appropriate \ef\ value to meet a target recall is highly challenging.   
Despite its practical importance, the problem of adaptively configuring \ef\ to meet recall requirements has received little attention. 
Most benchmarking efforts evaluate indexing methods by sweeping across various \ef\ values to generate latency-recall trade-off curves for the fixed query load of the benchmarks. While useful for comparing the relative performance of indexing structures, these evaluations do not offer a practical solution for configuring \ef\ in real-world deployments.
Specifically, it is infeasible  to iteratively increase \ef\ to reach a desired recall level for online query processing, as doing so would require repeated vector search and computing the ground truth for incoming query vectors. 
Moreover, although setting a large \ef\ can improve overall recall, it also incurs substantial computation overhead and may still fail to achieve high recall across all queries, as demonstrated in the experiments.

In this paper, we investigate the following two key questions:\\
    (i) \textit{Can we configure \ef\ to meet a user-specified target recall?} \\
    (ii)  \textit{Can we assign adaptive \ef\ values on a per-query basis, allocating a larger \ef\ to queries requiring more computation to achieve the target recall, and a smaller \ef\ to those for which a lower value suffices?}

\begin{figure*}
    \centering
    \includegraphics[width=\linewidth]{figures/overview.pdf}
    \caption{Overview of \texttt{Ada-\ef}. 
It consists of two stages: offline  and online computations.
In the offline stage, dataset-level statistics  are computed (\S~\ref{sec:sim_distribution}), followed by the construction of an \ef-estimation table (\S~\ref{sec:data_driven_ef}).
In the online stage, the search follows the standard HNSW process until the base layer, where adaptive-\ef\ search begins (\S~\ref{sec:adaptive_ef}): 
(1) Distance collection: exploring a limited number of nodes to compute a query score using offline statistics;
(2) Search with estimated \ef: using the score to select an \ef\ from the estimation table to approximately reach a declarative target recall.
}
    \label{fig:overview}
\end{figure*}

To address the above challenges, we propose \textbf{Adaptive-\ef\ (\texttt{Ada-\ef})}, a query-adaptive approach for \ef\ configuration, \textit{i.e.}, configuring an \ef\ value for each individual query vector to approximately achieve a  target recall at runtime.
\texttt{Ada-\ef}\ is lightweight, update-friendly, and purely rule-based, which allows it to generalize effectively across different datasets. Figure \ref{fig:overview} presents an overview of our approach.
The core idea is as follows:
Given a query vector, we first perform the standard HNSW search to reach the base layer and obtain the entry point. The subsequent online search process consists of two phases: (i) Distance collection; and (ii) 
Search with a deterministic \ef\ value.
In phase (i), the algorithm explores the graph and records the distances between the query vector and the visited nodes. 
Based on these distances collected, we design an \ef-estimator that takes as input the query vector, the collected distance list, and a user-defined target recall, and outputs an \ef\ value intended to achieve the target recall for the given query, which is used in phase (ii).

The \ef-estimator is the core component of our adaptive approach. 
Its design is based on our theoretical foundation on the similarity distance distribution between a query vector $q$ and a vector database: \textit{Cosine similarity and inner product among high-dimensional learned embeddings often approximate a Gaussian (normal) distribution}.
More importantly, the distribution can be estimated efficiently at runtime. 
Combined with the distances collected during phase (i) of the search process, the estimated distribution allows us to distinguish between queries that can achieve high recall with a small \ef\ and those that require a large \ef\ to achieve a target recall.

Empirically, we demonstrate that our \ef-adaptive approach is able to  achieve the user-defined target recall across query workloads. Additionally, it can improve the recall of queries that suffer from under-searching under a static \ef\ configuration, \textit{i.e.}, the tail cases shown in Figure~\ref{fig:recall_distribution}.
Importantly, although our method involves collecting distance statistics and estimating \ef\ for each query on the fly, the overall end-to-end query latency can be lower than that of the original HNSW search. This is attributed to the adaptive nature of our approach, which allocates computational effort more efficiently across queries, thereby mitigating the issue of over-searching.
We further show that, while our method requires an offline preprocessing step, the associated computational cost and storage overhead are negligible compared to the indexing overhead of HNSW.
Last but not least, by simulating a real-world production scenario, we demonstrate that our approach can effectively adapt to dynamic workloads involving both insertion and deletion, and that the corresponding offline updates are lightweight.

We make the following two major contributions in this paper:
\begin{itemize} 
    \item We introduce \texttt{Ada-\ef}, a distribution-aware approach that adaptively sets \ef\ for each query at runtime to approximately meet a user-specified target recall, while improving efficiency by avoiding both under- and over-searching. 
    To the best of our knowledge, this is the first work that  addresses \ef\ configuration for HNSW search.
    
    \item We establish a theoretical foundation for the distribution of similarity distances between a query vector and  a vector database. We demonstrate the practical applicability of this foundation in the context of searching high-dimensional learned embeddings. 
\end{itemize}

\section{Related Work}\label{sec:related_work}
We position our work within existing work. For a more comprehensive discussion, we refer readers to prior surveys~\cite{jafari2021surveylocalitysensitivehashing,wang2021survey,wang2023surveygt,azizi2025survey,pan2024survey}.

Various indexing techniques have been proposed (see \S\ref{sec:intro} for detail).
Among them, HNSW~\cite{Malkov2020HNSW} stands out as the most influential and widely adopted approach. 
HNSW has been extensively adopted (see \S\ref{sec:intro} for detail). 
Although graph-based indexes~\cite{dong2011kgraph,jin2014IEH,MALKOV2014NSW,iwasaki2015ngt,Harwood2016FANNG,fu2016efanna,chen2018sptag,li2019dpg,Malkov2020HNSW,Fu2019NSG,Vargas2019HCNNG,Subramanya2019Vamana,Fu2022NSSG,Azizi2023ELPIS,Azizi2023ELPIS} differ in their indexing algorithms for graph construction, 
their online search algorithms remain fundamentally the same across all methods~\cite{azizi2025survey,wang2023surveygt}, such as NSG~\cite{Fu2019NSG}, Vamana~\cite{Subramanya2019Vamana},  and SPANN~\cite{ChenW21}.
A few studies have explored ways to further optimize the search process. These include leveraging intra-query parallelism to accelerate search execution~\cite{10.1145/3572848.3577527} and approximating distance computations to improve efficiency while maintaining accuracy~\cite{10.1145/3543507.3583318}. 
Several recent studies have also explored the impact of hardware architectures on HNSW performance, including GPU acceleration~\cite{10597683}, CPU-based optimizations~\cite{wang2025accelerating}, hybrid CPU-GPU approaches~\cite{305214}, and SSD-based indexing~\cite{10.1145/3639269}.

The key distinction between our work and all prior graph-based approaches lies in the use of a fixed versus an adaptive exploration factor in the best-first search algorithm, \textit{i.e.}, the \ef\ parameter. Existing methods rely on a fixed \ef\ value, manually pre-configured by the user and applied uniformly across all queries during search. 
This static configuration lacks recall approximation and often leads to inefficiencies, \textit{i.e.}, over-searching for some queries and under-searching for others, as shown in our motivation examples and experiments.
In contrast, we introduce \texttt{Ada-\ef}, the first distribution-aware approach for adaptively configuring \ef\ on a per-query basis at runtime. 
By leveraging dataset- and query-specific statistics, this adaptive strategy enables a more fine-grained trade-off between accuracy and efficiency, and is particularly well-suited to real-world workloads, where data and query distributions are often non-uniform and highly skewed.
We focus  on HNSW, given its widespread adoption in practice. Our method operates on existing pre-built indexes without requiring any index modifications, enabling seamless integration into deployed infrastructures.
Furthermore, the proposed adaptive strategy is orthogonal to both algorithmic and hardware-level optimizations, making it compatible with and complementary to existing acceleration techniques.

Early termination in ANNS has been explored through both learning-based and heuristic approaches.
Learned Adaptive Early Termination (\texttt{LAET})~\cite{10.1145/3318464.3380600} predicts per-query stopping conditions using Gradient Boosting Decision Trees~\cite{4a848dd1-54e3-3c3c-83c3-04977ded2e71} and runtime features. \texttt{Tao}~\cite{yang2021taolearningframeworkadaptive} is a static-feature-only framework that replaces runtime features with the Local Intrinsic Dimension (LID)~\cite{10.1145/2783258.2783405}.
\texttt{DARTH}~\cite{10.1145/3749160} extends \texttt{LAET} by integrating a recall predictor within HNSW to achieve declarative recall via prediction intervals.
In contrast, Patience in Proximity (\texttt{PiP})~\cite{10.1007/978-3-031-88714-7_39} is a simple saturation-based heuristic that halts HNSW search once candidate improvements plateau.
Unlike \texttt{PiP}, which requires defining a fixed \ef\ value, \texttt{LAET}, \texttt{Tao}, and \texttt{DARTH} automatically determine termination points without preset \ef\ parameters, relying instead on model predictions. Among these methods, DARTH represents the state-of-the-art and is the only one supporting declarative recall, offering the interface as \texttt{Ada-\ef}.

Compared to \texttt{LAET}, \texttt{Tao},  \texttt{DARTH}, and \texttt{PiP}, \texttt{Ada-\ef} is entirely rule-based and has a novel design based on our foundation on similarity distribution (Theorem~\ref{theorem:inner_product}). 
It estimates the \ef\ value required to approximately reach a target recall.
Consequently, \texttt{Ada-\ef} incurs minimal offline computation (\S\ref{sec:exp_offline}) and supports efficient incremental updates of insertion and deletion (\S\ref{sec:incremental_update}). 
More importantly, \texttt{Ada-\ef} achieves robust query performance, reaching target recall across datasets, reducing latency, and improving recall on the hardest queries, as demonstrated in our experimental evaluation.

VBASE~\cite{zhang2023vbase} incorporates an early termination mechanism for hybrid search combining vector similarity and attribute filtering, enabling simultaneous evaluation of both components. However, its benefit is limited in pure vector search, where it still depends on a fixed exploration parameter \textit{E} (equivalent to \ef\ in HNSW). Extending our query-adaptive \ef\ approach to hybrid search remains a promising direction for future work.

\section{Preliminary}\label{sec:pre}
A vector dataset $\mathbf{V} = (\mathbf{v_1}, ..., \mathbf{v_n})$ is a collection of $n$ vectors, where each vector $\mathbf{v_i} = (v_1, ..., v_d)$ has dimensionality $d$. Each $\mathbf{v_i}$ is referred to as a data vector, which typically represents an embedded form of raw data, such as an image or a text passage. 
We represent the entire vector dataset $\mathbf{V}$ as an $n \times d$ matrix. 
The similarity between a query vector $\mathbf{q}$ and a data vector $\mathbf{v}$ is measured using a distance function $dist(\mathbf{q}, \mathbf{v})$, where a smaller value indicates greater similarity. 
For instance, cosine distance is a common choice for $dist$ in the case of learned vector embeddings in modern AI systems.

\begin{definition}[Full Distance List]
\label{def:FDL}
Given a query vector $\mathbf{q}$ and a dataset $\mathbf{V} = (\mathbf{v}_1, \ldots, \mathbf{v}_n) $, the Full Distance List (FDL) of $\mathbf{q}$ with respect to $\mathbf{V}$ is 
$FDL(\mathbf{q}, \mathbf{V}) = \left(dist(\mathbf{q}, \mathbf{v}_1), \ldots, dist(\mathbf{q}, \mathbf{v}_n)\right)$.
\end{definition}

Given a query vector $\mathbf{q}$ and a vector dataset $\mathbf{V}$, the objective of nearest neighbor search (NNS) is to find a set $\mathcal{R}$ of $k$ vectors in $\mathbf{V}$ that are closest to $\mathbf{q}$ under a given distance function $dist$. Formally, $|\mathcal{R}| = k$, and for any vector $\mathbf{v} \in \mathcal{R}$, there does not exist a vector $\mathbf{v}' \in \mathbf{V} \setminus \mathcal{R}$ such that $dist(\mathbf{q}, \mathbf{v}) > dist(\mathbf{q}, \mathbf{v}')$.

When both $n$ (the number of vectors) and $d$ (the dimensionality) are large, as is common in modern applications (\textit{e.g.}, millions of vectors with 768 dimensions generated by BERT-based embedding models), computing exact NNS (\textit{i.e.}, computing FDL) becomes prohibitively expensive.
To address this scalability issue, practical systems typically employ ANNS, which computes an approximate result set $\tilde{\mathcal{R}}$ that aims to closely match the true nearest neighbor set $\mathcal{R}$. The accuracy of ANNS is commonly evaluated using the recall metric, defined as $Recall@k = \frac{|\mathcal{R} \cap \tilde{\mathcal{R}}|}{k}$.

Vector indexes are specialized data structures designed to accelerate ANNS by scanning only a subset of data vectors in $\mathbf{V}$ to compute an approximate result set $\tilde{\mathcal{R}}$.
In this paper, we focus on ANNS using HNSW, the most widely adopted vector indexes in practice. 
Specifically, we consider scenarios where an HNSW index has already been constructed and focus on searching the index.

The key parameter in HNSW search is the size of the priority queue that maintains the search results at the base layer, commonly denoted as \ef.
This parameter controls the trade-off between recall and search latency, as illustrated in Figure~\ref{fig:recall_distribution}.
In practice, \ef\ is typically configured based on subjective estimation or through iterative empirical tuning. 
Once selected, the same \ef\ value is applied uniformly across the entire query workload, resulting in the absence of recall guarantees and inefficient search due to under- and over-searching.
To overcome these issues, we study the problem of query-adaptive \ef\ configuration defined below.

\begin{definition}
Given a HNSW index, a query vector $\mathbf{q}$, and a user-specified expected recall, the adaptive \ef\ for $\mathbf{q}$ is defined as the minimum \ef\ value such that the search result of $\mathbf{q}$  over the HNSW index achieves the expected recall.
\end{definition}
 
We propose a distribution-aware, data-driven approach to estimate adaptive \ef\ per query, which can be seamlessly integrated into an existing HNSW index.

\section{Adaptive EF Search}\label{sec:adaptive_ef}
We introduce \texttt{Ada-\ef}, an alternative to standard HNSW search at the base-layer proximity graph. The key distinction is that \texttt{Ada-\ef} does not require a user-specified \ef\ value as input. 
Instead, it dynamically determines an appropriate \ef\ during the search process.

\texttt{Ada-\ef} extends the standard HNSW search described in Section~\ref{sec:pre}, with several key distinctions elaborated below.
First, the input to \texttt{Ada-\ef} does not require a predefined \ef\ value; instead, \ef\ is initially set to $\infty$, allowing the best-first graph traversal to visit any nodes around the entry point $ep$. 
Additionally, the algorithm initializes a sampled distance list $D$ to record the distances between visited nodes and the query vector $\mathbf{q}$. 
The size of $D$ is bounded by $l$ that is the number of nodes reachable within 2-hops from the entry point $ep$ at the base layer.
Once $l$ distances have been collected, the algorithm invokes the \texttt{ESTIMATE-EF} function (Algorithm~\ref{algo:ef_estimate} introduced later), which takes the query vector $\mathbf{q}$, the list $D$, and the user-specified recall $r$ as input, and returns a dynamically estimated \ef\ value. 
Due to space constraint, we provide the pseudocode of the \texttt{Ada-\ef} search algorithm in the appendix.

The \ef-estimator, referred to as \texttt{ESTIMATE-EF}, serves as the core component of the \texttt{Ada-\ef} search algorithm. We provide a high-level overview of its design, with a detailed presentation in the subsequent sections.
The estimator is grounded in a theoretical analysis of the similarity distance distribution. 
Specifically, we demonstrate that given a query vector $\mathbf{q}$ and a dataset $\mathbf{V}$ it is feasible to approximate the distribution of the full distance list $FDL(\mathbf{q}, \mathbf{V})$ without computing each individual distance. The theoretical basis and methodology for estimating this distribution are detailed in Section~\ref{sec:sim_distribution}.
Given the estimated $FDL(\mathbf{q}, \mathbf{V})$ distribution, we can infer the $p$-th percentile distance among all the distances. This inference is lightweight and does not require computing any actual distances.
Then, we compute the number of candidate distances in the list $D$ that fall below this estimated $p$-th quantile threshold.  
For instance, if $p = 0.001$, we count how many of the candidate distances lie within the smallest $0.1\%$ of all distances in $FDL(\mathbf{q}, \mathbf{V})$.
This count is used to characterize the query difficulty. Specifically, for a given query vector $\mathbf{q}$, a high count, particularly one approaching $|D|$, indicates that many candidate neighbors are highly similar to $\mathbf{q}$, and thus a small \ef\ value is likely sufficient to achieve high recall. Conversely, a low count might suggest that a larger \ef\ is required to reach the same recall level.
We leverage this information to estimate an appropriate \ef\ value for each incoming query.
The technical details of this process are presented in Section~\ref{sec:data_driven_ef}.

\section{FDL Distribution}\label{sec:sim_distribution}
We analyze the FDL distribution under commonly used similarity metrics for high-dimensional learned embedding vectors, in particular those derived from Transformer-based models, including inner product, cosine similarity, and cosine distance.
Our analysis begins with the inner product and is then systematically extended to the others.
Our key finding is that, across all considered metrics, $FDL(\mathbf{q}, \mathbf{V})$ approximately follows a Gaussian (normal) distribution when dimensionality becomes large. We derive closed-form expressions for the mean and variance of this distribution.
We then discuss how to estimate the distribution efficiently at runtime.

\begin{figure*}
    \centering
    \includegraphics[width=0.49\linewidth]{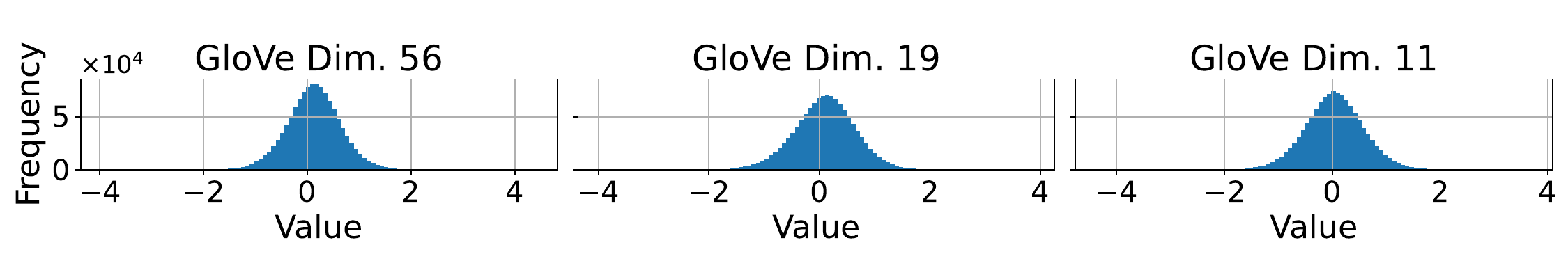}
    \hfill
    \includegraphics[width=0.495\linewidth]{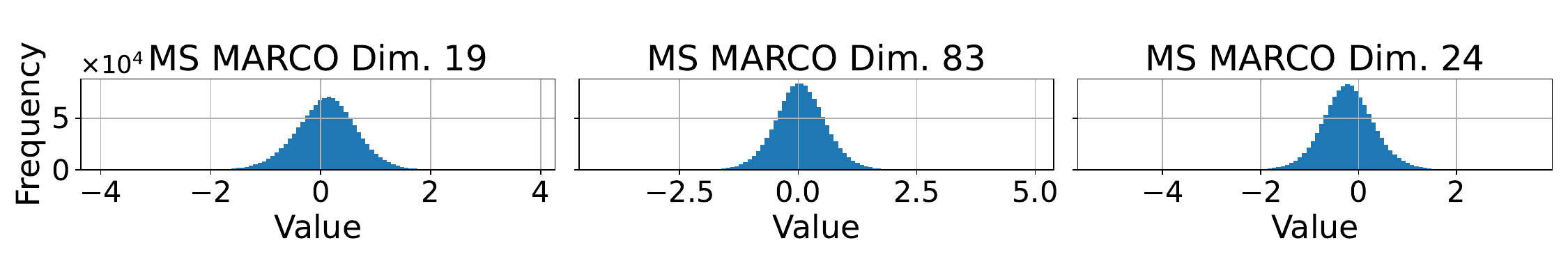}
    \caption{Distribution of individual embedding dimensions that are randomly sampled in GloVe and MS MARCO.}
    \label{fig:dimension_distribution}
\end{figure*}

\subsection{Inner Product}\label{sec:inner_product}
The inner product (IP) between a query vector $\mathbf{q} = (q_1, \dots, q_d)$ and a data vector $\mathbf{v} = (v_1, \dots, v_d)$ is defined as:
$\mathbf{q} \cdot \mathbf{v} = \sum_{i=1}^d q_i v_i$.
Given a dataset $\mathbf{V} = (\mathbf{v}_1, \ldots, \mathbf{v}_n)$, we define the FDL based on inner product as:
$FDL_{IP}(\mathbf{q}, \mathbf{V}) = (\mathbf{q}\cdot \mathbf{v}_1, \ldots, \mathbf{q}\cdot \mathbf{v}_n).$

We begin by introducing a property of $\mathbf{V}$. Under the assumption that this property holds, we then demonstrate that $FDL_{IP}(\mathbf{q}, \mathbf{V})$ 
follows a normal distribution.

\begin{definition}
    Let $\mathbf{V}$ be a vector dataset of $n$ data vectors, each of dimensionality $d$. 
    Let $\mathbf{v}=(v_1, \ldots, v_d)$ be a random data vector of $d$ random variables, where each $v_i$ represents the distribution of the values within the $i$-th column of $\mathbf{V}$.
    If the random variables $(v_1, \ldots, v_d)$ are independent and identically distributed (i.i.d.), then we say the dataset $\mathbf{V}$ is i.i.d across its dimensions.
\end{definition}

\begin{theorem}\label{theorem:inner_product}
 Let $\mathbf{q}=(q_1,\ldots,q_d)$ be a query vector, and let $\mathbf{V}$ be a dataset of data vectors, each of dimensionality $d$, and $\mathbf{V}$ is i.i.d. across its dimensions. 
 Then, the full distance list $FDL_{IP}(\mathbf{q}, \mathbf{V})$ converges in distribution to a normal distribution as $d \to \infty$: $\mathcal{N}\left(\mu_{IP}, \sigma^2_{IP} \right)$
 with the mean $\mu_{IP}$ and the variance $\sigma^2_{IP}$: 
    $\mu_{IP} = \sum_{i=1}^{d} q_i \text{E}[v_i];\ \sigma^2_{IP} = \sum_{i=1}^{d} q_i^2 \text{Var}(v_i)$.
\end{theorem}

\begin{proof}
Consider the inner product $\mathbf{q} \cdot \mathbf{v} = \sum_{i=1}^d q_i v_i$.
The query vector $\mathbf{q}$ is given during similarity search online, such that each $q_i$ is a constant at query time.
Then, the inner product is a sum of i.i.d random variables, where each is given by $q_i v_i$. According to the Central Limit Theorem, the sum follows a normal distribution  as $d$ becomes large.
For the mean and variance, we have: $\text{E}[\sum_{i=1}^d q_i v_i] =  \sum_{i=1}^dq_i \text{E}[v_i]$, and $ \text{Var}(\sum_{i=1}^d q_i v_i) =  \sum_{i=1}^d q^2_i\text{Var}( v_i)$.
\end{proof}

Theorem~\ref{theorem:inner_product} shows that $FDL_{IP}(\mathbf{q}, \mathbf{V})$ follows a normal distribution under the assumption that the components of $\mathbf{V}$ are i.i.d. across dimensions.
This assumption facilitates the application of the classical Central Limit Theorem in the proof of Theorem~\ref{theorem:inner_product}.
We provide empirical evidence that the identical distribution assumption is approximately valid in real-world datasets. As shown in Figure~\ref{fig:dimension_distribution}, the dimensions of GloVe and MS MARCO embeddings exhibit distributions that are approximately normal, with minor variations in mean and variance.
Similar distributional patterns have been empirically observed in real-world datasets \cite{arora-etal-2016-latent,mu2018allbutthetop,ethayarajh-2019-contextual}, and have also been explicitly encouraged through model design in recent work \cite{li-etal-2020-sentence}.
Therefore, we adopt the identical distribution assumption in this work.
For scenarios where the distributional assumption may not hold, we provide a detailed discussion in the appendix, leveraging Lindeberg’s condition~\cite{Klenke2020}.

\textit{Relaxing independent condition.}
Both the classical CLT and its relaxed form under Lindeberg’s condition require that the random variables involved are mutually independent, \textit{i.e.}, each pair $q_i v_i$ and $q_j v_j$ must be independent for $i \ne j$. 
In practice, however, data vectors are typically learned embeddings, where each dimension represents a latent feature. While these features are ideally independent,  in reality, dimensions may exhibit correlations.
Such correlations do not affect $\mu_{IP}$, but they do influence $\sigma^2_{IP}$. To account for these correlations, we adjust the variance as follows:
$$
\text{Var}\left(\sum_{i=1}^d q_i v_i\right) = \sum_{i=1}^d q_i^2 \text{Var}(v_i) + 2 \sum_{1 \leq i < j \leq d} q_i q_j \text{Cov}(v_i, v_j).
$$
Therefore, we model $FDL_{IP}(\mathbf{q}, \mathbf{V})$ as a normal distribution with the adjusted variance: 
\begin{equation}\label{equ:inner_product_distribution}
  \mathcal{N}\left( \mu_{IP}, \sigma^2_{IP} + \Delta_{IP} \right),   \Delta_{IP} = 2 \sum_{1 \leq i < j \leq d} q_i q_j \text{Cov}(X_i, X_j).
\end{equation}

\subsection{Cosine Similarity}\label{sec:cosine_similarity}
The cosine similarity (CS) between $\mathbf{q}$ and $\mathbf{v}$ is 
$CS(\mathbf{q}, \mathbf{v}) = \frac{\mathbf{q}\cdot \mathbf{v}} {\|\mathbf{q}\|\|\mathbf{v}\|}$.
Given a dataset $\mathbf{V} = (\mathbf{v}_1, \ldots, \mathbf{v}_n)$,  the full distance list based on CS is
$
FDL_{CS}(\mathbf{q}, \mathbf{V}) = (CS(\mathbf{q}, \mathbf{v_1}), \ldots, CS(\mathbf{q}, \mathbf{v_n})).
$

We analyze the distribution of $FDL_{CS}(\mathbf{q}, \mathbf{V})$ based on Theorem \ref{theorem:inner_product}.
Cosine similarity can be interpreted  as the normalized  inner product, \textit{i.e.}, $CS(\mathbf{q}, \mathbf{v}) = \hat{\mathbf{q}}\cdot \hat{\mathbf{v}}$, where $\hat{\mathbf{q}}=\frac{\mathbf{q}}{\|\mathbf{q}\|}$ and $\hat{\mathbf{v}}=\frac{\mathbf{v}}{\|\mathbf{v}\|}$.
Let $\hat{\mathbf{q}} = (\hat{q_1}, ...,\hat{q_d})$  and $\hat{\mathbf{v}} = (\hat{v_1}, ...,\hat{v_d})$, where $\hat{q_i}=\frac{q_i}{\|\mathbf{q}\|}$ and $\hat{v_i}=\frac{v_i}{\|\mathbf{v}\|}$.

According to Theorem~\ref{theorem:inner_product}, if the random variables $(\hat{v_1}, \ldots, \hat{v_d})$ representing the normalized column-wise distribution of $\mathbf{V}$ are i.i.d. with finite means and variances, then as dimension $d$ becomes large,  the distribution of $FDL_{CS}(\mathbf{q}, \mathbf{V})$ converges to a normal distribution as: 
$\mathcal{N}\left( \mu_{CS},\sigma^2_{CS}\right)$, with the following mean and variance: 
$$ \mu_{CS} = \sum_{i=1}^{d} \hat{q_i} \text{E}[\hat{v_i}]; \sigma^2_{CS} = \sum_{i=1}^{d} (\hat{q_i})^2 \text{Var}(\hat{v_i}).$$

Following the same analysis as in the case of IP, \textit{i.e.}, by relaxing the i.i.d. conditions,  $FDL_{CS}(\mathbf{q}, \mathbf{V})$  approximately follows: 
\begin{equation}\label{equ:consine_similarity}
 \mathcal{N}\left( \mu_{CS},\sigma^2_{CS} + \Delta_{CS}\right),    \Delta_{CS} = 2 \sum_{1 \leq i < j \leq d} \hat{q_i} \hat{q_j} \text{Cov}(\hat{v_i}, \hat{v_j}).
\end{equation}

\subsection{Cosine Distance}\label{sec:cosine_distance}
The cosine distance (CD) between $\mathbf{q}$ and $\mathbf{v}$ is defined as $CD(\mathbf{q}, \mathbf{v}) = 1 - CS(\mathbf{q}, \mathbf{v})$. 
Accordingly, the full distance list based on CD is defined as:
$
FDL_{CD}(\mathbf{q}, \mathbf{V}) = (1-CS(\mathbf{q}, \mathbf{v_1}), \ldots, 1-CS(\mathbf{q}, \mathbf{v_n})).
$
Since $FDL_{CS}(\mathbf{q}, \mathbf{V})$ 
is approximately normally distributed, and since an affine transformation of a normally distributed variable results in another normal distribution (with mean shifted and variance preserved), then  $FDL_{CD}(\mathbf{q}, \mathbf{V})$ approximately follows: 
\begin{equation}\label{equ:cosine_distance}
    \mathcal{N}(1-\mu_{CS}, \sigma^2_{CS} + \Delta_{CS}).
\end{equation}

\subsection{Computing FDL Distribution}\label{sec:fdl_computation}
Given a query vector $\mathbf{q}$, we aim to efficiently estimate the FDL distribution online by precomputing certain statistics of the dataset $\mathbf{V}$ offline.
The core idea is that for the mean and variance of the normal distribution that are formally expressed in Equations (\ref{equ:inner_product_distribution}), (\ref{equ:consine_similarity}), and (\ref{equ:cosine_distance}), the components depend solely on $\mathbf{V}$ and hence can be computed in advance.
At query time, these precomputed statistics are then combined with the query vector $\mathbf{q}$ to enable on-the-fly estimation of the FDL distribution. To illustrate the offline computation of $\mathbf{V}$'s statistics, we focus on the case of the inner product.

\textit{Offline computation}.
According to Equation (\ref{equ:inner_product_distribution}), the estimation of $FDL_{IP}$ requires the mean $\mu_{IP}$ and the combined variance term $\sigma^2_{IP} + \Delta_{IP}$. To facilitate efficient computation during query time, we precompute the following statistics of the dataset $\mathbf{V}$ in the offline phase. Let $v_i$ denote the $i$-th column of $\mathbf{V}$.
\begin{itemize} 
\item Mean Vector of $\mathbf{V}$: A vector of dimension $1 \times d$, where the $i$-th entry corresponds to $\text{E}[v_i]$, \textit{i.e.}, the mean of the $i$-th column of $\mathbf{V}$.
\item Covariance Matrix of $\mathbf{V}$: A matrix of dimension $d \times d$, where the $(i,j)$-th entry represents $\text{Cov}(v_i, v_j)$, \textit{i.e.}, the covariance between the $i$-th and $j$-th columns of $\mathbf{V}$. The diagonal entries of this matrix correspond to $\text{Var}(v_i)$, \textit{i.e.}, the variance of the $i$-th column.
\end{itemize}
Figure~\ref{fig:overview} illustrates the mean vector and the covariance matrix.

The mean vector is computed by scanning each column of $\mathbf{V}$ and calculating its mean.
To compute the covariance matrix, we proceed as follows.
Let $\mathbf{M}$ be a matrix where each row is a copy of the mean vector, \textit{i.e.}, $\mathbf{M}$ replicates the mean vector across all $n$ rows. 
The covariance matrix $\mathbf{\Sigma}$ of $\mathbf{V}$ is then given by:
$
\mathbf{\Sigma} = \frac{1}{n - 1}{\mathbf{(\mathbf{V} - \mathbf{M})}^\top\mathbf{(\mathbf{V} - \mathbf{M})}}.
$

\textit{Online computation}.
Given a query vector $\mathbf{q}$, in order to estimate $FDL_{IP}(\mathbf{q}, \mathbf{V})$, we use the mean vector and covariance matrix of the dataset $\mathbf{V}$ to compute $\mu_{IP}$ and $\sigma^2_{IP} + \Delta_{IP}$, as defined in Theorem~\ref{theorem:inner_product} and Equation~\ref{equ:inner_product_distribution}.
We compute $\mu_{IP}$ as the dot product between the query $\mathbf{q}$ and the mean of $\mathbf{V}$, and compute $\sigma^2_{IP} + \Delta_{IP}$ as $\mathbf{q} \mathbf{\Sigma} \mathbf{q}^\top$.

Although focused on the inner product, the approach extends to cosine similarity and distance. Both offline and online computations can be formulated as matrix operations, enabling efficient SIMD and parallel acceleration.

\section{Query-Aware EF Estimation}\label{sec:data_driven_ef}
We present the design of the \ef-estimator, \textit{i.e.}, the ESTIMATE-EF function.
The primary objective of the \ef-estimator is to \textit{differentiate between query vectors that can achieve the declarative target recall with a small \ef\ value and those that require a large \ef\ value to obtain the recall}.
To accomplish this, we utilize the estimated distribution of the FDL to design a query scoring model, \textit{i.e.}, a mechanism that assigns a score to each input query vector. 
We estimate \ef\ for each input query based on its query score.
To achieve this, we perform offline sampling of data vectors from the dataset $\mathbf{V}$ to build an \ef-estimation table. This table stores pairs of \ef\ values and their corresponding estimated recall values, indexed by query scores.
During the online search phase, the query scoring mechanism is applied to compute the score for the incoming query, and then the appropriate \ef\ value is retrieved from the \ef-estimation table based on this score. 
In essence, the query scoring mechanism distinguishes queries requiring different \ef\ values to meet the target recall, while the \ef-estimation table provides the actual \ef\ values needed to achieve it.
We present the query scoring mechanism in \S~\ref{sec:query_score}, the \ef-estimation table in \S~\ref{sec:ef_table}, and discuss how to perform incremental insertion and deletion in \S~\ref{sec:incremental_update}.

\subsection{Query Scoring Model}\label{sec:query_score}
The adaptive-\ef\ search visits a fixed number of nodes at the base layer of the HNSW index and collects a distance list $D$, which contains the distances between the query vector $\mathbf{q}$ and the visited nodes, as described in \S~\ref{sec:adaptive_ef}.
The query scoring model analyzes the list $D$ to assess the difficulty of computing the approximate nearest neighbors for $\mathbf{q}$. The design of the scoring model is guided by the following principle: \textit{if the majority of distances in $D$ fall among the smallest distances according to the distribution of $FDL(\mathbf{q}, \mathbf{V})$, 
then $\mathbf{q}$ likely requires only a small \ef\ value to achieve high recall. Otherwise, $\mathbf{q}$ may require a larger \ef\ value for specified recall.}

The query scoring model discretizes the normal distribution (parameterized by the estimated mean and variance of the FDL) into a fixed number of consecutive  quantile-based bins, each covering an equal portion of the FDL referred to as $\delta$. These bins are denoted as $(b_1, \dots, b_m)$, where $m$ is the total number of bins. Each bin is used to count the number of distances that fall within the quantile range $\delta$ of the distribution. For instance, if $\delta=0.001$, $b_1$ corresponds to distances smaller than the $0.001$ quantile of the FDL distribution, while $b_2$ captures distances between the $0.001$ and $0.002$ quantiles, and so on for the remaining bins.

The bins  $(b_1, \dots, b_m)$ are formally defined by threshold values derived from the quantiles of the FDL distribution. 
The threshold $\theta_i $ for separating $b_{i}$ and $b_{i+1}$  is computed as:
\begin{equation}\label{equ:bin_thresh}
\theta_i = \mu +  \sigma  \Phi^{-1}(\delta i), \text{ for } i = 1, \dots, m,    
\end{equation}
where $\Phi^{-1}(\cdot)$ denotes the inverse cumulative distribution function, 
$\delta$ multiplied by $i$ determines the quantile, and 
$\mu$ and $\sigma$ are the mean and the standard deviation of the normal distribution, respectively, \textit{e.g.}, for cosine distance, 
$\mu = 1-\mu_{CS}$ and $\sigma = \sqrt{\sigma^2_{CS} + \Delta_{CS}}$.
Each bin $b_i$ is then associated with a weight $w_i$ to indicate the importance of the bin, which is defined via an exponential decay function:
$w_i = 100 e^{-i+1},\text{ for } i = 1, \dots, m$.

Given a list $D$ containing the  distances  between the query vector and the visited nodes, we count how many falling into each bin:
\begin{equation}\label{equ:bin_counts}
c_i = \left| \left\{ \theta_{i-1} < d_j \leq \theta_i \mid  d_j \in D \right\} \right|.    
\end{equation}
The query score $s(\mathbf{q})$ is a weighted, normalized sum of bin counts:
\begin{equation}\label{equ:query_score}
s(\mathbf{q}) = \sum_{i=1}^{m} \left( w_i \frac{c_i}{|D|}  \right).   
\end{equation}
This query score serves as a proxy for query difficulty: a higher score indicates that more distances in $D$ fall within the favorable (low) quantile regions of the FDL distribution. Thus, a small \ef\ would be sufficient for high-score queries to achieve high recall. We provide an example to  explain the query scoring model in detail in the appendix.

We note that several query-difficulty metrics have been proposed, including LID~\cite{10.1145/2783258.2783405}, RC~\cite{10.5555/3042573.3042582}, Expansion~\cite{10.5555/3039686.3039702}, and Steiner-Hardness~\cite{10.14778/3704965.3704974}, which help characterize query hardness, but are impractical for online search. They require exact neighbor information or fine-grained distance statistics obtained through additional nearest-neighbor or multi-radius searches. In contrast, our FDL-based approach is lightweight and can estimate query difficulty directly during the online search process.

\subsection{Estimate EF}\label{sec:ef_table}
The query scoring model serves to quantitatively differentiate between queries that require a small \ef\ value and those that require a larger one. However, the model alone does not directly provide a concrete \ef\ value for a given query at runtime. To bridge this gap, we construct an \ef-estimation table offline, which records the relationship between query scores, \ef\ values, and the corresponding recall levels. At query time, we compute the score for each incoming query vector and use it to retrieve from the \ef-estimation table an \ef\ value that is expected to achieve the target recall. 

\textit{Offline \ef-estimation table construction.}
A key challenge in constructing the \ef-estimation table is the absence of query vectors during the offline phase. To address this, we uniformly sample a subset of data vectors (\textit{e.g.}, $200$ vectors) from the dataset $\mathbf{V}$ and use them as proxy query vectors. For each sampled vector, we compute its query score and group the vectors according to their scores. 
To reduce the granularity and number of score groups, the floating-point query scores are casted into integer ones.
For each score group, we evaluate recall using the original HNSW search under progressively increasing \ef\ values. If the observed average recall for a group falls below the user-defined target, a larger \ef\ value is selected and tested. This adaptive probing continues until the target recall is achieved or a predefined upper bound on \ef\ is reached. 
The resulting mappings between query score, \ef, and recall are stored in the \ef-estimation table. 

Figure~\ref{fig:overview} shows an example of \ef-estimation table.
Specifically, the table consists of two columns: \texttt{Score} and \texttt{EF\_Recall}. The \texttt{Score} column represents the query score group, while the \texttt{EF\_Recall} column contains a list of $(\texttt{EF}, \texttt{Recall})$ pairs, sorted in ascending order of \ef. Each entry in this list captures the recall achieved for a specific \ef\ value within the corresponding score group.

\textit{Weighted Average \ef}.
We compute the weighted average \ef\ (WAE) as a summary of \ef-estimation table, weighted by the number of queries in each score group (\textit{i.e.}, an integer query score): $\text{WAE} = \frac{1}{G} \sum_{i} g_i \cdot \text{ef}_i$, where $g_i$ is the number of queries in score group $i$, $\text{ef}_i$ is the smallest \textit{ef} that achieves the target recall for that group, and $G$ is the total number of sampled data vectors. This metric captures the average search effort required across queries of varying difficulty to reach the target recall and serves as an initialization point for \ef\ estimation.

\begin{algorithm} [t]
\caption{ESTIMATE-EF$(\mathbf{q}, D, r)$}\label{algo:ef_estimate}
\begin{flushleft}
\textbf{Input:} query vector $\mathbf{q}$, distance list $D$, expected recall $r$ \\
\textbf{Output:} estimated exploration factor \ef
\end{flushleft}
\begin{algorithmic}[1]
\State $\mu \gets \text{compute the estimated mean of }FDL(\mathbf{q}, \mathbf{V})$ \Comment{$\S$ \ref{sec:fdl_computation}}
\State $\sigma \gets \text{compute the estimated std of }FDL(\mathbf{q}, \mathbf{V})$ \Comment{$\S$ \ref{sec:fdl_computation}}
\State $(b_1,\ldots,b_m)\gets \text{compute the bins}$
\Comment{Eq. (\ref{equ:bin_thresh})}
\State $(c_1,\ldots, c_m)\gets$ \text{compute the bin counts}
\Comment{Eq. (\ref{equ:bin_counts})}
\State $s \gets \text{compute the query score}$
\Comment{Eq. (\ref{equ:query_score})}
\State $row \gets \text{get the row in \ef-estimation table with the score}$
\State $\ef \gets \text{get the largest EF in the } row$
\For{each (EF, Recall) in the $row$}
    \If{$\text{Recall} \geq r$}
        \State $\ef \gets $max(EF, WAE)
        \State \textbf{break}
    \EndIf
\EndFor
\State \textbf{return} \ef
\end{algorithmic}
\end{algorithm}

\textit{Online \ef\ estimation}.
Algorithm~\ref{algo:ef_estimate} presents the end-to-end procedure for estimating \ef\ for a given query vector $\mathbf{q}$ to achieve a target recall $r$. The process begins by estimating the parameters of the distribution $FDL(\mathbf{q}, \mathbf{V})$, specifically the mean $\mu$ and standard deviation $\sigma$ (lines 1–2). These computations are grounded in the theoretical foundations described in \S~\ref{sec:fdl_computation}, including Equations~(\ref{equ:inner_product_distribution}), (\ref{equ:consine_similarity}), and (\ref{equ:cosine_distance}). For efficient on-the-fly computation, the algorithm utilizes the mean vector and covariance matrix of the dataset $\mathbf{V}$, both of which are precomputed during the offline phase.
Lines 3–5 focus on computing the query score based on the input distance list $D$ and the estimated parameters $\mu$ and $\sigma$. The algorithm first constructs a set of quantile-based bins $(b_1, \ldots, b_m)$ (line 3) and then determines the bin counts $(c_1, \ldots, c_m)$ by analyzing how the distances in $D$ fall into these bins (line 4). These counts are subsequently aggregated to compute a scalar query score $s$ (line 5), which serves as a proxy for the difficulty of the query.
After computing the score, the algorithm retrieves the corresponding row from the precomputed \ef-estimation table, which associates each score group with a list of $(EF, Recall)$ pairs sorted in ascending order of \ef\ (lines 6–7). 
It then searches this list to find the smallest \ef\ value that satisfies the target recall requirement (lines 8–11). If the selected value is smaller than WAE, it is increased to WAE and returned as the estimated \ef; otherwise, if no such value exists, the algorithm returns the largest  \ef\ available for the specific score.

We note the query-aware \ef\ estimation framework is built on the FDL distribution. We have derived FDL distributions for Inner Product, Cosine Similarity, and Cosine Distance, while the Euclidean (L2) case remains open due to the squared terms in its formulation. Once established, the framework could be applied to L2 as well.

\subsection{Incremental Updates}\label{sec:incremental_update}
In practice, vector indexes such as HNSW are expected to support dynamic vector insertion and deletion, \textit{i.e.}, the ability to insert into and delete from the index.
We outline how our approach accommodates such updates.
Offline computation in our method consists of three main components:
(i) dataset-level statistics, including the mean vector and covariance matrix (see Section~\ref{sec:fdl_computation});
(ii) a sampled subset of data vectors and computing their  ground truth;
(iii) the \ef-estimation table.
To support vector updates, we observe that components (i) and (ii) can be incrementally updated. Once these are updated, component (iii), \textit{i.e.}, the \ef-estimation table can be recomputed with respect to the updated HNSW index.
Updating the ground truth of sampled vectors is straightforward.
For insertions, we compute distances to newly added vectors and update the Top-$k$ neighbors.
For deletions, we incrementally remove deleted vectors from the ground truth and refresh the Top-$k$ neighbors accordingly.
We now briefly describe how to incrementally update the dataset statistics, beginning with the case of insertions.
Let $\mathbf{V}$ denote the original dataset of $n$ vectors, and let $\mathbf{V}'$ represent the set of $n'$ newly inserted vectors.
The updated dataset is then given by $\mathbf{V}'' = \mathbf{V} \cup \mathbf{V}'$, containing $n'' = n + n'$ vectors.
We first compute the mean vector $\mathbf{M}'$ and covariance matrix $\mathbf{\Sigma}'$ of $\mathbf{V}'$, and then merge them with the existing $\mathbf{M}$ and $\mathbf{\Sigma}$ of $\mathbf{V}$ as follows:
$
\mathbf{M}''= \frac{n\mathbf{M} + n'\mathbf{M}'}{n''}$, and     
$
\mathbf{\Sigma}'' = \frac{1}{n '' - 1} \left[ (n - 1)\mathbf{\Sigma} + (n' - 1)\mathbf{\Sigma}'  + \frac{n n'}{ n''} (\mathbf{M} - \mathbf{M}')^\top(\mathbf{M} - \mathbf{M}') \right].$
Consider deleting $\mathbf{V}'$ from $\mathbf{V}''$ to obtain $\mathbf{M}$ and $\mathbf{\Sigma}$ of $\mathbf{V}$: $\mathbf{M}= \frac{n''\mathbf{M}'' - n'\mathbf{M}'}{n}$, and      
$
\mathbf{\Sigma} = \frac{1}{n  - 1} \left[ (n'' - 1)\mathbf{\Sigma}'' - (n' - 1)\mathbf{\Sigma}'  - \frac{n 'n''}{n} (\mathbf{M}'' - \mathbf{M}')^\top(\mathbf{M}'' - \mathbf{M}') \right].$
We note that such online updates to statistical summaries are widely adopted in practice, \textit{e.g.}, in Spark MLlib~\cite{JMLR:v17:15-237}.

\section{Experimental Evaluation}
We evaluate online effectiveness (\S~\ref{sec:exp_online}) and offline computation (\S~\ref{sec:exp_offline}) using six real-world and two synthetic datasets against five baselines, including state-of-the-art adaptive methods.
We report sensitivity analysis  (\S~\ref{sec:sensitivity}), update performance (\S~\ref{sec:exp_insertion}), and ablation study (\S~\ref{sec:abaltion_study}). 

\subsection{Experiment Setup}
\textit{Real-world datasets.}
We use six real-world datasets, summarized in Table~\ref{table:datasets}.
From the early ANNS benchmark~\cite{anns_bench}, we include GloVe-100 (word)~\cite{pennington-etal-2014-glove} and DeepImage (image)~\cite{7780595}. 
To reflect more recent trends, we incorporate several modern datasets that use advanced embedding techniques.
Specifically, we include the MS MARCO V1 Passage Ranking dataset~\cite{bajaj2018msmarcohumangenerated}, a widely used benchmark for passage reranking, which is embedded using OpenAI's ada2~\cite{openai-ada2}. This embedding captures semantic relationships between passages and queries and is particularly suited for dense retrieval tasks. This version is sourced from the Anserini framework~\cite{10.1145/3077136.3080721,msmarco_openai}.
We also use MS MARCO V2.1, drawn from the TREC-RAG 2024 corpus~\cite{trec-rag}, embedded using Cohere’s Embed English V3. This model produces dense semantic embeddings of passages, optimized for retrieval-augmented generation (RAG) settings and downstream QA applications.
Finally, we evaluate our approach in a multi-modal retrieval setting using the LAION dataset~\cite{laion400}, which consists of text-image pairs embedded with OpenAI’s CLIP~\cite{pmlr-v139-radford21a}. It aligns visual and textual modalities in a shared embedding space, enabling semantic similarity search across modalities. We consider two retrieval tasks: image-to-image retrieval and text-to-image retrieval, referred to as Laion-I2I and Laion-T2I, respectively. Both tasks utilize the same set of image embeddings as the database, but differ in the query modality: Laion-I2I uses image queries, while Laion-T2I uses text queries.
Each dataset comes with a set of query vectors, except for LAION for which we randomly sample 10000 image embedding vectors and 10000 text embedding vectors as query vectors, respectively.

\begin{table}[t]
    \centering
    \caption{Overview of real-world datasets.}
    \label{table:datasets}
    \resizebox{0.9\linewidth}{!}{    \begin{tabular}{|c|r|r|r|}
    \hline
        \textbf{Dataset}&\textbf{Data Size}&\textbf{Dimension}&\textbf{Query Size}\\ \hline\hline         
         GloVe-100~\cite{pennington-etal-2014-glove}&1,183,514 & 100& 10,000 \\ \hline
         DeepImage~\cite{7780595}   &9,990,000 & 96 & 10,000 \\ \hline
         MS MARCO V1~\cite{msmarco_openai} & 8,841,823 &1536 &6980 \\ \hline
         MS MARCO V2.1~\cite{msmarco_cohere} & 18,380,565 & 1024 & 1677\\ \hline     
         Laion-I2I~\cite{laion400} & 30,646,115 & 512 & 10,000 \\ \hline
         Laion-T2I~\cite{laion400} & 30,646,115 & 512 & 10,000\\ \hline         
    \end{tabular}}
\end{table}

\textit{Synthetic datasets}. To evaluate the effectiveness of our adaptive approach, we include two synthetic datasets, Uniform Cluster and Zipfian Cluster. Both datasets contain 10 million vectors in 100 dimensions, organized into 5000 Gaussian distributed clusters. The primary difference between them lies in the distribution of cluster sizes. In Uniform Cluster, all clusters have the same size, with each containing exactly 2000 vectors. In contrast, Zipfian Cluster follows a Zipfian distribution with an exponent of $1$, resulting in highly skewed cluster sizes. The largest cluster contains more than one million vectors, and the sizes decrease exponentially. For each dataset, we sample 10000 vectors to serve as query vectors for evaluating retrieval performance.

\textit{Implementation}.
We integrate our approach into \texttt{HNSWlib}~\cite{hnswlib}, the official implementation of HNSW developed by its original authors. We use Eigen~\cite{eigen} for efficient matrix operations involved in dataset statistics computation.
Our method, referred to as \textbf{Adaptive-\ef} (\textbf{Ada-\ef}), extends the HNSW index without modification.
Our source code\footnote{\textbf{GitHub Repository:} \url{https://github.com/chaozhang-cs/hnsw-ada-ef}} is publicly available online.

\textit{Baselines}.
We include the following approaches as baselines:
Learned Adaptive Early Termination (\textbf{\texttt{LAET}})~\cite{10.1145/3318464.3380600},
\textbf{\texttt{DARTH}}~\cite{10.1145/3749160},
Patience in Proximity (\textbf{\texttt{PiP}})~\cite{10.1007/978-3-031-88714-7_39},
\textbf{\texttt{HNSW (HNSWlib)}}~\cite{hnswlib}, and
\textbf{\texttt{HNSW (FAISS)}}~\cite{faiss}.
Both \texttt{LAET} and \texttt{DARTH} are learning-based adaptive vector search methods built on Gradient Boosting Decision Trees (GBDTs)~\cite{4a848dd1-54e3-3c3c-83c3-04977ded2e71}.
They share similar feature designs and both require offline preparation of training data and model training.
The key difference is that \texttt{DARTH} periodically predicts the current recall during search to determine whether the declarative target recall is met, whereas \texttt{LAET} performs a single prediction to estimate the required number of distance computations.
We use the open-source implementation of \texttt{DARTH}, which also provides a declarative recall variant of \texttt{LAET}.
\texttt{PiP} is an early-termination method that requires no offline computation, which terminates the search if the current results have not been updated for a fixed amount of consecutive steps. We implement it on top of \texttt{HNSWlib}.
Finally, we include two widely used HNSW implementations, \texttt{HNSW (HNSWlib)} and \texttt{HNSW (FAISS)}, to evaluate the benefits of adaptive approaches.
We note that \texttt{DARTH} represents the state of the art among these baselines~\cite{10.1145/3749160}.

\textit{Parameters}.
For the construction of HNSW indexes on each dataset, we set $M=16$ (maximum number of outgoing degree) and $efConstruction = 500$ (search parameter during indexing).
For Top-$k$ ANNS, we follow standard settings in existing benchmarks. Specifically, we set $k = 1000$ for both the MS MARCO and LAION datasets, following recommendations from the MS MARCO benchmark~\cite{msmarco_benchmar} and the SISAP Indexing Challenge~\cite{laion_indexing_challenge}. 
This setting mirrors typical retrieval pipelines in modern applications, such as RAG systems~\cite{gao2023retrieval}, where a recall stage first retrieves a broad set of candidate items via Top-$k$ ANNS, followed by a reranking stage that refines these candidates for improved relevance.
For the remaining datasets from the ANNS benchmark suite~\cite{anns_benchmark}, we adopt the commonly used setting of $k = 100$. 
The influence of varying $k$ is analyzed in Section~\ref{sec:sensitivity}.
We use cosine distance in all experiments, as it is a widely adopted choice for ANNS over high-dimensional Transformer-based learned embeddings.
For \texttt{DARTH} and \texttt{LAET}, we train the models using 10000 learn vectors, following the recommended configuration~\cite{10.1145/3749160}.
The prediction intervals in \texttt{DARTH} and the number of distance computations at which \texttt{LAET} collects features for training or predicting follow the same setup as in \texttt{DARTH}, \textit{i.e.}, extracting the statistics of the training data.
For \texttt{PiP}, we set the saturation threshold to $0.95$, as recommended in the original paper, and the patience parameter to $30$. Since the configuration of patience was not explicitly specified in \texttt{PiP}~\cite{10.1007/978-3-031-88714-7_39}, we adopt a conservative value to ensure stable convergence.
We note that Ada-\ef\ requires no parameter setting, except in the ablation studies in \S~\ref{sec:abaltion_study}, which examine its design choices.

\begin{figure*}
 \centering      
 \includegraphics[width=\linewidth]{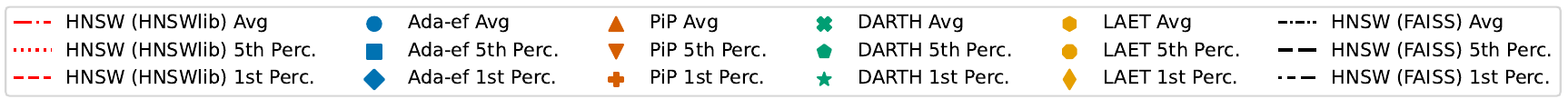}\\
  \begin{minipage}{0.245\textwidth}
    \centering
    \includegraphics[width=\linewidth]{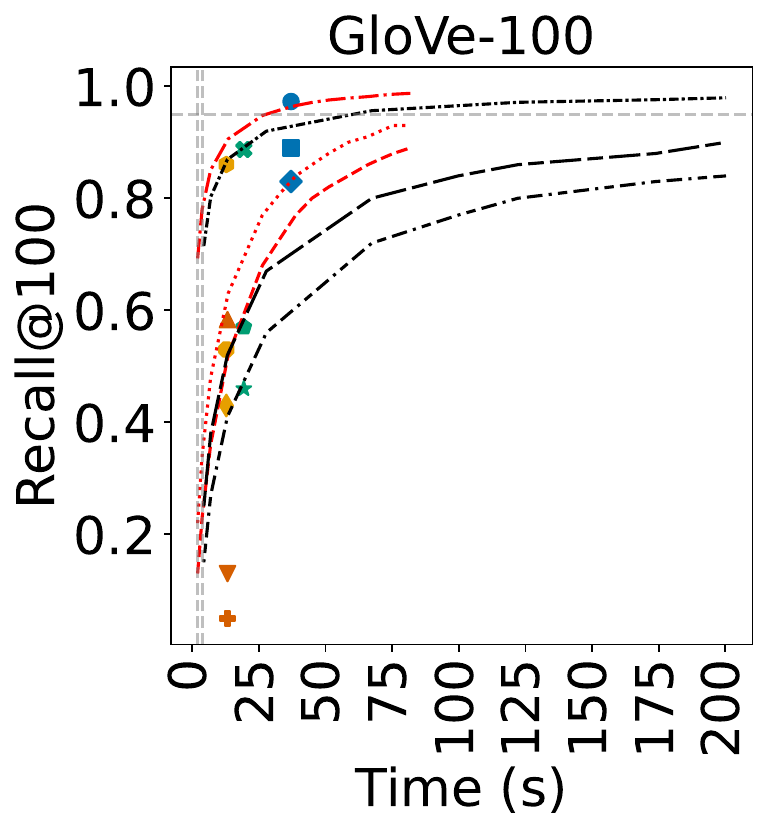}
  \end{minipage}%
  \begin{minipage}{0.252\textwidth}
    \centering
    \includegraphics[width=\linewidth]{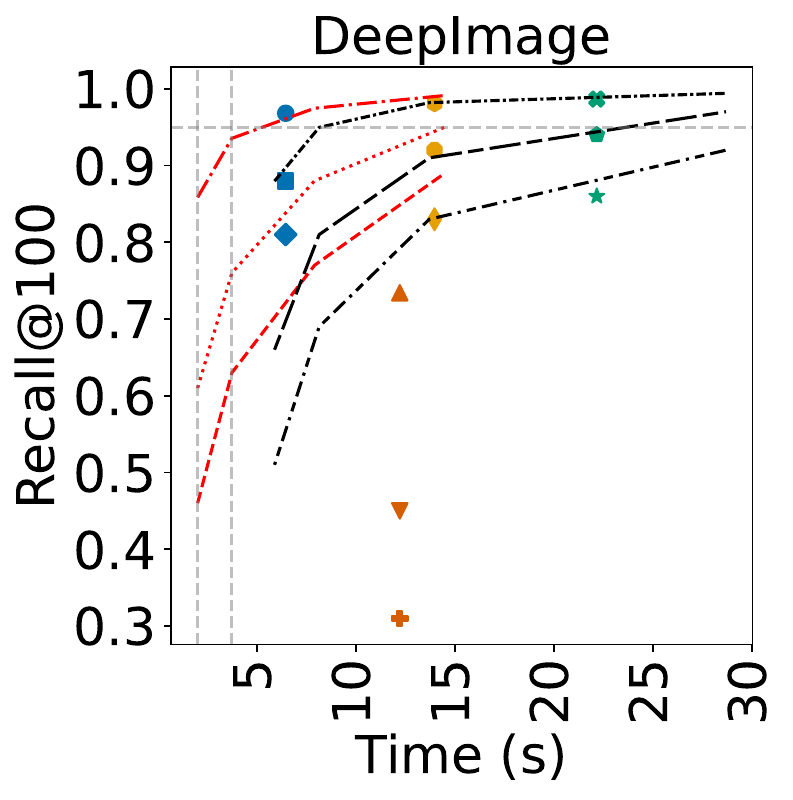}
  \end{minipage}%
  \begin{minipage}{0.245\textwidth}
    \centering
        \vspace*{+0.25cm}
    \includegraphics[width=\linewidth]{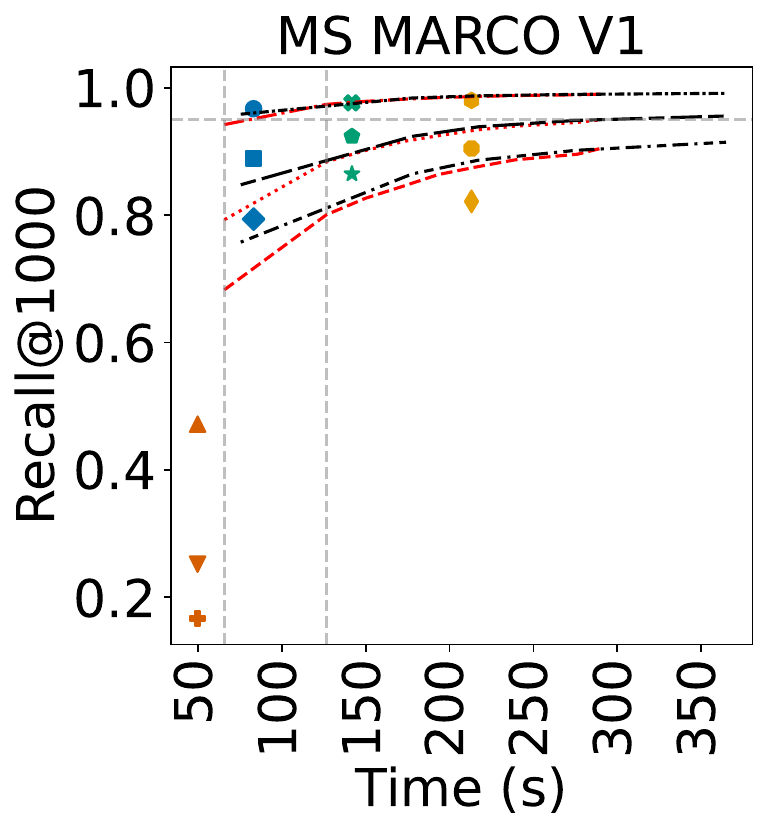}
  \end{minipage}%
  \begin{minipage}{0.25\textwidth}
    \centering
            \vspace*{+0.15cm}
    \includegraphics[width=\linewidth]{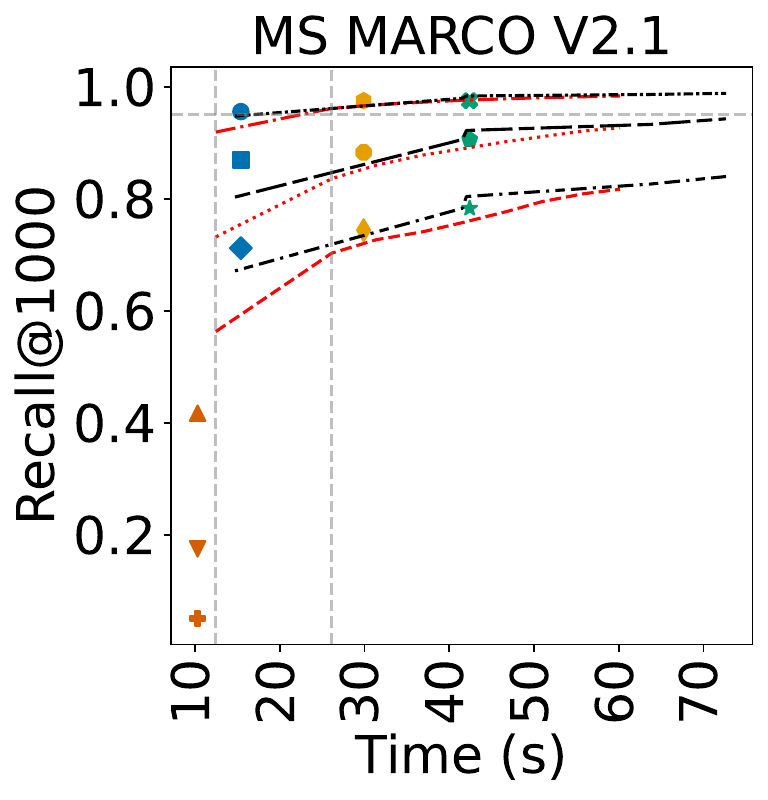}
  \end{minipage}%
  
    \begin{minipage}{0.245\textwidth}
    \centering
    \includegraphics[width=\linewidth]{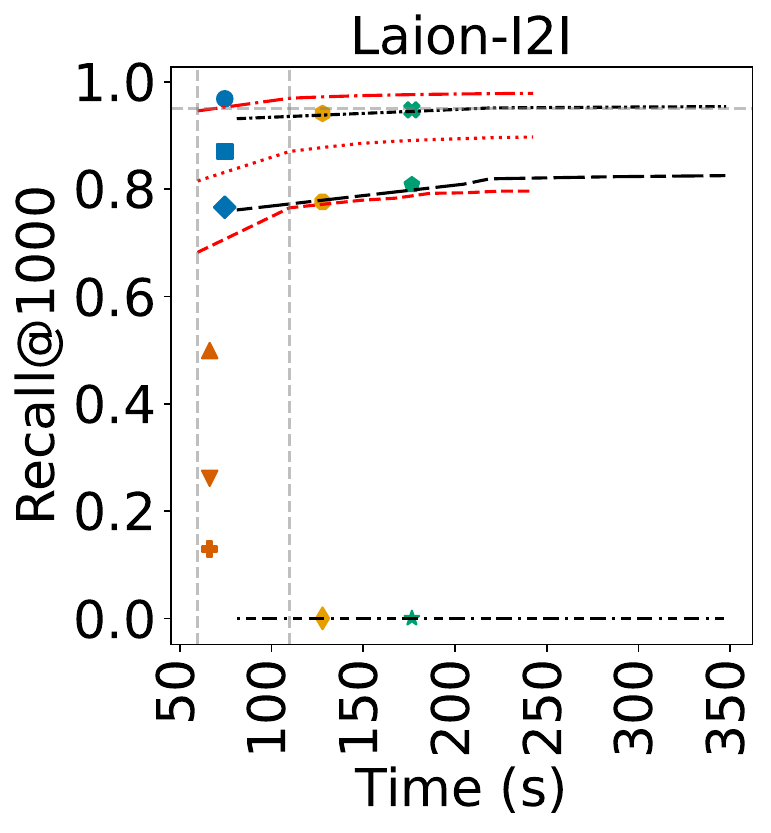}
  \end{minipage}%
  \begin{minipage}{0.245\textwidth}
    \centering
    \includegraphics[width=\linewidth]{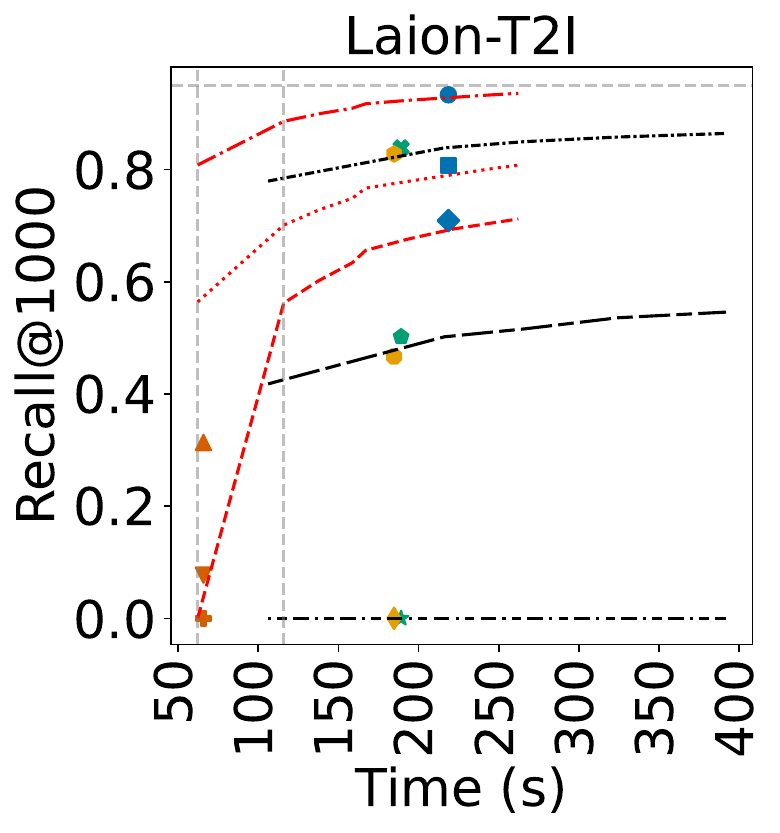}
  \end{minipage}
\begin{minipage}{0.255\textwidth}
    \centering
    \includegraphics[width=\linewidth]{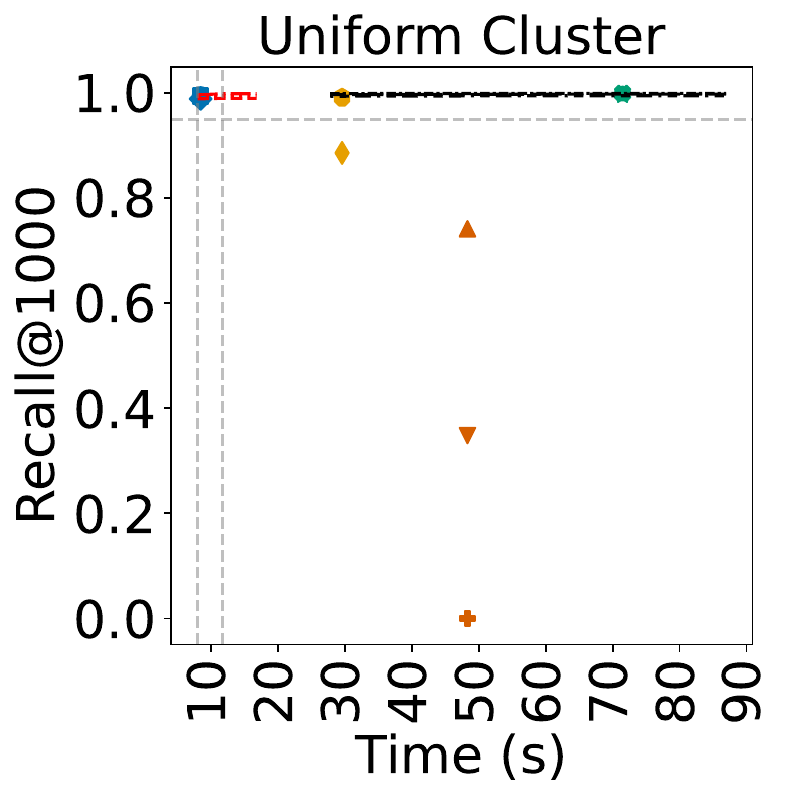}
  \end{minipage}
  \begin{minipage}{0.245\textwidth}
    \centering
    \includegraphics[width=\linewidth]{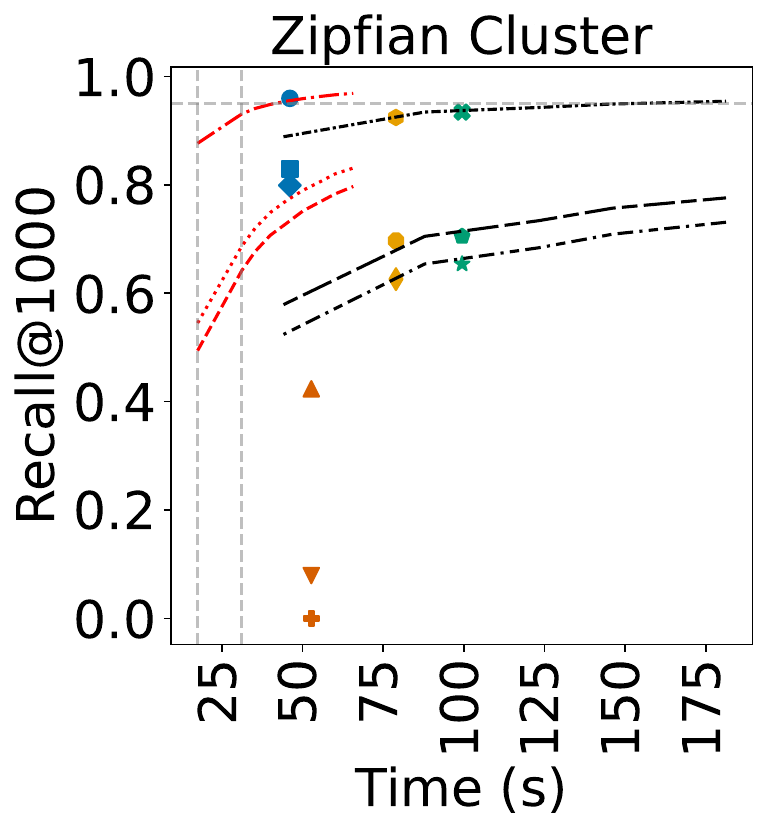}
  \end{minipage}
    \caption{
    Search performance on real and synthetic datasets. 
    }\label{fig:online_searching}
    \label{fig:offline_time}
\end{figure*}
\begin{figure*}
    \centering
    \includegraphics[width=\linewidth]{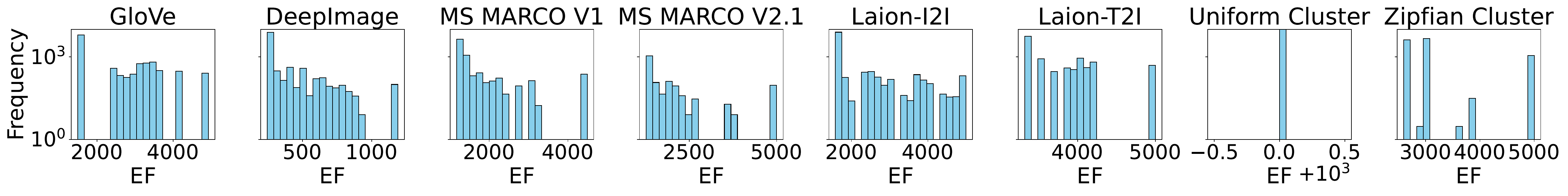}
    \caption{Distribution of \ef\ values dynamically assigned by \texttt{Ada-\ef}  across queries in each dataset  (log scale).}
    \label{fig:ef_distributions}
\end{figure*}

\textit{Settings}.
All experiments are conducted on a shared server running Ubuntu 22.04, equipped with 80 physical CPU cores (Intel Xeon Platinum 8380 @ 2.30 GHz), 160 hardware threads, and 1 TB of DDR4-3200 MHz RDIMM main memory. 
All implementations are written in C++, leveraging SIMD intrinsics for vectorized computations. The code is compiled with \texttt{g++ 12.3.0}, using the optimization flag \texttt{-O3} and enabling AVX-512, via setting \texttt{FAISS\_OPT\_LEVEL} as \texttt{avx512} for FAISS-based methods—\texttt{HNSW (FAISS)}, \texttt{DARTH}, and \texttt{LAET}—or setting \texttt{-march=native} for HNSWlib-based ones—\texttt{HNSW (HNSWlib)}, \texttt{Ada-ef}, and \texttt{PiP}. All offline experiments are conducted using 40 threads. For online searching, we follow the common practice of using a single thread~\cite{10.1145/3749160}. Note that queries can be processed in inter-query parallelism to fully utilize all available CPU cores.

\subsection{Online Searching Performance}\label{sec:exp_online}
We present the search results in Figure~\ref{fig:online_searching}, which reports both recall and end-to-end query workload execution time.
For \texttt{DARTH}, \texttt{LAET}, and \texttt{Ada-ef}, we set the target recall to $0.95$ across all datasets.
For \texttt{PiP}, we set \ef\ to 300 for Top-100 ANNS or 2000 for Top-1000 ANNS.
For \texttt{HNSW (HNSWlib)} and \texttt{HNSW (FAISS)}, we execute the query workload multiple times, gradually increasing \ef\ starting from \ef\ = $k$ for Top-$k$ ANNS, until either the average recall reaches $0.99$ or \ef\ reaches a upper bound of 5000, and we report the execution time and corresponding recall.
We note that this experiment focuses on simulating a real-time scenario where queries are served on the fly, and \texttt{HNSW (HNSWlib)} and \texttt{HNSW (FAISS)} results under varying \ef\ values are used as references because tuning \ef\ values requires ground-truth  for incoming queries.
In Figure~\ref{fig:online_searching}, we also highlight key reference points to aid interpretation. Specifically, we mark the cases of \ef\ = $k$ and \ef\ = $2k$ using two vertical dotted gray lines; the former is one of the commonly used settings in practice. The target recall of 0.95 is indicated with a horizontal dashed line.
To better assess search quality, we report three recall statistics: average recall, 1st percentile, and 5th percentile. While average recall reflects the overall workload performance, the 1st and 5th percentiles capture performance on the hardest queries. These metrics help diagnose under-searching and over-searching behaviors.
Figure~\ref{fig:ef_distributions} shows the distribution of \ef\ values dynamically configured by \texttt{Ada-\ef}.

For \texttt{HNSW (HNSWlib)} and \texttt{HNSW (FAISS)} results, different datasets require different \ef\ values to achieve high recall, and a fixed setting of \ef\ = $k$ is insufficient. Moreover, increasing \ef\ beyond a certain threshold results in diminishing returns, \textit{i.e.}, recall improves only marginally while query time increases substantially, highlighting the over-searching issue.

For \texttt{PiP}, the achieved recall is substantially lower than that of the other methods, highlighting the limitation of its fixed-threshold early termination strategy in maintaining practical effectiveness.

For the learning-based adaptive approaches, \texttt{DARTH} and \texttt{LAET}, both methods reach the declarative target recall of 0.95 on five out of the eight evaluated datasets. However, on these datasets, they tend to over-search, entering a plateau region where additional computation produced negligible recall gains, \textit{e.g.}, MS MARCO V1.
For the remaining datasets, their recall fails to meet the target and is, in some cases, considerably lower, \textit{e.g.}, GloVe-100. Overall, \texttt{DARTH} generally achieves higher recall than \texttt{LAET}, but at the expense of substantially higher query latency.

\texttt{Ada-ef} consistently achieves the best performance across all datasets.
It dynamically configures the \ef\ value on a per-query basis to approximately meet the declarative target recall, effectively avoiding over-searching by skipping unnecessarily large \ef\ values that offer only marginal recall improvements at the cost of significantly increased latency.
At the same time, it mitigates under-searching by improving recall for harder queries, as reflected in the enhanced 1st and 5th percentile recall metrics, which are substantially higher than those of \texttt{HNSW (HNSWlib)}.
Compared to the state-of-the-art approach \texttt{DARTH}, \texttt{Ada-ef} achieves up to a $4\times$ reduction in workload latency for the same level of average recall.
In addition, the distributions of dynamically configured \ef\ values for each query, shown in Figure~\ref{fig:ef_distributions}, exhibit a long-tail pattern, \textit{i.e.}, smaller \ef\ values are assigned to the majority of queries, while a minority receive larger values. This fine-grained dynamic configuration substantially improves the worst-case recall.

In the multimodal setting, \textit{i.e.}, Laion-I2I and Laion-T2I, although the dataset vectors (image embeddings) are identical, the differing modalities of the query vectors lead to substantial variations in search performance.
The text-to-image scenario is considerably more challenging than the image-to-image case: setting \ef\ to $k$ yields a recall close to 0.95 for Laion-I2I, but only around 0.8 for Laion-T2I. This discrepancy arises primarily from the latent distribution gap between visual and textual modalities, which makes cross-modal retrieval inherently more difficult.
Existing approaches achieve recall levels well below the declarative target of 0.95, whereas \texttt{Ada-\ef} adaptively adjusts \ef\ based on the joint distribution of data and query vectors, attaining recall close to the target without incurring excessive computation.

\begin{table*}[t]
\centering
\caption{Offline computation time (seconds) with breakdowns for \texttt{Ada-ef}, \texttt{DARTH}, and \texttt{LAET}.}
\label{table:offline_time}
\resizebox{\textwidth}{!}{
\begin{tabular}{|l|r|rrrr|r|rrrr|rrrr|}
\hline
\multirow{2}{*}{\textbf{Dataset}} 
& \multirow{2}{*}{\makecell{\textbf{\texttt{HNSW}}\\\textbf{\texttt{(HNSWlib)}}}} 
& \multicolumn{4}{c|}{\textbf{\texttt{Ada-ef}}} 
& \multirow{2}{*}{\makecell{\textbf{\texttt{HNSW}}\\\textbf{\texttt{(FAISS)}}}} 
& \multicolumn{4}{c|}{\textbf{\texttt{DARTH}}} 
& \multicolumn{4}{c|}{\textbf{\texttt{LAET}}} \\ \cline{3-6} \cline{8-15}
&  & \textbf{Stats} & \textbf{Samp.} & \textbf{EF-Est.} & \textbf{Total}
&  & \textbf{LVec GT} & \textbf{TData}& \textbf{Train} & \textbf{Total}
& \textbf{LVec GT} & \textbf{TData}& \textbf{Train} & \textbf{Total} \\ \hline \hline

DeepImage        &439.536 &0.098 &6.905 &0.678&7.681&  602.636&179.465&37.521&157.235&374.221&179.465
&37.521
& 0.424&217.410\\ \hline
GloVe-100        &49.238  &0.014 &0.794 &2.125 &2.933  & 70.937&22.820&40.618&147.395&210.833&22.820
&40.618
&0.439&63.877\\ \hline
MS MARCO V1      &2013.632&25.271&54.472&6.327 &86.070 & 3148.286&2894.502&1738.706&284.351&4917.558&2894.502
&1738.706
&2.348&4635.556\\ \hline
MS MARCO V2.1    &2263.155&13.814&80.031&8.206&102.051& 4619.897&4699.835&1819.703&381.704&6901.242&4699.835
&1819.703
&1.909&6521.447\\ \hline
Laion-I2I        &4110.403&4.625 &71.228&3.199 &79.052 & 5155.223&3076.920&1793.698&226.168&5096.786&3076.920
&1793.698
&0.833&4871.451\\ \hline
Laion-T2I        &4110.403&4.625 &69.048&6.021 &79.694 & 5155.223&3307.787&1688.520&274.387&5270.693&3307.787
&1688.520
&0.866&4997.173\\ \hline
Uniform Cluster  &198.340 &0.132 &7.217 &0.402 &7.751  & 241.179&193.547&141.618&32.157&367.321&193.547
&141.618
&0.280&335.445\\ \hline
Zipfian Cluster  &436.964 &0.235 &7.183 &2.860 &10.278 & 522.806&210.620&1660.215&284.941&2155.776&210.620&1660.215&0.325&1871.160\\ \hline
\end{tabular}}
\end{table*}

\begin{table*}[t]
\centering
\caption{Offline computation memory usage (bytes) with breakdowns for \texttt{Ada-ef}, \texttt{DARTH}, and \texttt{LAET}.}
\label{table:offline_memory}
\resizebox{0.95\textwidth}{!}{
\begin{tabular}{|l|r|rrrr|r|rrrr|rrrr|}
\hline
\multirow{2}{*}{\textbf{Dataset}} 
& \multirow{2}{*}{\makecell{\textbf{\texttt{HNSW}}\\\textbf{\texttt{(HNSWlib)}}}} 
& \multicolumn{4}{c|}{\textbf{\texttt{Ada-ef}}} 
& \multirow{2}{*}{\makecell{\textbf{\texttt{HNSW}}\\\textbf{\texttt{(FAISS)}}}} 
& \multicolumn{4}{c|}{\textbf{\texttt{DARTH}}} 
& \multicolumn{4}{c|}{\textbf{\texttt{LAET}}} \\ \cline{3-6} \cline{8-15}
&  & \textbf{Stats} & \textbf{Samp.} & \textbf{EF-Est.} & \textbf{Total}
& & \textbf{LVec GT} & \textbf{TData} & \textbf{Train}& \textbf{Total}
& \textbf{LVec GT} & \textbf{TData} & \textbf{Train}& \textbf{Total} \\ \hline \hline

DeepImage        &5G      &37K &154K &2K   &0.2M  & 5G&7.8M  &2G   &342K &2G   
&7.8M  &2G   
&294K&2G   
\\ \hline
GloVe-100        &620M   &40K &157K &3.5K &0.21M & 0.615G&7.8M  &2.4G &348K &2.4G 
&7.8M  &2.4G 
&296K&
2.4G\\ \hline
MS MARCO V1      &52G     &9.1M&2M   &1K   &11.2M & 52G&98M   &4.7G &346K &4.8G&98M   &4.7G 
&397K&4.8G\\ \hline
MS MARCO V2.1    &73G     &4.1M&1.6M &1K   &5.8M  & 73G&79M   &5.2G &345K &5.3G&79M   &5.2G 
&360K&
5.3G\\ \hline
Laion-I2I        &63G     &1.1M&1.2M &2.2K &3.4M  & 63G&59M   &5G   &342K&5G   
&59M   &5G   
&325K&5G\\ \hline
Laion-T2I        &63G     &1.1M&1.2M &2.1K &3.4M  & 63G&59M   &5.6G&345K&5.7G&59M   &
5.6G&321K&5.7G\\ \hline
Uniform Cluster  &5.2G    &40K &860K &0.6K &0.9M  & 5.1G&42.9M &373M &323K&416M&42.9M &373M 
&280K
&416M\\ \hline
Zipfian Cluster  &5.2G    &40K &860K &1K   &0.9M  & 5.1G&42.9M &4.3G &346K &4.4G&42.9M &4.3G &295K&
4.4G\\ \hline
\end{tabular}}
\end{table*}

The two synthetic datasets, Uniform Cluster and Zipfian Cluster, share identical characteristics except for their cluster size distributions: Uniform Cluster consists of equally sized clusters, whereas Zipfian Cluster exhibits highly skewed cluster sizes following a Zipfian distribution. This difference leads to distinct search behaviors. Achieving high recall is substantially easier in the Uniform case than in the Zipfian case.
All existing methods struggle under the Zipfian setting, attaining lower recall while incurring significantly higher search time.
In contrast, \texttt{Ada-\ef} demonstrates robust performance, substantially improving both the 5th and 1st percentile recall and achieving the target recall without unnecessary computation.
These results highlight the practical effectiveness of \texttt{Ada-\ef}, particularly in real-world scenarios where data and query distributions are often skewed rather than uniform.

\begin{figure}
    \centering
    \includegraphics[width=\linewidth]{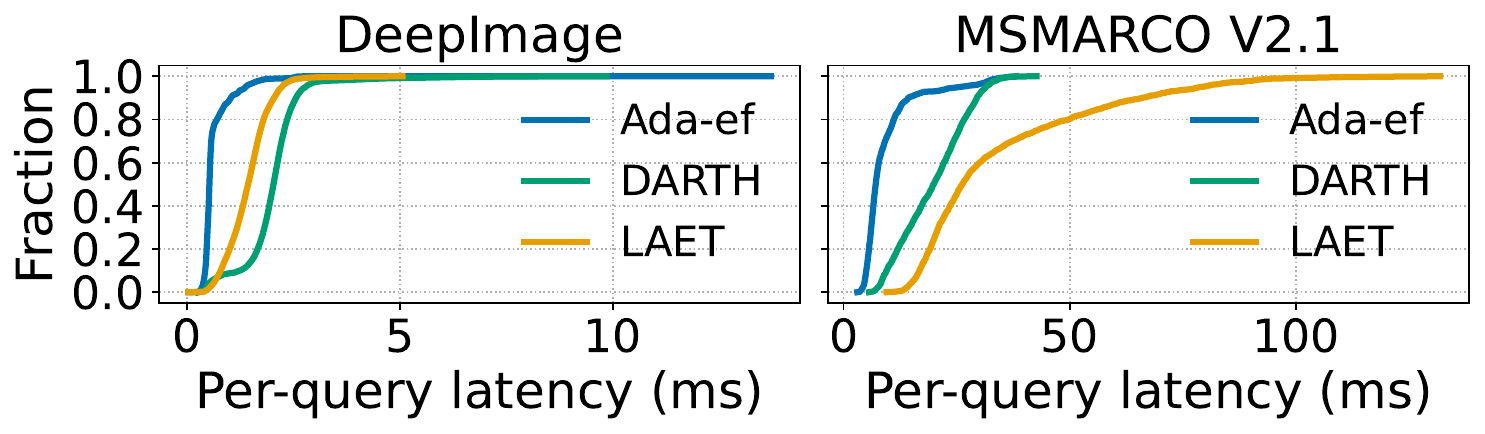}
    \caption{CDFs of per-query latencies.}\label{fig:latency_cdf} 
\end{figure}

We use DeepImage and MS MARCO V2.1 as representatives to analyze the per-query latency distribution. The cumulative distribution functions (CDFs) are shown in Figure~\ref{fig:latency_cdf}.
As illustrated, \texttt{Ada-\ef} achieves low latency for the majority of queries while spending more time on a small subset of difficult ones. 
This behavior indicates that \texttt{Ada-\ef} effectively avoids both over-searching and under-searching, thereby reaching the target recall more efficiently and improving recall performance for the hardest queries.

\begin{figure*}
    \centering
    \includegraphics[width=0.7\linewidth]{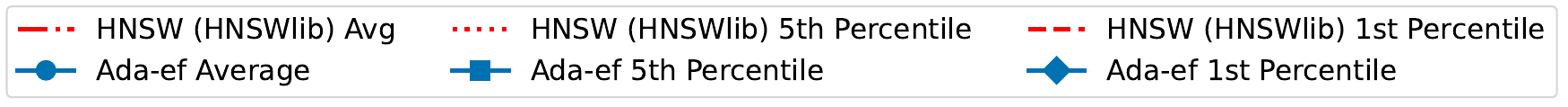} 
    
    \includegraphics[width=0.33\linewidth]{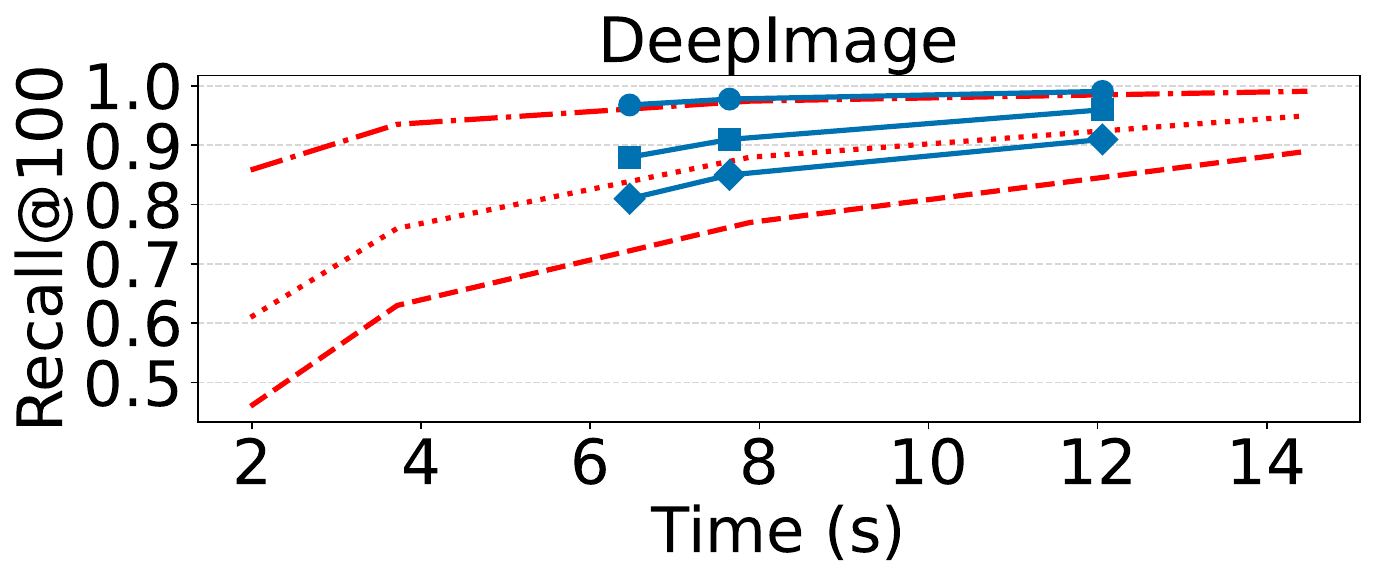}
    \includegraphics[width=0.33\linewidth]{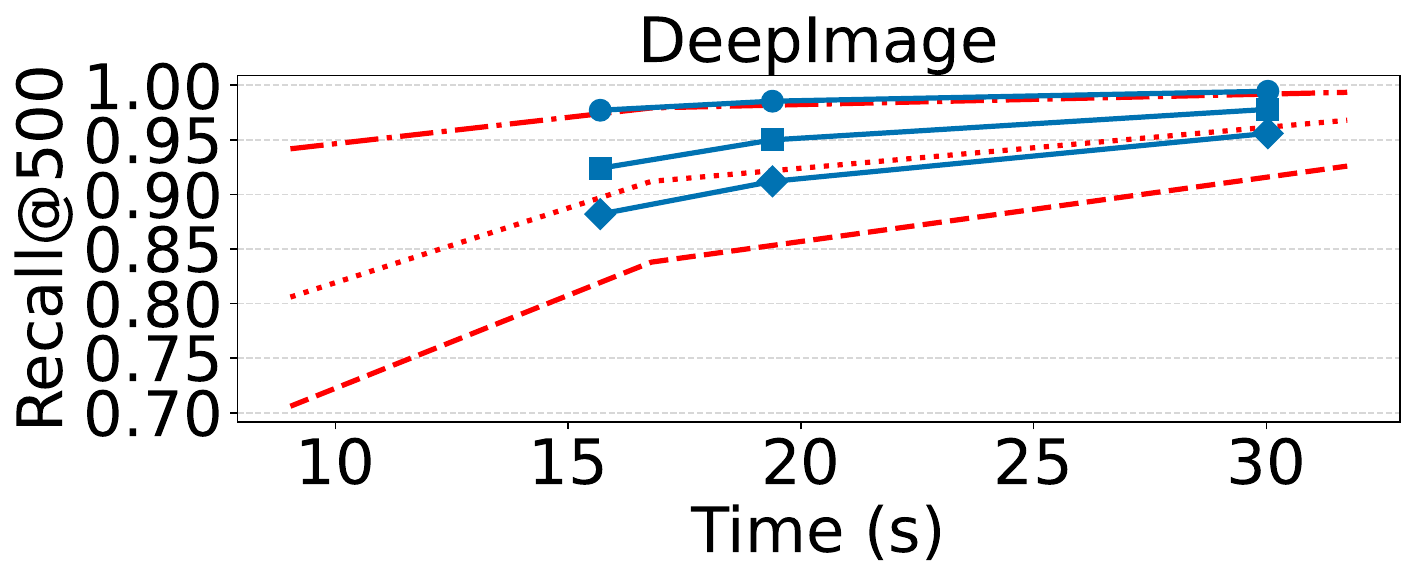}
    \includegraphics[width=0.33\linewidth]{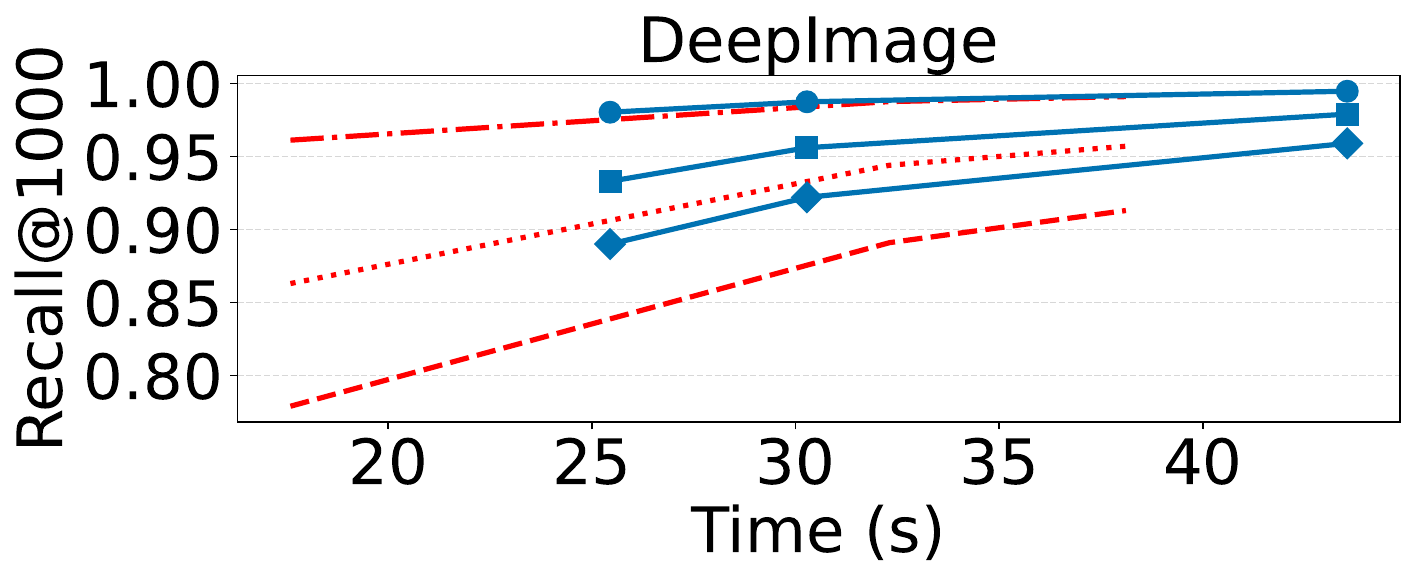}
    \includegraphics[width=0.33\linewidth]{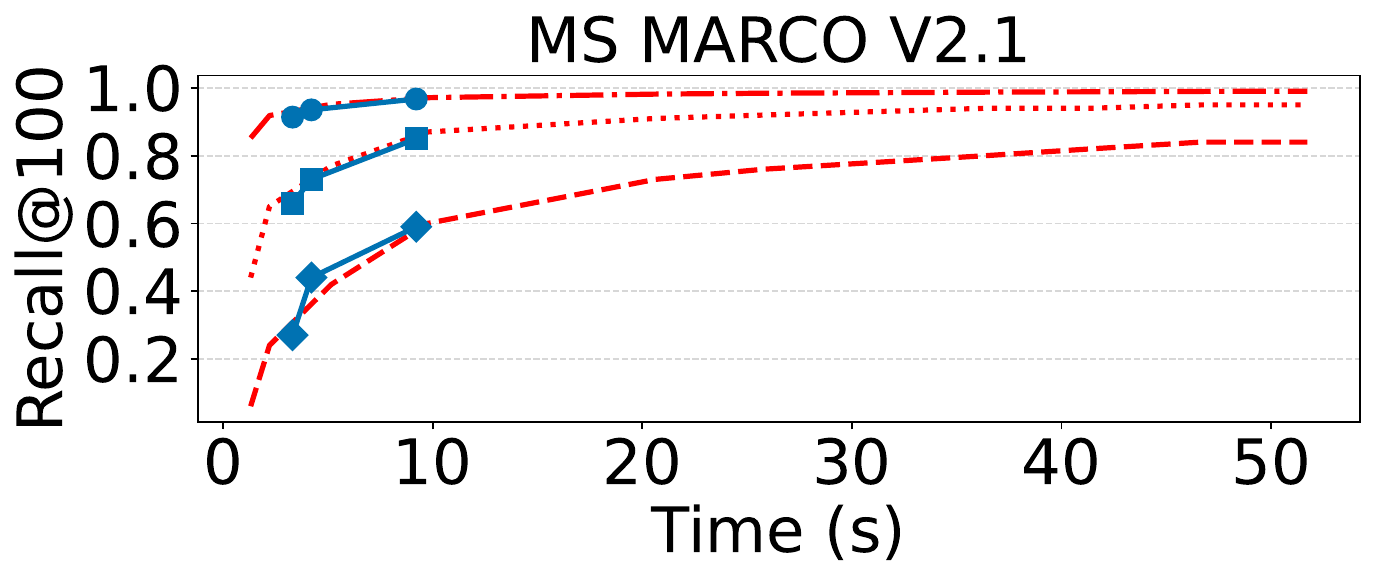}
    \includegraphics[width=0.33\linewidth]{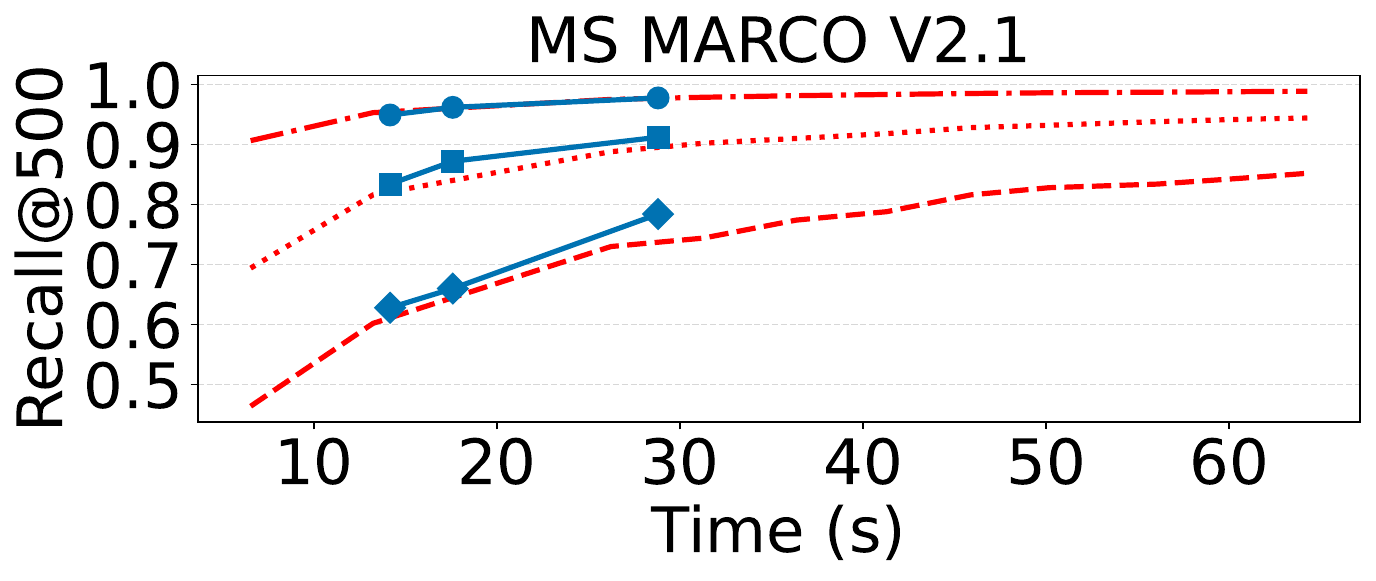}
    \includegraphics[width=0.33\linewidth]{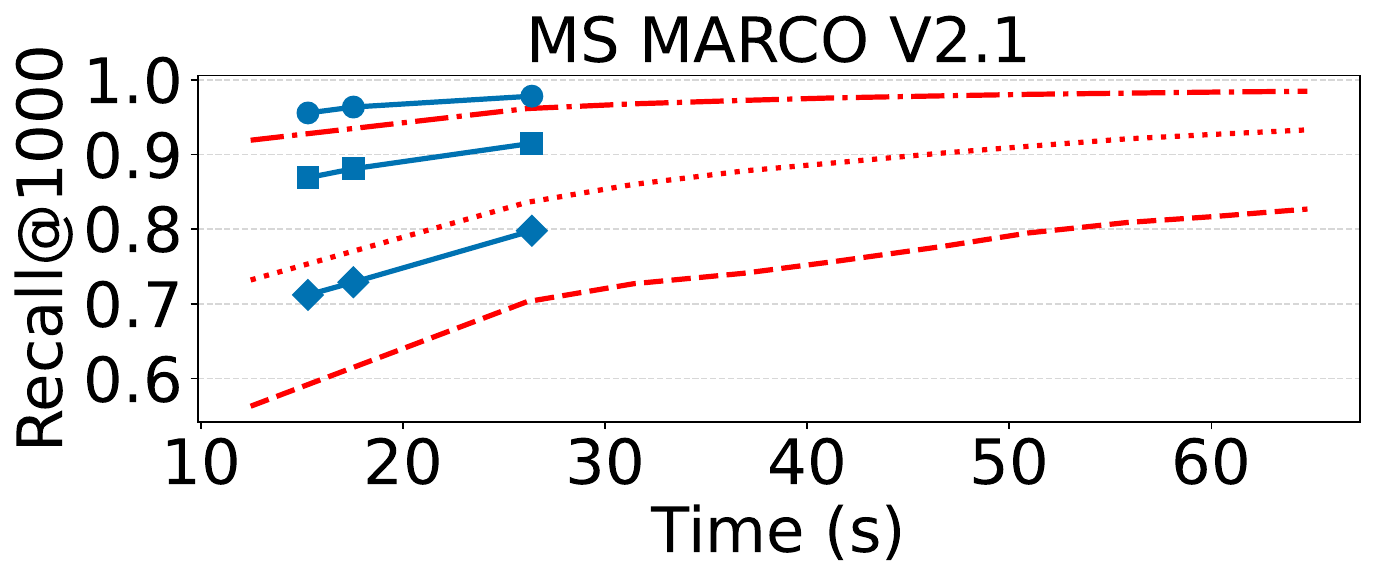}
    \caption{Sensitivity analysis with different values of $k$ for Top-k ANNS.}
    \label{fig:exp_sensitivity_er}
\end{figure*}

\subsection{Offline Computing Performance}\label{sec:exp_offline}
Table~\ref{table:offline_time} reports the offline computation time. The \texttt{HNSW (HNSWlib)} and \texttt{HNSW (FAISS)} columns show the index construction time.
For \texttt{DARTH} and \texttt{LAET}, the offline computation consists of three steps: (i) computing the ground truth for sampled 10000 learn vectors, (ii) generating training data using these vectors, and (iii) training the model with the generated data. These steps are denoted as \texttt{LVec GT}, \texttt{TData}, and \texttt{Train}, respectively.
The offline computation of \texttt{Ada-\ef} includes: (i) computing dataset-level statistics (\texttt{Stats}), (ii) sampling 200 data vectors and computing their ground truth (\texttt{Samp.}), and (iii) constructing the \ef-estimation table (\texttt{EF-Est.}) with an expected recall of 0.95 and an \ef\ upper bound of 5000.
Table~\ref{table:offline_memory} presents the corresponding memory usage. The \texttt{HNSW (HNSWlib)} and \texttt{HNSW (FAISS)} columns indicate index size; the \texttt{DARTH} and \texttt{LAET} columns report the memory footprint of their data and models; and the \texttt{Ada-\ef} columns show the total memory usage across its components.
Note that \texttt{PiP} does not require offline computation and is therefore omitted from this experiment.

For the learning-based methods, \texttt{DARTH} and \texttt{LAET}, the major offline computation overhead is computing the ground-truth for 10000 learning vectors (\texttt{LVec GT}), which can exceed the index construction time. 
In addition, the training data generation stage (\texttt{TData}) also incurs a substantial cost.
We note that the training time of \texttt{LAET} is considerably lower than that of \texttt{DARTH}, since \texttt{LAET} extracts features once per learn vector, whereas \texttt{DARTH} performs this process multiple times for each learn vector.

In contrast, the offline computation time of \texttt{Ada-\ef} is negligible compared to HNSW index construction, typically less than 5\% of the indexing time.
Compared with \texttt{DARTH} and \texttt{LAET}, \texttt{Ada-\ef} achieves, on average, a $50\times$ reduction in total offline computation time. Its total time is even lower than the least costly offline stage of \texttt{DARTH} (model training).
These results demonstrate that \texttt{Ada-\ef} can operate as a lightweight add-on to existing HNSW indexes, providing distribution-aware, adaptive \ef\ configuration that improves recall while avoiding both over- and under-searching.

Consider the breakdown of \texttt{Ada-\ef}'s offline computation cost. 
We observe that while \texttt{Samp.} and \texttt{Stats} dominate the offline computation, the cost of constructing the \texttt{EF-Est.} table remains negligible. 
This design highlights the generality and reusability of \texttt{Ada-\ef}. Specifically, both \texttt{Samp.} and \texttt{Stats} are agnostic to the value of $k$ and the declarative target recall, meaning that once computed, they can be reused across different configurations, such as in the experiments of Section~\ref{sec:sensitivity}, where only the \texttt{EF-Est.} table needs to be recomputed, making the additional computation negligible.
In contrast, for \texttt{DARTH} and \texttt{LAET}, both the \texttt{TData} generation and model training must be repeated for different $k$ values, resulting in significant offline overhead.
Furthermore, both \texttt{Stats} and \texttt{Samp.} can be efficiently updated incrementally when the dataset is updated (\S~\ref{sec:exp_insertion}), demonstrating that \texttt{Ada-\ef} is well-suited for practical, evolving real-world scenarios.

Table~\ref{table:offline_memory} reports the offline memory usage with detailed breakdowns for each method.
For learning-based approaches, the training data (\texttt{TData}) dominates the overall memory footprint. This component primarily consists of large tabular datasets containing the features used to train the GBDT models, which can reach hundreds of millions of rows.
In contrast, \texttt{Ada-\ef} requires minimal offline memory, achieving an average $100\times$ reduction in memory footprint.

\subsection{Sensitivity Analysis}\label{sec:sensitivity}
We analyze the performance of \texttt{Ada-\ef} under varying configurations, including different values of $k$ for Top-$k$ ANNS and different target recall levels.
Specifically, we vary $k$ from 100 to 1000, and for each value, we run \texttt{Ada-\ef} with three target recall settings: 0.95, 0.97, and 0.99. 
For comparison, HNSW is executed with increasing \ef\ values until either the average recall reaches 0.99 or the maximum \ef\ of 5000 is reached.
We conduct experiments on two representative datasets: DeepImage and MS MARCO V2.1, and we present the sensitivity analysis results in Figure~\ref{fig:exp_sensitivity_er}.

Overall, increasing the target recall in \texttt{Ada-\ef}  consistently improves average recall as well as the 5th and 1st percentile recall across different $k$ values and datasets.
Ada-\ef\ demonstrates robust and reliable behavior, achieving or remaining close to the specified target recall while mitigating both over- and under-searching.
Notably, on the DeepImage dataset with $k=1000$, \texttt{Ada-\ef}  with a target recall of 0.99 incurs higher query latency. 
However, this additional computation is adaptively allocated to the most difficult queries, as evidenced by the substantial improvements in the 1st and 5th percentile recall metrics.

\subsection{Incremental Insertion and Deletion}\label{sec:exp_insertion}
We use DeepImage and MS MARCO V2.1 as the representative datasets to simulate a practical scenario in which data vectors are inserted into or deleted from the current dataset.
Specifically,  each original dataset is partitioned  into two subsets: the first serves as existing data, and the remaining as updated data,  whose size denoted as the batch size (BS) simulates real-world batch updates.
We consider two batch sizes: 10\% and 50\% of the original data.

In the insertion experiments, the update data is inserted into the existing data to reconstruct the original dataset. In this setting, we first build an HNSW index over the existing data and then insert the new data to measure the index update time.
For the deletion experiments, we remove the update data from the original dataset to obtain the existing subset. Since \texttt{HNSW (HNSWlib)} does not support in-place deletions (the same holds for \texttt{HNSW (FAISS)}), we instead report the indexing time for rebuilding the index over the existing data. 
For \texttt{Ada-\ef}, we adopt the incremental update strategy described in Section~\ref{sec:incremental_update} to capture the effects of dataset changes. Specifically, for both insertion and deletion scenarios, we incrementally update (i) the dataset statistics and (ii) the ground truth of the sampled subset, and use these updated components to reconstruct the \ef-estimation table. The corresponding update times are reported in Tables~\ref{table:inc_update_time} and~\ref{table:deletion_time}, respectively.

\begin{table}[t]
    \centering
    \caption{Insertion time (s).}\label{table:inc_update_time}
    \resizebox{0.95\linewidth}{!}{
        \begin{tabular}{|l|r|r|r|r|r|r|} 
            \hline
            \textbf{Dataset} & \textbf{BS}    &\textbf{\texttt{HNSW (HNSWlib)}}& \textbf{\texttt{Ada-ef}}   & \textbf{Stats}& \textbf{Samp.}& \textbf{EF-Est.} \\ \hline \hline
            DeepImage        &  $10\%$ &   58.047& 0.740& 0.039&0.393& 0.308\\ \hline
            DeepImage        &  $50\%$ &   247.416& 2.072& 0.154&1.647& 0.271\\ \hline
            MS MARCO V2.1    & $10\%$  &  274.259& 16.589& 2.075&6.513& 8.001\\ \hline
            MS MARCO V2.1    & $50\%$  &  1213.185& 46.175& 9.613&28.318& 8.244\\ \hline
        \end{tabular}
    }
\end{table}
\begin{table}[t]
\centering
\caption{Query time (QT) and recall of \texttt{Ada-\ef} after  insertion.}\label{table:inc_update_search}
\resizebox{\linewidth}{!}{
\begin{tabular}{|l|r|l|r|r|r|r|}
\hline
\textbf{Dataset}    &   \textbf{Method}     &\textbf{BS}& \textbf{QT (s)} & \textbf{Avg. Recall} & \textbf{P5 Recall}& \textbf{P1 Recall}\\ \hline \hline 
    DeepImage       & Stale Ada-\ef         & $10\%$& 6.223& 0.964& 0.860&0.780\\\hline     
    DeepImage       & Incr. Ada-\ef         &$10\%$& 7.463& 0.973& 0.900& 0.830\\\hline    
    DeepImage       & Stale Ada-\ef         & $50\%$ & 4.828& 0.945& 0.820&0.720\\\hline     
    DeepImage       & Incr. Ada-\ef         &$50\%$ & 6.307&0.967&  0.880& 0.800\\\hline 
    DeepImage       & Reco. Ada-\ef         &N/A& 6.437& 0.968& 0.880&0.810\\\hline  
    MS MARCO V2.1   & Stale Ada-\ef         &$10\%$ & 15.262& 0.952& 0.865&0.789\\ \hline    
    MS MARCO V2.1   & Incr. Ada-\ef &$10\%$& 16.265&    0.957& 0.873& 0.802\\ \hline
    MS MARCO V2.1   & Stale Ada-\ef         &$50\%$ & 14.089& 0.950& 0.848&0.765\\ \hline
    MS MARCO V2.1   & Incr. Ada-\ef &$50\%$ & 14.230&    0.951& 0.833& 0.625\\ \hline 
    MS MARCO V2.1   & Reco. Ada-\ef         &N/A& 15.406& 0.955& 0.869&0.712\\ \hline
\end{tabular}}
\end{table}
\begin{table}[t]
    \centering
    \caption{Deletion time of \texttt{Ada-\ef} and indexing time   (s).}\label{table:deletion_time}
    \resizebox{0.95\linewidth}{!}{
        \begin{tabular}{|l|r|r|r|r|r|r|} 
            \hline
            \textbf{Dataset} & \textbf{BS}    &\textbf{\texttt{HNSW (HNSWlib)}}& \textbf{\texttt{Ada-ef}}   & \textbf{Stats}& \textbf{Samp.}& \textbf{EF-Est.} \\ \hline \hline
            DeepImage        &  $10\%$ &   417.936& 1.566& 0.036&1.258& 0.272\\ \hline
            DeepImage        &  $50\%$ &   197.927& 3.352& 0.175&2.923& 0.254\\ \hline
            MS MARCO V2.1    & $10\%$  &  2079.652& 13.538& 2.209&4.270& 7.059\\ \hline
            MS MARCO V2.1    & $50\%$  &  1063.294& 24.144& 10.135&7.721& 6.288\\ \hline
        \end{tabular}
    }
\end{table}
\begin{table}[t]
\centering
\caption{Query time (QT) and recall of \texttt{Ada-\ef} after  deletion.}\label{table:deletion_search}
\resizebox{\linewidth}{!}{
\begin{tabular}{|l|r|l|r|r|r|r|}
\hline
\textbf{Dataset}    &   \textbf{Method}     &\textbf{BS}& \textbf{QT (s)} & \textbf{Avg. Recall} & \textbf{P5 Recall}& \textbf{P1 Recall}\\ \hline 
    DeepImage       & Stale Ada-\ef         & $10\%$& 6.297& 0.969& 0.890&0.810\\\hline     
    DeepImage       & Incr. Ada-\ef         &$10\%$& 6.457& 0.970& 0.890& 0.820\\\hline
 DeepImage       & Reco. Ada-\ef         & $10\%$& 6.089& 0.966& 0.870&0.790\\\hline
    DeepImage       & Stale Ada-\ef         & $50\%$ & 6.025& 0.979& 0.920&0.860\\\hline     
    DeepImage       & Incr. Ada-\ef         &$50\%$ & 5.427&0.974&  0.910& 0.850\\\hline 
    DeepImage       & Reco. Ada-\ef         &$50\%$& 4.373& 0.960& 0.860&0.780\\\hline  
    MS MARCO V2.1   & Stale Ada-\ef         &$10\%$ & 16.235& 0.954& 0.870&0.708\\ \hline
    MS MARCO V2.1   & Incr. Ada-\ef &$10\%$& 14.405&    0.948&  0.820& 0.648\\\hline
 MS MARCO V2.1   & Reco. Ada-\ef         & $10\%$& 15.723& 0.953& 0.864&0.782\\\hline 
    MS MARCO V2.1   & Stale Ada-\ef         &$50\%$ & 16.821& 0.960& 0.886&0.678\\ \hline
    MS MARCO V2.1   & Incr. Ada-\ef &$50\%$ & 15.762&    0.959& 0.883&  0.823\\ \hline 
    MS MARCO V2.1   & Reco. Ada-\ef         &$50\%$& 14.793& 0.956& 0.871&0.770\\ \hline
\end{tabular}}
\end{table}

To evaluate the search performance of \texttt{Ada-\ef}, we consider three variants:
(i) Stale \texttt{Ada-\ef}, which uses the unmodified (pre-update) \texttt{Ada-\ef};
(ii) Incr. \texttt{Ada-\ef}, which applies the incrementally updated \texttt{Ada-\ef} described above; and
(iii) Reco. \texttt{Ada-\ef}, which is fully recomputed from scratch.
Their search performance results are reported in Tables~\ref{table:inc_update_search} and~\ref{table:deletion_search}, respectively.

As shown in Tables~\ref{table:inc_update_time} and~\ref{table:deletion_time}, the update time required by \texttt{Ada-\ef} is modest compared to the HNSW index update time in the insertion scenario and the index rebuilding time in the deletion scenario.

For search performance, in the insertion case, Stale \texttt{Ada-\ef} maintains robust performance but may fall slightly below the target recall, particularly with large batch sizes (\textit{e.g.}, 50\% insertion on DeepImage).
In the deletion case, Stale \texttt{Ada-\ef} consistently achieves recall above the target of 0.95 across all datasets, though at the expense of slightly higher search time.
The incrementally updated version, Incr. \texttt{Ada-\ef}, effectively resolves this issue, maintaining target recall while avoiding over-searching, and achieves performance comparable to the fully recomputed Reco. \texttt{Ada-\ef}.
In summary, Stale \texttt{Ada-\ef} performs robustly under small batch updates, whereas Incr. \texttt{Ada-\ef} is preferable for large batch updates, offering a better balance between recall stability and computational efficiency.

\begin{table}[t]
    \centering
    \caption{Impact of the distance list size $|D|$ on the offline computation time and online query time of \texttt{Ada-ef}.}\label{table:impact_D}
    \resizebox{\linewidth}{!}{
        \begin{tabular}{|l|r|r|r|r|r|r|} 
            \hline
            \textbf{Dataset} & $|D|$&\textbf{EF-Est. (s)}& \textbf{QT (s)}& \textbf{Avg. Recall} & \textbf{P5 Recall}& \textbf{P1 Recall}\\ \hline 
            DeepImage        &  1-hop&   0.349& 6.524& 0.968&0.870& 0.770\\\hline
 DeepImage& 2-hop& 0.282& 6.902& 0.968& 0.880&0.810\\\hline 
            DeepImage        &  3-hop&   3.741& 92.410& 0.998&0.990& 0.960\\\hline
 MS MARCO V2.1& 1-hop& 3.081& 12.890& 0.943& 0.783&0.626\\ \hline
            MS MARCO V2.1    & 2-hop&  3.171& 15.622& 0.955&0.869& 0.712\\ \hline
            MS MARCO V2.1    & 3-hop&  7.540& 30.871& 0.968&0.849& 0.702\\ \hline
        \end{tabular}
    }
\end{table}
\begin{table}[t]
    \centering
    \caption{Impact of the number of sampled data vectors on the offline computation time and online query time of \texttt{Ada-ef}.}\label{table:impact_samplings}
    \resizebox{\linewidth}{!}{
        \begin{tabular}{|l|r|l|r|r|r|r|r|} 
            \hline
            \textbf{Dataset} & \textbf{Num.} & \textbf{Samp. (s)} &\textbf{EF-Est. (s)}& \textbf{QT (s)}& \textbf{Avg. Recall} & \textbf{P5 Recall}& \textbf{P1 Recall}\\ \hline 
            DeepImage        &  200&    6.827&0.273& 6.420& 0.968& 0.880& 0.810\\\hline
 DeepImage& 1000&  23.704&1.463& 6.342& 0.966& 0.880&0.820\\\hline 
            DeepImage        &  5000&    99.100&8.657& 4.713& 0.940&0.770& 0.650\\\hline
 MS MARCO V2.1& 200
&  84.240&3.155& 15.348& 0.955& 0.869&0.712\\ \hline
            MS MARCO V2.1    & 1000
&   102.390&14.455& 13.401& 0.948&0.824&  0.636\\ \hline
            MS MARCO V2.1    & 5000&   260.469&83.036& 13.610& 0.950&0.845&  0.662\\ \hline
        \end{tabular}
    }
\end{table}
\begin{table}[t]
    \centering
    \caption{Impact of the design of weighted function on the offline computation time and online query time of \texttt{Ada-ef}.}\label{table:impact_weighted_function}
    \resizebox{\linewidth}{!}{
        \begin{tabular}{|l|r|r|r|r|r|r|} 
            \hline
            \textbf{Dataset} & \textbf{Decay}&\textbf{EF-Est. (s)}& \textbf{QT (s)}& \textbf{Avg. Recall} & \textbf{P5 Recall}& \textbf{P1 Recall}\\ \hline 
            DeepImage        &  None&   0.649& 5.707& 0.951&0.800& 0.680\\\hline
 DeepImage& Linear & 0.301& 5.419& 0.946& 0.790&0.660\\\hline 
            DeepImage        &  Exp&   0.286& 6.950& 0.968&0.880& 0.81\\\hline
 MS MARCO V2.1& None
& 2.663& 9.594& 0.919& 0.732&0.563\\ \hline
            MS MARCO V2.1    & Linear 
&  2.723& 12.858& 0.944&0.804& 0.659\\ \hline
            MS MARCO V2.1    & Exp&  3.239& 15.987& 0.955&0.869& 0.712\\ \hline
        \end{tabular}
    }
\end{table}

\subsection{Ablation Study on Component Designs}\label{sec:abaltion_study}
We conduct ablation studies on two representative datasets, DeepImage and MS MARCO V2.1, to analyze the effectiveness of different component designs in \texttt{Ada-\ef}.

The first ablation study evaluates the impact of the distance list size $|D|$, considering three configurations: neighbors within 1-hop, 2-hop, and 3-hop. The results are shown in Table~\ref{table:impact_D}.
In general, the 3-hop configuration substantially increases both query latency and \texttt{EF-Est.} table computation time, as the search must traverse all neighbors within three hops from the entry point on the base layer. The 1-hop configuration suffers from low recall, particularly on high-dimensional datasets such as MS MARCO V2.1. The 2-hop design achieves the best balance between recall and latency, and is thus adopted as the default configuration.

The second study investigates the effect of sampling size, with three settings: 200, 1000, and 5000 sampled vectors. The results are presented in Table~\ref{table:impact_samplings}.
Larger sampling sizes significantly increase offline computation time without consistently improving online search performance. Sampling 200 vectors provides the best overall trade-off between accuracy and efficiency.

The third study examines the influence of the decay function, which assigns weights used in Eq.~(\ref{equ:query_score}). We evaluate three alternatives: no decay function (None), a linear decay function (Linear), and an exponential decay function (Exp). The results are shown in Table~\ref{table:impact_weighted_function}.
The None setting assigns equal weights to all bins in Eq.~(\ref{equ:query_score}), failing to distinguish between easy and hard queries and resulting in lower recall, especially on MS MARCO V2.1. The Linear function gives higher weights to closer neighbors, partially improving discrimination but still less effectively than Exp. The Exponential decay function yields the best performance, avoiding both over- and under-searching, as evidenced by improvements in average recall as well as 1st- and 5th-percentile recall.

\section{Conclusion}
HNSW has been widely adopted in practice for efficient ANNS at scale. However, its search performance is highly sensitive to the choice of the \ef\ parameter, which governs the trade-off between efficiency and recall. The commonly used fixed \ef\ configuration suffers from practical limitations, including lack of recall guarantees and issues related to over- or under-searching across diverse queries.
In this paper, we identify these limitations and propose \texttt{Ada-\ef}, a distribution-aware approach that adaptively assigns an \ef\ value to each incoming query at runtime. 
\texttt{Ada-\ef} enables the search to approximately meet user-specified target recall while minimizing unnecessary computation, thus mitigating both under- and over-searching.
\texttt{Ada-\ef} is seamlessly integrable with existing HNSW indexes, requiring no modification to the index structure, and is friendly to updates, making it suitable for real-time and evolving workloads.
At the core of \texttt{Ada-\ef} lies a theoretical foundation based on similarity distance distribution, along with a tailored query scoring model designed to effectively characterize query difficulty. 
On real-world embeddings produced by advanced Transformer models from OpenAI and Cohere, \texttt{Ada-\ef} attains the declarative recall while mitigating over- and under-searching. Compared to state-of-the-art learning-based adaptive baselines, \texttt{Ada-\ef} achieves up to 4$\times$ reduction in online query latency, 50$\times$ reduction in offline computation, and 100$\times$ reduction in offline memory usage.

\bibliographystyle{ACM-Reference-Format}
\bibliography{publications,bibliography}


\begin{thebibliography}{101}


\ifx \showCODEN    \undefined \def \showCODEN     #1{\unskip}     \fi
\ifx \showDOI      \undefined \def \showDOI       #1{#1}\fi
\ifx \showISBNx    \undefined \def \showISBNx     #1{\unskip}     \fi
\ifx \showISBNxiii \undefined \def \showISBNxiii  #1{\unskip}     \fi
\ifx \showISSN     \undefined \def \showISSN      #1{\unskip}     \fi
\ifx \showLCCN     \undefined \def \showLCCN      #1{\unskip}     \fi
\ifx \shownote     \undefined \def \shownote      #1{#1}          \fi
\ifx \showarticletitle \undefined \def \showarticletitle #1{#1}   \fi
\ifx \showURL      \undefined \def \showURL       {\relax}        \fi
\providecommand\bibfield[2]{#2}
\providecommand\bibinfo[2]{#2}
\providecommand\natexlab[1]{#1}
\providecommand\showeprint[2][]{arXiv:#2}

\bibitem[\protect\citeauthoryear{??}{fai}{[n.d.]}]%
        {faiss}
 \bibinfo{year}{[n.d.]}\natexlab{}.
\newblock \bibinfo{booktitle}{\emph{FAISS: A Library for Efficient Similarity
  Search and Clustering of Dense Vectors}}.
\newblock
\urldef\tempurl%
\url{https://github.com/facebookresearch/faiss}
\showURL{%
\tempurl}
\newblock
\shownote{Accessed: 2025-10-21.}


\bibitem[\protect\citeauthoryear{??}{hns}{[n.d.]}]%
        {hnswlib}
 \bibinfo{year}{[n.d.]}\natexlab{}.
\newblock \bibinfo{booktitle}{\emph{Hnswlib: Fast Approximate Nearest Neighbor
  Search}}.
\newblock
\urldef\tempurl%
\url{https://github.com/nmslib/hnswlib}
\showURL{%
\tempurl}
\newblock
\shownote{Accessed: 2025-10-21.}


\bibitem[\protect\citeauthoryear{??}{eig}{2006}]%
        {eigen}
 \bibinfo{year}{2006}\natexlab{}.
\newblock \bibinfo{booktitle}{\emph{Eigen is a C++ template library for linear
  algebra: matrices, vectors, numerical solvers, and related algorithms}}.
\newblock
\urldef\tempurl%
\url{https://eigen.tuxfamily.org/index.php?title=Main_Page}
\showURL{%
\tempurl}
\newblock
\shownote{Accessed: 2025-01-10.}


\bibitem[\protect\citeauthoryear{??}{ann}{2019}]%
        {anns_benchmark}
 \bibinfo{year}{2019}\natexlab{}.
\newblock \bibinfo{booktitle}{\emph{ann-benchmarks}}.
\newblock
\urldef\tempurl%
\url{https://github.com/erikbern/ann-benchmarks}
\showURL{%
\tempurl}


\bibitem[\protect\citeauthoryear{??}{pin}{2019}]%
        {pinecone_website}
 \bibinfo{year}{2019}\natexlab{}.
\newblock \bibinfo{booktitle}{\emph{Pinecone}}.
\newblock
\urldef\tempurl%
\url{https://www.pinecone.io/}
\showURL{%
\tempurl}
\newblock
\shownote{Accessed: 2025-05-19.}


\bibitem[\protect\citeauthoryear{??}{wea}{2019}]%
        {weaviate_website}
 \bibinfo{year}{2019}\natexlab{}.
\newblock \bibinfo{booktitle}{\emph{Weaviate}}.
\newblock
\urldef\tempurl%
\url{https://weaviate.io/}
\showURL{%
\tempurl}
\newblock
\shownote{Accessed: 2025-03-19.}


\bibitem[\protect\citeauthoryear{??}{msm}{2020}]%
        {msmarco_benchmar}
 \bibinfo{year}{2020}\natexlab{}.
\newblock \bibinfo{booktitle}{\emph{MS MARCO}}.
\newblock
\urldef\tempurl%
\url{https://microsoft.github.io/msmarco}
\showURL{%
\tempurl}


\bibitem[\protect\citeauthoryear{??}{ope}{2021}]%
        {opensearch_lucene}
 \bibinfo{year}{2021}\natexlab{}.
\newblock \bibinfo{booktitle}{\emph{OpenSearch Project}}.
\newblock
\urldef\tempurl%
\url{https://github.com/opensearch-project}
\showURL{%
\tempurl}


\bibitem[\protect\citeauthoryear{??}{pgv}{2021}]%
        {pgvector_github}
 \bibinfo{year}{2021}\natexlab{}.
\newblock \bibinfo{booktitle}{\emph{pgvector: Open-source vector similarity
  search for Postgres}}.
\newblock


\bibitem[\protect\citeauthoryear{??}{luc}{2022}]%
        {lucene_hnswgraph}
 \bibinfo{year}{2022}\natexlab{}.
\newblock \bibinfo{booktitle}{\emph{HnswGraph (Lucene 9.8.0 core API)}}.
\newblock
\urldef\tempurl%
\url{https://lucene.apache.org/core/9_8_0/core/org/apache/lucene/util/hnsw/HnswGraph.html}
\showURL{%
\tempurl}


\bibitem[\protect\citeauthoryear{??}{msm}{2024a}]%
        {msmarco_openai}
 \bibinfo{year}{2024}\natexlab{a}.
\newblock \bibinfo{booktitle}{\emph{Anserini: OpenAI-ada2 Embeddings for MS
  MARCO Passage}}.
\newblock
\urldef\tempurl%
\url{https://github.com/castorini/anserini/blob/master/docs/experiments-msmarco-passage-openai-ada2.md?utm_source=chatgpt.com}
\showURL{%
\tempurl}
\newblock
\shownote{Accessed: 2025-03-20.}


\bibitem[\protect\citeauthoryear{??}{lai}{2024}]%
        {laion_indexing_challenge}
 \bibinfo{year}{2024}\natexlab{}.
\newblock \bibinfo{booktitle}{\emph{SISAP 2024 Indexing Challenge}}.
\newblock
\urldef\tempurl%
\url{https://sisap-challenges.github.io/2025/index.html}
\showURL{%
\tempurl}


\bibitem[\protect\citeauthoryear{??}{tre}{2024}]%
        {trec-rag}
 \bibinfo{year}{2024}\natexlab{}.
\newblock \bibinfo{booktitle}{\emph{TREC 2024 RAG Corpus}}.
\newblock
\urldef\tempurl%
\url{https://trec-rag.github.io/annoucements/2024-corpus-finalization}
\showURL{%
\tempurl}


\bibitem[\protect\citeauthoryear{??}{msm}{2024b}]%
        {msmarco_cohere}
 \bibinfo{year}{2024}\natexlab{b}.
\newblock \bibinfo{booktitle}{\emph{TREC-RAG 2024 Corpus (MSMARCO 2.1) -
  Encoded with Cohere Embed English v3}}.
\newblock
\urldef\tempurl%
\url{https://huggingface.co/datasets/Cohere/msmarco-v2.1-embed-english-v3}
\showURL{%
\tempurl}
\newblock
\shownote{Accessed: 2024-12-10.}


\bibitem[\protect\citeauthoryear{Ahle, Aum\"{u}ller, and Pagh}{Ahle
  et~al\mbox{.}}{2017}]%
        {10.5555/3039686.3039702}
\bibfield{author}{\bibinfo{person}{Thomas~D. Ahle}, \bibinfo{person}{Martin
  Aum\"{u}ller}, {and} \bibinfo{person}{Rasmus Pagh}.}
  \bibinfo{year}{2017}\natexlab{}.
\newblock \showarticletitle{Parameter-free locality sensitive hashing for
  spherical range reporting}. In \bibinfo{booktitle}{\emph{Proc. 28th Annual
  ACM-SIAM Symp. on Discrete Algorithms}} \emph{(\bibinfo{series}{SODA '17})}.
  \bibinfo{pages}{239–256}.
\newblock


\bibitem[\protect\citeauthoryear{Amsaleg, Chelly, Furon, Girard, Houle,
  Kawarabayashi, and Nett}{Amsaleg et~al\mbox{.}}{2015}]%
        {10.1145/2783258.2783405}
\bibfield{author}{\bibinfo{person}{Laurent Amsaleg}, \bibinfo{person}{Oussama
  Chelly}, \bibinfo{person}{Teddy Furon}, \bibinfo{person}{St\'{e}phane
  Girard}, \bibinfo{person}{Michael~E. Houle}, \bibinfo{person}{Ken-ichi
  Kawarabayashi}, {and} \bibinfo{person}{Michael Nett}.}
  \bibinfo{year}{2015}\natexlab{}.
\newblock \showarticletitle{Estimating Local Intrinsic Dimensionality}. In
  \bibinfo{booktitle}{\emph{Proc. 21st ACM SIGKDD Int. Conf. on Knowledge
  Discovery and Data Mining}}. \bibinfo{pages}{29–38}.
\newblock
\showISBNx{9781450336642}
\urldef\tempurl%
\url{https://doi.org/10.1145/2783258.2783405}
\showDOI{\tempurl}


\bibitem[\protect\citeauthoryear{Andr\'{e}, Kermarrec, and
  Le~Scouarnec}{Andr\'{e} et~al\mbox{.}}{2015}]%
        {10.14778/2856318.2856324}
\bibfield{author}{\bibinfo{person}{Fabien Andr\'{e}},
  \bibinfo{person}{Anne-Marie Kermarrec}, {and} \bibinfo{person}{Nicolas
  Le~Scouarnec}.} \bibinfo{year}{2015}\natexlab{}.
\newblock \showarticletitle{Cache locality is not enough: high-performance
  nearest neighbor search with product quantization fast scan}.
\newblock \bibinfo{journal}{\emph{Proc. VLDB Endowment}} \bibinfo{volume}{9},
  \bibinfo{number}{4} (\bibinfo{date}{Dec.} \bibinfo{year}{2015}),
  \bibinfo{pages}{288–299}.
\newblock
\showISSN{2150-8097}
\urldef\tempurl%
\url{https://doi.org/10.14778/2856318.2856324}
\showDOI{\tempurl}


\bibitem[\protect\citeauthoryear{Arora, Sinha, Kumar, and Bhattacharya}{Arora
  et~al\mbox{.}}{2018}]%
        {arora18hdindex}
\bibfield{author}{\bibinfo{person}{Akhil Arora}, \bibinfo{person}{Sakshi
  Sinha}, \bibinfo{person}{Piyush Kumar}, {and} \bibinfo{person}{Arnab
  Bhattacharya}.} \bibinfo{year}{2018}\natexlab{}.
\newblock \showarticletitle{HD-index: pushing the scalability-accuracy boundary
  for approximate kNN search in high-dimensional spaces}.
\newblock \bibinfo{journal}{\emph{Proc. VLDB Endowment}} \bibinfo{volume}{11},
  \bibinfo{number}{8} (\bibinfo{date}{April} \bibinfo{year}{2018}),
  \bibinfo{pages}{906–919}.
\newblock
\showISSN{2150-8097}
\urldef\tempurl%
\url{https://doi.org/10.14778/3204028.3204034}
\showDOI{\tempurl}


\bibitem[\protect\citeauthoryear{Arora, Li, Liang, Ma, and Risteski}{Arora
  et~al\mbox{.}}{2016}]%
        {arora-etal-2016-latent}
\bibfield{author}{\bibinfo{person}{Sanjeev Arora}, \bibinfo{person}{Yuanzhi
  Li}, \bibinfo{person}{Yingyu Liang}, \bibinfo{person}{Tengyu Ma}, {and}
  \bibinfo{person}{Andrej Risteski}.} \bibinfo{year}{2016}\natexlab{}.
\newblock \showarticletitle{A Latent Variable Model Approach to {PMI}-based
  Word Embeddings}.
\newblock \bibinfo{journal}{\emph{Transactions of the Association for
  Computational Linguistics}}  \bibinfo{volume}{4} (\bibinfo{year}{2016}),
  \bibinfo{pages}{385--399}.
\newblock
\urldef\tempurl%
\url{https://doi.org/10.1162/tacl_a_00106}
\showDOI{\tempurl}


\bibitem[\protect\citeauthoryear{Aumüller, Bernhardsson, and
  Faithfull}{Aumüller et~al\mbox{.}}{2018}]%
        {anns_bench}
\bibfield{author}{\bibinfo{person}{Martin Aumüller}, \bibinfo{person}{Erik
  Bernhardsson}, {and} \bibinfo{person}{Alexander Faithfull}.}
  \bibinfo{year}{2018}\natexlab{}.
\newblock \bibinfo{title}{ANN-Benchmarks: A Benchmarking Tool for Approximate
  Nearest Neighbor Algorithms}.
\newblock
\newblock
\showeprint[arxiv]{1807.05614}~[cs.IR]
\urldef\tempurl%
\url{https://arxiv.org/abs/1807.05614}
\showURL{%
\tempurl}


\bibitem[\protect\citeauthoryear{Azizi, Echihabi, and Palpanas}{Azizi
  et~al\mbox{.}}{2023}]%
        {Azizi2023ELPIS}
\bibfield{author}{\bibinfo{person}{Ilias Azizi}, \bibinfo{person}{Karima
  Echihabi}, {and} \bibinfo{person}{Themis Palpanas}.}
  \bibinfo{year}{2023}\natexlab{}.
\newblock \showarticletitle{ELPIS: Graph-Based Similarity Search for Scalable
  Data Science}.
\newblock \bibinfo{journal}{\emph{Proc. VLDB Endowment}} \bibinfo{volume}{16},
  \bibinfo{number}{6} (\bibinfo{date}{Feb.} \bibinfo{year}{2023}),
  \bibinfo{pages}{1548–1559}.
\newblock
\showISSN{2150-8097}
\urldef\tempurl%
\url{https://doi.org/10.14778/3583140.3583166}
\showDOI{\tempurl}


\bibitem[\protect\citeauthoryear{Azizi, Echihabi, and Palpanas}{Azizi
  et~al\mbox{.}}{2025}]%
        {azizi2025survey}
\bibfield{author}{\bibinfo{person}{Ilias Azizi}, \bibinfo{person}{Karima
  Echihabi}, {and} \bibinfo{person}{Themis Palpanas}.}
  \bibinfo{year}{2025}\natexlab{}.
\newblock \showarticletitle{Graph-Based Vector Search: An Experimental
  Evaluation of the State-of-the-Art}.
\newblock \bibinfo{journal}{\emph{Proc. ACM Manag. Data}} \bibinfo{volume}{3},
  \bibinfo{number}{1}, Article \bibinfo{articleno}{43} (\bibinfo{date}{Feb.}
  \bibinfo{year}{2025}), \bibinfo{numpages}{31}~pages.
\newblock
\urldef\tempurl%
\url{https://doi.org/10.1145/3709693}
\showDOI{\tempurl}


\bibitem[\protect\citeauthoryear{Ba, Kiros, and Hinton}{Ba
  et~al\mbox{.}}{2016}]%
        {ba2016layer}
\bibfield{author}{\bibinfo{person}{Jimmy~Lei Ba}, \bibinfo{person}{Jamie~Ryan
  Kiros}, {and} \bibinfo{person}{Geoffrey~E Hinton}.}
  \bibinfo{year}{2016}\natexlab{}.
\newblock \showarticletitle{Layer normalization}.
\newblock \bibinfo{journal}{\emph{arXiv preprint arXiv:1607.06450}}
  (\bibinfo{year}{2016}).
\newblock


\bibitem[\protect\citeauthoryear{Bajaj, Campos, Craswell, Deng, Gao, Liu,
  Majumder, McNamara, Mitra, Nguyen, Rosenberg, Song, Stoica, Tiwary, and
  Wang}{Bajaj et~al\mbox{.}}{2018}]%
        {bajaj2018msmarcohumangenerated}
\bibfield{author}{\bibinfo{person}{Payal Bajaj}, \bibinfo{person}{Daniel
  Campos}, \bibinfo{person}{Nick Craswell}, \bibinfo{person}{Li Deng},
  \bibinfo{person}{Jianfeng Gao}, \bibinfo{person}{Xiaodong Liu},
  \bibinfo{person}{Rangan Majumder}, \bibinfo{person}{Andrew McNamara},
  \bibinfo{person}{Bhaskar Mitra}, \bibinfo{person}{Tri Nguyen},
  \bibinfo{person}{Mir Rosenberg}, \bibinfo{person}{Xia Song},
  \bibinfo{person}{Alina Stoica}, \bibinfo{person}{Saurabh Tiwary}, {and}
  \bibinfo{person}{Tong Wang}.} \bibinfo{year}{2018}\natexlab{}.
\newblock \bibinfo{title}{MS MARCO: A Human Generated MAchine Reading
  COmprehension Dataset}.
\newblock
\newblock
\showeprint[arxiv]{1611.09268}~[cs.CL]
\urldef\tempurl%
\url{https://arxiv.org/abs/1611.09268}
\showURL{%
\tempurl}


\bibitem[\protect\citeauthoryear{Bernhardsson}{Bernhardsson}{2017}]%
        {bernhardsson2017annoy}
\bibfield{author}{\bibinfo{person}{Erik Bernhardsson}.}
  \bibinfo{year}{2017}\natexlab{}.
\newblock \bibinfo{booktitle}{\emph{Annoy: approximate nearest neighbors in
  C++/Python optimized for memory usage and loading/saving to disk}}.
\newblock
\urldef\tempurl%
\url{https://github.com/spotify/annoy}
\showURL{%
\tempurl}


\bibitem[\protect\citeauthoryear{Bogolin, Croitoru, Jin, Liu, and
  Albanie}{Bogolin et~al\mbox{.}}{2022}]%
        {9878857}
\bibfield{author}{\bibinfo{person}{Simion-Vlad Bogolin}, \bibinfo{person}{Ioana
  Croitoru}, \bibinfo{person}{Hailin Jin}, \bibinfo{person}{Yang Liu}, {and}
  \bibinfo{person}{Samuel Albanie}.} \bibinfo{year}{2022}\natexlab{}.
\newblock \showarticletitle{Cross Modal Retrieval with Querybank
  Normalisation}. In \bibinfo{booktitle}{\emph{Proc. IEEE Int. Conf. on
  Computer Vision and Pattern Recognition}}. \bibinfo{pages}{5184--5195}.
\newblock
\urldef\tempurl%
\url{https://doi.org/10.1109/CVPR52688.2022.00513}
\showDOI{\tempurl}


\bibitem[\protect\citeauthoryear{Chatzakis, Papakonstantinou, and
  Palpanas}{Chatzakis et~al\mbox{.}}{2025}]%
        {10.1145/3749160}
\bibfield{author}{\bibinfo{person}{Manos Chatzakis}, \bibinfo{person}{Yannis
  Papakonstantinou}, {and} \bibinfo{person}{Themis Palpanas}.}
  \bibinfo{year}{2025}\natexlab{}.
\newblock \showarticletitle{DARTH: Declarative Recall Through Early Termination
  for Approximate Nearest Neighbor Search}.
\newblock \bibinfo{journal}{\emph{Proc. ACM Manag. Data}} \bibinfo{volume}{3},
  \bibinfo{number}{4} (\bibinfo{date}{Sept.} \bibinfo{year}{2025}), 26.
\newblock
\urldef\tempurl%
\url{https://doi.org/10.1145/3749160}
\showDOI{\tempurl}


\bibitem[\protect\citeauthoryear{Chen, Jin, Zhang, Podolsky, Wu, Wang, Hanson,
  Sun, Walzer, and Wang}{Chen et~al\mbox{.}}{2024}]%
        {10.14778/3685800.3685805}
\bibfield{author}{\bibinfo{person}{Cheng Chen}, \bibinfo{person}{Chenzhe Jin},
  \bibinfo{person}{Yunan Zhang}, \bibinfo{person}{Sasha Podolsky},
  \bibinfo{person}{Chun Wu}, \bibinfo{person}{Szu-Po Wang},
  \bibinfo{person}{Eric Hanson}, \bibinfo{person}{Zhou Sun},
  \bibinfo{person}{Robert Walzer}, {and} \bibinfo{person}{Jianguo Wang}.}
  \bibinfo{year}{2024}\natexlab{}.
\newblock \showarticletitle{SingleStore-V: An Integrated Vector Database System
  in SingleStore}.
\newblock \bibinfo{journal}{\emph{Proc. VLDB Endowment}} \bibinfo{volume}{17},
  \bibinfo{number}{12} (\bibinfo{date}{Aug.} \bibinfo{year}{2024}),
  \bibinfo{pages}{3772–3785}.
\newblock
\showISSN{2150-8097}
\urldef\tempurl%
\url{https://doi.org/10.14778/3685800.3685805}
\showDOI{\tempurl}


\bibitem[\protect\citeauthoryear{Chen, Chang, Jiang, Yu, Dhillon, and
  Hsieh}{Chen et~al\mbox{.}}{2023}]%
        {10.1145/3543507.3583318}
\bibfield{author}{\bibinfo{person}{Patrick Chen}, \bibinfo{person}{Wei-Cheng
  Chang}, \bibinfo{person}{Jyun-Yu Jiang}, \bibinfo{person}{Hsiang-Fu Yu},
  \bibinfo{person}{Inderjit Dhillon}, {and} \bibinfo{person}{Cho-Jui Hsieh}.}
  \bibinfo{year}{2023}\natexlab{}.
\newblock \showarticletitle{FINGER: Fast Inference for Graph-based Approximate
  Nearest Neighbor Search}. In \bibinfo{booktitle}{\emph{ProC. Web Conf.
  2023}}. \bibinfo{pages}{3225–3235}.
\newblock
\showISBNx{9781450394161}
\urldef\tempurl%
\url{https://doi.org/10.1145/3543507.3583318}
\showDOI{\tempurl}


\bibitem[\protect\citeauthoryear{Chen, Wang, Li, Ren, Li, Zhu, Li, Liu, Zhang,
  and Wang}{Chen et~al\mbox{.}}{2018}]%
        {chen2018sptag}
\bibfield{author}{\bibinfo{person}{Qi Chen}, \bibinfo{person}{Haidong Wang},
  \bibinfo{person}{Mingqin Li}, \bibinfo{person}{Gang Ren},
  \bibinfo{person}{Scarlett Li}, \bibinfo{person}{Jeffery Zhu},
  \bibinfo{person}{Jason Li}, \bibinfo{person}{Chuanjie Liu},
  \bibinfo{person}{Lintao Zhang}, {and} \bibinfo{person}{Jingdong Wang}.}
  \bibinfo{year}{2018}\natexlab{}.
\newblock \bibinfo{title}{SPTAG: A library for fast approximate nearest
  neighbor search}.
\newblock
\newblock
\urldef\tempurl%
\url{https://github.com/Microsoft/SPTAG}
\showURL{%
\tempurl}


\bibitem[\protect\citeauthoryear{Chen, Zhao, Wang, Li, Liu, Li, Yang, and
  Wang}{Chen et~al\mbox{.}}{2021}]%
        {ChenW21}
\bibfield{author}{\bibinfo{person}{Qi Chen}, \bibinfo{person}{Bing Zhao},
  \bibinfo{person}{Haidong Wang}, \bibinfo{person}{Mingqin Li},
  \bibinfo{person}{Chuanjie Liu}, \bibinfo{person}{Zengzhong Li},
  \bibinfo{person}{Mao Yang}, {and} \bibinfo{person}{Jingdong Wang}.}
  \bibinfo{year}{2021}\natexlab{}.
\newblock \showarticletitle{SPANN: Highly-efficient Billion-scale Approximate
  Nearest Neighbor Search}. In \bibinfo{booktitle}{\emph{Advances in Neural
  Information Processing Syst., Proc. 35th Annual Conf. on Neural Information
  Processing Syst.}}
\newblock


\bibitem[\protect\citeauthoryear{Devlin, Chang, Lee, and Toutanova}{Devlin
  et~al\mbox{.}}{2019}]%
        {devlin-etal-2019-bert}
\bibfield{author}{\bibinfo{person}{Jacob Devlin}, \bibinfo{person}{Ming-Wei
  Chang}, \bibinfo{person}{Kenton Lee}, {and} \bibinfo{person}{Kristina
  Toutanova}.} \bibinfo{year}{2019}\natexlab{}.
\newblock \showarticletitle{{BERT}: Pre-training of Deep Bidirectional
  Transformers for Language Understanding}. In
  \bibinfo{booktitle}{\emph{Proceedings of the 2019 Conference of the North
  {A}merican Chapter of the Association for Computational Linguistics: Human
  Language Technologies, Volume 1 (Long and Short Papers)}}.
  \bibinfo{pages}{4171--4186}.
\newblock
\urldef\tempurl%
\url{https://doi.org/10.18653/v1/N19-1423}
\showDOI{\tempurl}


\bibitem[\protect\citeauthoryear{Dong}{Dong}{2011}]%
        {dong2011kgraph}
\bibfield{author}{\bibinfo{person}{Wei Dong}.} \bibinfo{year}{2011}\natexlab{}.
\newblock \bibinfo{booktitle}{\emph{KGraph: A Library for Approximate Nearest
  Neighbor Search}}.
\newblock
\urldef\tempurl%
\url{https://github.com/aaalgo/kgraph}
\showURL{%
\tempurl}


\bibitem[\protect\citeauthoryear{Dosovitskiy, Beyer, Kolesnikov, Weissenborn,
  Zhai, Unterthiner, Dehghani, Minderer, Heigold, Gelly, Uszkoreit, and
  Houlsby}{Dosovitskiy et~al\mbox{.}}{2021}]%
        {dosovitskiy2021an}
\bibfield{author}{\bibinfo{person}{Alexey Dosovitskiy}, \bibinfo{person}{Lucas
  Beyer}, \bibinfo{person}{Alexander Kolesnikov}, \bibinfo{person}{Dirk
  Weissenborn}, \bibinfo{person}{Xiaohua Zhai}, \bibinfo{person}{Thomas
  Unterthiner}, \bibinfo{person}{Mostafa Dehghani}, \bibinfo{person}{Matthias
  Minderer}, \bibinfo{person}{Georg Heigold}, \bibinfo{person}{Sylvain Gelly},
  \bibinfo{person}{Jakob Uszkoreit}, {and} \bibinfo{person}{Neil Houlsby}.}
  \bibinfo{year}{2021}\natexlab{}.
\newblock \showarticletitle{An Image is Worth 16x16 Words: Transformers for
  Image Recognition at Scale}. In \bibinfo{booktitle}{\emph{Proc. 9th Int.
  Conf. on Learning Representations}}.
\newblock
\urldef\tempurl%
\url{https://openreview.net/forum?id=YicbFdNTTy}
\showURL{%
\tempurl}


\bibitem[\protect\citeauthoryear{Douze, Guzhva, Deng, Johnson, Szilvasy,
  Mazaré, Lomeli, Hosseini, and Jégou}{Douze et~al\mbox{.}}{2024}]%
        {douze2024faiss}
\bibfield{author}{\bibinfo{person}{Matthijs Douze}, \bibinfo{person}{Alexandr
  Guzhva}, \bibinfo{person}{Chengqi Deng}, \bibinfo{person}{Jeff Johnson},
  \bibinfo{person}{Gergely Szilvasy}, \bibinfo{person}{Pierre-Emmanuel
  Mazaré}, \bibinfo{person}{Maria Lomeli}, \bibinfo{person}{Lucas Hosseini},
  {and} \bibinfo{person}{Hervé Jégou}.} \bibinfo{year}{2024}\natexlab{}.
\newblock \showarticletitle{The Faiss library}.
\newblock  (\bibinfo{year}{2024}).
\newblock
\showeprint[arxiv]{2401.08281}~[cs.LG]


\bibitem[\protect\citeauthoryear{Elliott}{Elliott}{2024}]%
        {practical_ef_recall}
\bibfield{author}{\bibinfo{person}{Owen Elliott}.}
  \bibinfo{year}{2024}\natexlab{}.
\newblock \bibinfo{booktitle}{\emph{Understanding Recall in HNSW Search}}.
\newblock
\urldef\tempurl%
\url{https://www.marqo.ai/blog/understanding-recall-in-hnsw-search}
\showURL{%
\tempurl}


\bibitem[\protect\citeauthoryear{Ethayarajh}{Ethayarajh}{2019}]%
        {ethayarajh-2019-contextual}
\bibfield{author}{\bibinfo{person}{Kawin Ethayarajh}.}
  \bibinfo{year}{2019}\natexlab{}.
\newblock \showarticletitle{How Contextual are Contextualized Word
  Representations? {C}omparing the Geometry of {BERT}, {ELM}o, and {GPT}-2
  Embeddings}. In \bibinfo{booktitle}{\emph{Proceedings of the 2019 Conference
  on Empirical Methods in Natural Language Processing and the 9th International
  Joint Conference on Natural Language Processing (EMNLP-IJCNLP)}}.
  \bibinfo{pages}{55--65}.
\newblock
\urldef\tempurl%
\url{https://doi.org/10.18653/v1/D19-1006}
\showDOI{\tempurl}


\bibitem[\protect\citeauthoryear{Feng, Jin, Chen, Liu, and
  Saliho{\u{g}}lu}{Feng et~al\mbox{.}}{2023}]%
        {feng2023kuzu}
\bibfield{author}{\bibinfo{person}{Xiyang Feng}, \bibinfo{person}{Guodong Jin},
  \bibinfo{person}{Ziyi Chen}, \bibinfo{person}{Chang Liu}, {and}
  \bibinfo{person}{Semih Saliho{\u{g}}lu}.} \bibinfo{year}{2023}\natexlab{}.
\newblock \showarticletitle{K{\`u}zu graph database management system}. In
  \bibinfo{booktitle}{\emph{The Conf. on Innovative Data Systems Research}},
  Vol.~\bibinfo{volume}{7}. \bibinfo{pages}{25--35}.
\newblock


\bibitem[\protect\citeauthoryear{Friedman}{Friedman}{2001}]%
        {4a848dd1-54e3-3c3c-83c3-04977ded2e71}
\bibfield{author}{\bibinfo{person}{Jerome~H. Friedman}.}
  \bibinfo{year}{2001}\natexlab{}.
\newblock \showarticletitle{Greedy Function Approximation: A Gradient Boosting
  Machine}.
\newblock \bibinfo{journal}{\emph{The Annals of Statistics}}
  \bibinfo{volume}{29}, \bibinfo{number}{5} (\bibinfo{year}{2001}),
  \bibinfo{pages}{1189--1232}.
\newblock
\showISSN{00905364, 21688966}
\urldef\tempurl%
\url{http://www.jstor.org/stable/2699986}
\showURL{%
\tempurl}


\bibitem[\protect\citeauthoryear{Fu and Cai}{Fu and Cai}{2016}]%
        {fu2016efanna}
\bibfield{author}{\bibinfo{person}{Cong Fu} {and} \bibinfo{person}{Deng Cai}.}
  \bibinfo{year}{2016}\natexlab{}.
\newblock \bibinfo{title}{EFANNA : An Extremely Fast Approximate Nearest
  Neighbor Search Algorithm Based on kNN Graph}.
\newblock
\newblock
\showeprint[arxiv]{1609.07228}~[cs.CV]
\urldef\tempurl%
\url{https://arxiv.org/abs/1609.07228}
\showURL{%
\tempurl}


\bibitem[\protect\citeauthoryear{Fu, Wang, and Cai}{Fu et~al\mbox{.}}{2022}]%
        {Fu2022NSSG}
\bibfield{author}{\bibinfo{person}{Cong Fu}, \bibinfo{person}{Changxu Wang},
  {and} \bibinfo{person}{Deng Cai}.} \bibinfo{year}{2022}\natexlab{}.
\newblock \showarticletitle{High Dimensional Similarity Search With Satellite
  System Graph: Efficiency, Scalability, and Unindexed Query Compatibility}.
\newblock \bibinfo{journal}{\emph{IEEE Trans. Pattern Analy. Machine Intell.}}
  \bibinfo{volume}{44}, \bibinfo{number}{8} (\bibinfo{year}{2022}),
  \bibinfo{pages}{4139--4150}.
\newblock
\urldef\tempurl%
\url{https://doi.org/10.1109/TPAMI.2021.3067706}
\showDOI{\tempurl}


\bibitem[\protect\citeauthoryear{Fu, Xiang, Wang, and Cai}{Fu
  et~al\mbox{.}}{2019}]%
        {Fu2019NSG}
\bibfield{author}{\bibinfo{person}{Cong Fu}, \bibinfo{person}{Chao Xiang},
  \bibinfo{person}{Changxu Wang}, {and} \bibinfo{person}{Deng Cai}.}
  \bibinfo{year}{2019}\natexlab{}.
\newblock \showarticletitle{Fast approximate nearest neighbor search with the
  navigating spreading-out graph}.
\newblock \bibinfo{journal}{\emph{Proc. VLDB Endowment}} \bibinfo{volume}{12},
  \bibinfo{number}{5} (\bibinfo{date}{Jan.} \bibinfo{year}{2019}),
  \bibinfo{pages}{461–474}.
\newblock
\showISSN{2150-8097}
\urldef\tempurl%
\url{https://doi.org/10.14778/3303753.3303754}
\showDOI{\tempurl}


\bibitem[\protect\citeauthoryear{Gabrielsson}{Gabrielsson}{2024}]%
        {gabrielsson2024vectorsimilarity}
\bibfield{author}{\bibinfo{person}{Max Gabrielsson}.}
  \bibinfo{year}{2024}\natexlab{}.
\newblock \bibinfo{booktitle}{\emph{Vector Similarity Search in DuckDB}}.
\newblock
\urldef\tempurl%
\url{https://duckdb.org/2024/05/03/vector-similarity-search-vss.html}
\showURL{%
\tempurl}


\bibitem[\protect\citeauthoryear{Gao, Xiong, Gao, Jia, Pan, Bi, Dai, Sun, Wang,
  and Wang}{Gao et~al\mbox{.}}{2023}]%
        {gao2023retrieval}
\bibfield{author}{\bibinfo{person}{Yunfan Gao}, \bibinfo{person}{Yun Xiong},
  \bibinfo{person}{Xinyu Gao}, \bibinfo{person}{Kangxiang Jia},
  \bibinfo{person}{Jinliu Pan}, \bibinfo{person}{Yuxi Bi}, \bibinfo{person}{Yi
  Dai}, \bibinfo{person}{Jiawei Sun}, \bibinfo{person}{Haofen Wang}, {and}
  \bibinfo{person}{Haofen Wang}.} \bibinfo{year}{2023}\natexlab{}.
\newblock \showarticletitle{Retrieval-augmented generation for large language
  models: A survey}.
\newblock \bibinfo{journal}{\emph{arXiv preprint arXiv:2312.10997}}
  \bibinfo{volume}{2} (\bibinfo{year}{2023}).
\newblock


\bibitem[\protect\citeauthoryear{Gong, Wang, Ogihara, and Xu}{Gong
  et~al\mbox{.}}{2020}]%
        {10.14778/3397230.3397243}
\bibfield{author}{\bibinfo{person}{Long Gong}, \bibinfo{person}{Huayi Wang},
  \bibinfo{person}{Mitsunori Ogihara}, {and} \bibinfo{person}{Jun Xu}.}
  \bibinfo{year}{2020}\natexlab{}.
\newblock \showarticletitle{iDEC: indexable distance estimating codes for
  approximate nearest neighbor search}.
\newblock \bibinfo{journal}{\emph{Proc. VLDB Endowment}} \bibinfo{volume}{13},
  \bibinfo{number}{9} (\bibinfo{date}{May} \bibinfo{year}{2020}),
  \bibinfo{pages}{1483–1497}.
\newblock
\showISSN{2150-8097}
\urldef\tempurl%
\url{https://doi.org/10.14778/3397230.3397243}
\showDOI{\tempurl}


\bibitem[\protect\citeauthoryear{Greene, Sanders, Weng, and Neelakantan}{Greene
  et~al\mbox{.}}{2022}]%
        {openai-ada2}
\bibfield{author}{\bibinfo{person}{Ryan Greene}, \bibinfo{person}{Ted Sanders},
  \bibinfo{person}{Lilian Weng}, {and} \bibinfo{person}{Arvind Neelakantan}.}
  \bibinfo{year}{2022}\natexlab{}.
\newblock \bibinfo{booktitle}{\emph{New and improved embedding model:
  text-embedding-ada-002}}.
\newblock
\urldef\tempurl%
\url{https://openai.com/index/new-and-improved-embedding-model}
\showURL{%
\tempurl}


\bibitem[\protect\citeauthoryear{Harwood and Drummond}{Harwood and
  Drummond}{2016}]%
        {Harwood2016FANNG}
\bibfield{author}{\bibinfo{person}{Ben Harwood} {and} \bibinfo{person}{Tom
  Drummond}.} \bibinfo{year}{2016}\natexlab{}.
\newblock \showarticletitle{FANNG: Fast Approximate Nearest Neighbour Graphs}.
  In \bibinfo{booktitle}{\emph{Proc. IEEE Int. Conf. on Computer Vision and
  Pattern Recognition}}.
\newblock


\bibitem[\protect\citeauthoryear{He, Kumar, and Chang}{He
  et~al\mbox{.}}{2012}]%
        {10.5555/3042573.3042582}
\bibfield{author}{\bibinfo{person}{Junfeng He}, \bibinfo{person}{Sanjiv Kumar},
  {and} \bibinfo{person}{Shih-Fu Chang}.} \bibinfo{year}{2012}\natexlab{}.
\newblock \showarticletitle{On the difficulty of nearest neighbor search}. In
  \bibinfo{booktitle}{\emph{Proc. 29th Int. Conf. on Machine Learning}}.
  \bibinfo{pages}{41–48}.
\newblock


\bibitem[\protect\citeauthoryear{Huang, Feng, Zhang, Fang, and Ng}{Huang
  et~al\mbox{.}}{2015}]%
        {10.14778/2850469.2850470}
\bibfield{author}{\bibinfo{person}{Qiang Huang}, \bibinfo{person}{Jianlin
  Feng}, \bibinfo{person}{Yikai Zhang}, \bibinfo{person}{Qiong Fang}, {and}
  \bibinfo{person}{Wilfred Ng}.} \bibinfo{year}{2015}\natexlab{}.
\newblock \showarticletitle{Query-aware locality-sensitive hashing for
  approximate nearest neighbor search}.
\newblock \bibinfo{journal}{\emph{Proc. VLDB Endowment}} \bibinfo{volume}{9},
  \bibinfo{number}{1} (\bibinfo{date}{Sept.} \bibinfo{year}{2015}),
  \bibinfo{pages}{1–12}.
\newblock
\showISSN{2150-8097}
\urldef\tempurl%
\url{https://doi.org/10.14778/2850469.2850470}
\showDOI{\tempurl}


\bibitem[\protect\citeauthoryear{Indyk and Motwani}{Indyk and Motwani}{1998}]%
        {10.1145/276698.276876}
\bibfield{author}{\bibinfo{person}{Piotr Indyk} {and} \bibinfo{person}{Rajeev
  Motwani}.} \bibinfo{year}{1998}\natexlab{}.
\newblock \showarticletitle{Approximate nearest neighbors: towards removing the
  curse of dimensionality}. In \bibinfo{booktitle}{\emph{Proceedings of the
  Thirtieth Annual ACM Symposium on Theory of Computing}}.
  \bibinfo{pages}{604–613}.
\newblock
\showISBNx{0897919629}
\urldef\tempurl%
\url{https://doi.org/10.1145/276698.276876}
\showDOI{\tempurl}


\bibitem[\protect\citeauthoryear{Iwasaki}{Iwasaki}{2015}]%
        {iwasaki2015ngt}
\bibfield{author}{\bibinfo{person}{Masajiro Iwasaki}.}
  \bibinfo{year}{2015}\natexlab{}.
\newblock \bibinfo{booktitle}{\emph{Neighborhood Graph and Tree for Indexing
  Highdimensional Data}}.
\newblock Yahoo Japan Corporation.
\newblock
\newblock
\shownote{https://github.com/yahoojapan/NGT.}


\bibitem[\protect\citeauthoryear{Jafari, Maurya, Nagarkar, Islam, and
  Crushev}{Jafari et~al\mbox{.}}{2021}]%
        {jafari2021surveylocalitysensitivehashing}
\bibfield{author}{\bibinfo{person}{Omid Jafari}, \bibinfo{person}{Preeti
  Maurya}, \bibinfo{person}{Parth Nagarkar},
  \bibinfo{person}{Khandker~Mushfiqul Islam}, {and}
  \bibinfo{person}{Chidambaram Crushev}.} \bibinfo{year}{2021}\natexlab{}.
\newblock \bibinfo{title}{A Survey on Locality Sensitive Hashing Algorithms and
  their Applications}.
\newblock
\newblock
\showeprint[arxiv]{2102.08942}~[cs.DB]
\urldef\tempurl%
\url{https://arxiv.org/abs/2102.08942}
\showURL{%
\tempurl}


\bibitem[\protect\citeauthoryear{Jayaram~Subramanya, Devvrit, Simhadri,
  Krishnawamy, and Kadekodi}{Jayaram~Subramanya et~al\mbox{.}}{2019}]%
        {Subramanya2019Vamana}
\bibfield{author}{\bibinfo{person}{Suhas Jayaram~Subramanya},
  \bibinfo{person}{Fnu Devvrit}, \bibinfo{person}{Harsha~Vardhan Simhadri},
  \bibinfo{person}{Ravishankar Krishnawamy}, {and} \bibinfo{person}{Rohan
  Kadekodi}.} \bibinfo{year}{2019}\natexlab{}.
\newblock \showarticletitle{DiskANN: Fast Accurate Billion-point Nearest
  Neighbor Search on a Single Node}. In \bibinfo{booktitle}{\emph{Advances in
  Neural Information Processing Syst., Proc. 33rd Annual Conf. on Neural
  Information Processing Syst.}}, Vol.~\bibinfo{volume}{32}.
\newblock
\urldef\tempurl%
\url{https://proceedings.neurips.cc/paper_files/paper/2019/file/09853c7fb1d3f8ee67a61b6bf4a7f8e6-Paper.pdf}
\showURL{%
\tempurl}


\bibitem[\protect\citeauthoryear{Jin, Zhang, Hu, Lin, Cai, and He}{Jin
  et~al\mbox{.}}{2014}]%
        {jin2014IEH}
\bibfield{author}{\bibinfo{person}{Zhongming Jin}, \bibinfo{person}{Debing
  Zhang}, \bibinfo{person}{Yao Hu}, \bibinfo{person}{Shiding Lin},
  \bibinfo{person}{Deng Cai}, {and} \bibinfo{person}{Xiaofei He}.}
  \bibinfo{year}{2014}\natexlab{}.
\newblock \showarticletitle{Fast and Accurate Hashing Via Iterative Nearest
  Neighbors Expansion}.
\newblock \bibinfo{journal}{\emph{IEEE Trans. Cybernetics}}
  \bibinfo{volume}{44}, \bibinfo{number}{11} (\bibinfo{year}{2014}),
  \bibinfo{pages}{2167--2177}.
\newblock
\urldef\tempurl%
\url{https://doi.org/10.1109/TCYB.2014.2302018}
\showDOI{\tempurl}


\bibitem[\protect\citeauthoryear{Jégou, Douze, and Schmid}{Jégou
  et~al\mbox{.}}{2011}]%
        {5432202}
\bibfield{author}{\bibinfo{person}{Herve Jégou}, \bibinfo{person}{Matthijs
  Douze}, {and} \bibinfo{person}{Cordelia Schmid}.}
  \bibinfo{year}{2011}\natexlab{}.
\newblock \showarticletitle{Product Quantization for Nearest Neighbor Search}.
\newblock \bibinfo{journal}{\emph{IEEE Trans. Pattern Analy. Machine Intell.}}
  \bibinfo{volume}{33}, \bibinfo{number}{1} (\bibinfo{year}{2011}),
  \bibinfo{pages}{117--128}.
\newblock
\urldef\tempurl%
\url{https://doi.org/10.1109/TPAMI.2010.57}
\showDOI{\tempurl}


\bibitem[\protect\citeauthoryear{Karpukhin, Oguz, Min, Lewis, Wu, Edunov, Chen,
  and Yih}{Karpukhin et~al\mbox{.}}{2020}]%
        {karpukhin-etal-2020-dense}
\bibfield{author}{\bibinfo{person}{Vladimir Karpukhin}, \bibinfo{person}{Barlas
  Oguz}, \bibinfo{person}{Sewon Min}, \bibinfo{person}{Patrick Lewis},
  \bibinfo{person}{Ledell Wu}, \bibinfo{person}{Sergey Edunov},
  \bibinfo{person}{Danqi Chen}, {and} \bibinfo{person}{Wen-tau Yih}.}
  \bibinfo{year}{2020}\natexlab{}.
\newblock \showarticletitle{Dense Passage Retrieval for Open-Domain Question
  Answering}. In \bibinfo{booktitle}{\emph{Proceedings of the 2020 Conference
  on Empirical Methods in Natural Language Processing (EMNLP)}}.
  \bibinfo{pages}{6769--6781}.
\newblock
\urldef\tempurl%
\url{https://doi.org/10.18653/v1/2020.emnlp-main.550}
\showDOI{\tempurl}


\bibitem[\protect\citeauthoryear{Klenke}{Klenke}{2020}]%
        {Klenke2020}
\bibfield{author}{\bibinfo{person}{Achim Klenke}.}
  \bibinfo{year}{2020}\natexlab{}.
\newblock \bibinfo{booktitle}{\emph{Characteristic Functions and the Central
  Limit Theorem}}.
\newblock \bibinfo{publisher}{Springer International Publishing},
  \bibinfo{address}{Cham}, \bibinfo{pages}{327--366}.
\newblock
\showISBNx{978-3-030-56402-5}
\urldef\tempurl%
\url{https://doi.org/10.1007/978-3-030-56402-5_15}
\showDOI{\tempurl}


\bibitem[\protect\citeauthoryear{Li, Zhou, He, Wang, Yang, and Li}{Li
  et~al\mbox{.}}{2020c}]%
        {li-etal-2020-sentence}
\bibfield{author}{\bibinfo{person}{Bohan Li}, \bibinfo{person}{Hao Zhou},
  \bibinfo{person}{Junxian He}, \bibinfo{person}{Mingxuan Wang},
  \bibinfo{person}{Yiming Yang}, {and} \bibinfo{person}{Lei Li}.}
  \bibinfo{year}{2020}\natexlab{c}.
\newblock \showarticletitle{On the Sentence Embeddings from Pre-trained
  Language Models}. In \bibinfo{booktitle}{\emph{Proceedings of the 2020
  Conference on Empirical Methods in Natural Language Processing (EMNLP)}}.
  \bibinfo{pages}{9119--9130}.
\newblock
\urldef\tempurl%
\url{https://doi.org/10.18653/v1/2020.emnlp-main.733}
\showDOI{\tempurl}


\bibitem[\protect\citeauthoryear{Li, Zhang, Andersen, and He}{Li
  et~al\mbox{.}}{2020a}]%
        {10.1145/3318464.3380600}
\bibfield{author}{\bibinfo{person}{Conglong Li}, \bibinfo{person}{Minjia
  Zhang}, \bibinfo{person}{David~G. Andersen}, {and} \bibinfo{person}{Yuxiong
  He}.} \bibinfo{year}{2020}\natexlab{a}.
\newblock \showarticletitle{Improving Approximate Nearest Neighbor Search
  through Learned Adaptive Early Termination}. In
  \bibinfo{booktitle}{\emph{Proc. ACM SIGMOD Int. Conf. on Management of
  Data}}. \bibinfo{pages}{2539–2554}.
\newblock
\showISBNx{9781450367356}
\urldef\tempurl%
\url{https://doi.org/10.1145/3318464.3380600}
\showDOI{\tempurl}


\bibitem[\protect\citeauthoryear{Li, Ai, Zhan, Mao, Liu, Liu, and Cao}{Li
  et~al\mbox{.}}{2023}]%
        {10.1145/3539618.3591651}
\bibfield{author}{\bibinfo{person}{Haitao Li}, \bibinfo{person}{Qingyao Ai},
  \bibinfo{person}{Jingtao Zhan}, \bibinfo{person}{Jiaxin Mao},
  \bibinfo{person}{Yiqun Liu}, \bibinfo{person}{Zheng Liu}, {and}
  \bibinfo{person}{Zhao Cao}.} \bibinfo{year}{2023}\natexlab{}.
\newblock \showarticletitle{Constructing Tree-based Index for Efficient and
  Effective Dense Retrieval}. In \bibinfo{booktitle}{\emph{Proc. 46th Annual
  Int. ACM SIGIR Conf. on Research and Development in Information Retrieval}}.
  \bibinfo{pages}{131–140}.
\newblock
\showISBNx{9781450394086}
\urldef\tempurl%
\url{https://doi.org/10.1145/3539618.3591651}
\showDOI{\tempurl}


\bibitem[\protect\citeauthoryear{Li, Zhang, Sun, Wang, Li, Zhang, and Lin}{Li
  et~al\mbox{.}}{2020b}]%
        {li2019dpg}
\bibfield{author}{\bibinfo{person}{Wen Li}, \bibinfo{person}{Ying Zhang},
  \bibinfo{person}{Yifang Sun}, \bibinfo{person}{Wei Wang},
  \bibinfo{person}{Mingjie Li}, \bibinfo{person}{Wenjie Zhang}, {and}
  \bibinfo{person}{Xuemin Lin}.} \bibinfo{year}{2020}\natexlab{b}.
\newblock \showarticletitle{Approximate Nearest Neighbor Search on High
  Dimensional Data — Experiments, Analyses, and Improvement}.
\newblock \bibinfo{journal}{\emph{IEEE Trans. Knowl. and Data Eng.}}
  \bibinfo{volume}{32}, \bibinfo{number}{8} (\bibinfo{year}{2020}),
  \bibinfo{pages}{1475--1488}.
\newblock
\urldef\tempurl%
\url{https://doi.org/10.1109/TKDE.2019.2909204}
\showDOI{\tempurl}


\bibitem[\protect\citeauthoryear{Liu, Zeng, Chen, Ainihaer, Ramasami, Chen, Xu,
  Wu, and Wang}{Liu et~al\mbox{.}}{2025}]%
        {TigerVector}
\bibfield{author}{\bibinfo{person}{Shige Liu}, \bibinfo{person}{Zhifang Zeng},
  \bibinfo{person}{Li Chen}, \bibinfo{person}{Adil Ainihaer},
  \bibinfo{person}{Arun Ramasami}, \bibinfo{person}{Songting Chen},
  \bibinfo{person}{Yu Xu}, \bibinfo{person}{Mingxi Wu}, {and}
  \bibinfo{person}{Jianguo Wang}.} \bibinfo{year}{2025}\natexlab{}.
\newblock \bibinfo{title}{TigerVector: Supporting Vector Search in Graph
  Databases for Advanced RAGs}.
\newblock
\newblock
\showeprint[arxiv]{2501.11216}~[cs.DB]
\urldef\tempurl%
\url{https://arxiv.org/abs/2501.11216}
\showURL{%
\tempurl}


\bibitem[\protect\citeauthoryear{Lu, Wang, Wang, and Kudo}{Lu
  et~al\mbox{.}}{2020}]%
        {10.14778/3397230.3397240}
\bibfield{author}{\bibinfo{person}{Kejing Lu}, \bibinfo{person}{Hongya Wang},
  \bibinfo{person}{Wei Wang}, {and} \bibinfo{person}{Mineichi Kudo}.}
  \bibinfo{year}{2020}\natexlab{}.
\newblock \showarticletitle{VHP: approximate nearest neighbor search via
  virtual hypersphere partitioning}.
\newblock \bibinfo{journal}{\emph{Proc. VLDB Endowment}} \bibinfo{volume}{13},
  \bibinfo{number}{9} (\bibinfo{date}{May} \bibinfo{year}{2020}),
  \bibinfo{pages}{1443–1455}.
\newblock
\showISSN{2150-8097}
\urldef\tempurl%
\url{https://doi.org/10.14778/3397230.3397240}
\showDOI{\tempurl}


\bibitem[\protect\citeauthoryear{Malkov, Ponomarenko, Logvinov, and
  Krylov}{Malkov et~al\mbox{.}}{2014}]%
        {MALKOV2014NSW}
\bibfield{author}{\bibinfo{person}{Yury Malkov}, \bibinfo{person}{Alexander
  Ponomarenko}, \bibinfo{person}{Andrey Logvinov}, {and}
  \bibinfo{person}{Vladimir Krylov}.} \bibinfo{year}{2014}\natexlab{}.
\newblock \showarticletitle{Approximate nearest neighbor algorithm based on
  navigable small world graphs}.
\newblock \bibinfo{journal}{\emph{Information Systems}}  \bibinfo{volume}{45}
  (\bibinfo{year}{2014}), \bibinfo{pages}{61--68}.
\newblock
\showISSN{0306-4379}
\urldef\tempurl%
\url{https://doi.org/10.1016/j.is.2013.10.006}
\showDOI{\tempurl}


\bibitem[\protect\citeauthoryear{Malkov and Yashunin}{Malkov and
  Yashunin}{2020}]%
        {Malkov2020HNSW}
\bibfield{author}{\bibinfo{person}{Yu~A. Malkov} {and} \bibinfo{person}{D.~A.
  Yashunin}.} \bibinfo{year}{2020}\natexlab{}.
\newblock \showarticletitle{Efficient and Robust Approximate Nearest Neighbor
  Search Using Hierarchical Navigable Small World Graphs}.
\newblock \bibinfo{journal}{\emph{IEEE Trans. Pattern Analy. Machine Intell.}}
  \bibinfo{volume}{42}, \bibinfo{number}{4} (\bibinfo{year}{2020}),
  \bibinfo{pages}{824--836}.
\newblock
\urldef\tempurl%
\url{https://doi.org/10.1109/TPAMI.2018.2889473}
\showDOI{\tempurl}


\bibitem[\protect\citeauthoryear{Meng, Bradley, Yavuz, Sparks, Venkataraman,
  Liu, Freeman, Tsai, Amde, Owen, Xin, Xin, Franklin, Zadeh, Zaharia, and
  Talwalkar}{Meng et~al\mbox{.}}{2016}]%
        {JMLR:v17:15-237}
\bibfield{author}{\bibinfo{person}{Xiangrui Meng}, \bibinfo{person}{Joseph
  Bradley}, \bibinfo{person}{Burak Yavuz}, \bibinfo{person}{Evan Sparks},
  \bibinfo{person}{Shivaram Venkataraman}, \bibinfo{person}{Davies Liu},
  \bibinfo{person}{Jeremy Freeman}, \bibinfo{person}{DB Tsai},
  \bibinfo{person}{Manish Amde}, \bibinfo{person}{Sean Owen},
  \bibinfo{person}{Doris Xin}, \bibinfo{person}{Reynold Xin},
  \bibinfo{person}{Michael~J. Franklin}, \bibinfo{person}{Reza Zadeh},
  \bibinfo{person}{Matei Zaharia}, {and} \bibinfo{person}{Ameet Talwalkar}.}
  \bibinfo{year}{2016}\natexlab{}.
\newblock \showarticletitle{MLlib: Machine Learning in Apache Spark}.
\newblock \bibinfo{journal}{\emph{Journal of Machine Learning Research}}
  \bibinfo{volume}{17}, \bibinfo{number}{34} (\bibinfo{year}{2016}),
  \bibinfo{pages}{1--7}.
\newblock
\urldef\tempurl%
\url{http://jmlr.org/papers/v17/15-237.html}
\showURL{%
\tempurl}


\bibitem[\protect\citeauthoryear{Mikolov, Sutskever, Chen, Corrado, and
  Dean}{Mikolov et~al\mbox{.}}{2013}]%
        {NIPS2013_9aa42b31}
\bibfield{author}{\bibinfo{person}{Tomas Mikolov}, \bibinfo{person}{Ilya
  Sutskever}, \bibinfo{person}{Kai Chen}, \bibinfo{person}{Greg~S Corrado},
  {and} \bibinfo{person}{Jeff Dean}.} \bibinfo{year}{2013}\natexlab{}.
\newblock \showarticletitle{Distributed Representations of Words and Phrases
  and their Compositionality}. In \bibinfo{booktitle}{\emph{Advances in Neural
  Information Processing Syst., Proc. 27th Annual Conf. on Neural Information
  Processing Syst.}}, \bibfield{editor}{\bibinfo{person}{C.J. Burges},
  \bibinfo{person}{L.~Bottou}, \bibinfo{person}{M.~Welling},
  \bibinfo{person}{Z.~Ghahramani}, {and} \bibinfo{person}{K.Q. Weinberger}}
  (Eds.), Vol.~\bibinfo{volume}{26}. \bibinfo{publisher}{Curran Associates,
  Inc.}
\newblock
\urldef\tempurl%
\url{https://proceedings.neurips.cc/paper_files/paper/2013/file/9aa42b31882ec039965f3c4923ce901b-Paper.pdf}
\showURL{%
\tempurl}


\bibitem[\protect\citeauthoryear{Mu and Viswanath}{Mu and Viswanath}{2018}]%
        {mu2018allbutthetop}
\bibfield{author}{\bibinfo{person}{Jiaqi Mu} {and} \bibinfo{person}{Pramod
  Viswanath}.} \bibinfo{year}{2018}\natexlab{}.
\newblock \showarticletitle{All-but-the-Top: Simple and Effective
  Postprocessing for Word Representations}. In \bibinfo{booktitle}{\emph{Proc.
  6th Int. Conf. on Learning Representations}}.
\newblock


\bibitem[\protect\citeauthoryear{Neelakantan, Xu, Puri, Radford, Han, Tworek,
  Yuan, Tezak, Kim, Hallacy, Heidecke, Shyam, Power, Nekoul, Sastry, Krueger,
  Schnurr, Such, Hsu, Thompson, Khan, Sherbakov, Jang, Welinder, and
  Weng}{Neelakantan et~al\mbox{.}}{2022}]%
        {neelakantan2022textcodeembeddingscontrastive}
\bibfield{author}{\bibinfo{person}{Arvind Neelakantan}, \bibinfo{person}{Tao
  Xu}, \bibinfo{person}{Raul Puri}, \bibinfo{person}{Alec Radford},
  \bibinfo{person}{Jesse~Michael Han}, \bibinfo{person}{Jerry Tworek},
  \bibinfo{person}{Qiming Yuan}, \bibinfo{person}{Nikolas Tezak},
  \bibinfo{person}{Jong~Wook Kim}, \bibinfo{person}{Chris Hallacy},
  \bibinfo{person}{Johannes Heidecke}, \bibinfo{person}{Pranav Shyam},
  \bibinfo{person}{Boris Power}, \bibinfo{person}{Tyna~Eloundou Nekoul},
  \bibinfo{person}{Girish Sastry}, \bibinfo{person}{Gretchen Krueger},
  \bibinfo{person}{David Schnurr}, \bibinfo{person}{Felipe~Petroski Such},
  \bibinfo{person}{Kenny Hsu}, \bibinfo{person}{Madeleine Thompson},
  \bibinfo{person}{Tabarak Khan}, \bibinfo{person}{Toki Sherbakov},
  \bibinfo{person}{Joanne Jang}, \bibinfo{person}{Peter Welinder}, {and}
  \bibinfo{person}{Lilian Weng}.} \bibinfo{year}{2022}\natexlab{}.
\newblock \bibinfo{title}{Text and Code Embeddings by Contrastive
  Pre-Training}.
\newblock
\newblock
\showeprint[arxiv]{2201.10005}~[cs.CL]
\urldef\tempurl%
\url{https://arxiv.org/abs/2201.10005}
\showURL{%
\tempurl}


\bibitem[\protect\citeauthoryear{Nielsen and Hansen}{Nielsen and
  Hansen}{2024}]%
        {pmlr-v233-nielsen24a}
\bibfield{author}{\bibinfo{person}{Beatrix Miranda~Ginn Nielsen} {and}
  \bibinfo{person}{Lars~Kai Hansen}.} \bibinfo{year}{2024}\natexlab{}.
\newblock \showarticletitle{Hubness Reduction Improves Sentence-{BERT} Semantic
  Spaces}. In \bibinfo{booktitle}{\emph{Proceedings of the 5th Northern Lights
  Deep Learning Conference ({NLDL})}} \emph{(\bibinfo{series}{Proceedings of
  Machine Learning Research})}, \bibfield{editor}{\bibinfo{person}{Tetiana
  Lutchyn}, \bibinfo{person}{Adín Ramírez~Rivera}, {and}
  \bibinfo{person}{Benjamin Ricaud}} (Eds.), Vol.~\bibinfo{volume}{233}.
  \bibinfo{pages}{181--204}.
\newblock


\bibitem[\protect\citeauthoryear{Ootomo, Naruse, Nolet, Wang, Feher, and
  Wang}{Ootomo et~al\mbox{.}}{2024}]%
        {10597683}
\bibfield{author}{\bibinfo{person}{Hiroyuki Ootomo}, \bibinfo{person}{Akira
  Naruse}, \bibinfo{person}{Corey Nolet}, \bibinfo{person}{Ray Wang},
  \bibinfo{person}{Tamas Feher}, {and} \bibinfo{person}{Yong Wang}.}
  \bibinfo{year}{2024}\natexlab{}.
\newblock \showarticletitle{CAGRA: Highly Parallel Graph Construction and
  Approximate Nearest Neighbor Search for GPUs}. In
  \bibinfo{booktitle}{\emph{Proc. 36th Int. Conf. on Data Engineering}}.
  \bibinfo{pages}{4236--4247}.
\newblock
\urldef\tempurl%
\url{https://doi.org/10.1109/ICDE60146.2024.00323}
\showDOI{\tempurl}


\bibitem[\protect\citeauthoryear{Pan, Wang, and Li}{Pan et~al\mbox{.}}{2024}]%
        {pan2024survey}
\bibfield{author}{\bibinfo{person}{James~Jie Pan}, \bibinfo{person}{Jianguo
  Wang}, {and} \bibinfo{person}{Guoliang Li}.} \bibinfo{year}{2024}\natexlab{}.
\newblock \showarticletitle{Survey of vector database management systems}.
\newblock \bibinfo{journal}{\emph{VLDB J.}} \bibinfo{volume}{33},
  \bibinfo{number}{5} (\bibinfo{date}{July} \bibinfo{year}{2024}),
  \bibinfo{pages}{1591–1615}.
\newblock
\showISSN{1066-8888}
\urldef\tempurl%
\url{https://doi.org/10.1007/s00778-024-00864-x}
\showDOI{\tempurl}


\bibitem[\protect\citeauthoryear{Pan, Wang, Wang, and Liu}{Pan
  et~al\mbox{.}}{2020}]%
        {PAN202059}
\bibfield{author}{\bibinfo{person}{Zhibin Pan}, \bibinfo{person}{Liangzhuang
  Wang}, \bibinfo{person}{Yang Wang}, {and} \bibinfo{person}{Yuchen Liu}.}
  \bibinfo{year}{2020}\natexlab{}.
\newblock \showarticletitle{Product quantization with dual codebooks for
  approximate nearest neighbor search}.
\newblock \bibinfo{journal}{\emph{Neurocomputing}}  \bibinfo{volume}{401}
  (\bibinfo{year}{2020}), \bibinfo{pages}{59--68}.
\newblock
\showISSN{0925-2312}
\urldef\tempurl%
\url{https://doi.org/10.1016/j.neucom.2020.03.016}
\showDOI{\tempurl}


\bibitem[\protect\citeauthoryear{Peng, Zhang, Li, Jin, and Ren}{Peng
  et~al\mbox{.}}{2023}]%
        {10.1145/3572848.3577527}
\bibfield{author}{\bibinfo{person}{Zhen Peng}, \bibinfo{person}{Minjia Zhang},
  \bibinfo{person}{Kai Li}, \bibinfo{person}{Ruoming Jin}, {and}
  \bibinfo{person}{Bin Ren}.} \bibinfo{year}{2023}\natexlab{}.
\newblock \showarticletitle{iQAN: Fast and Accurate Vector Search with
  Efficient Intra-Query Parallelism on Multi-Core Architectures}. In
  \bibinfo{booktitle}{\emph{Proc. 28th ACM SIGPLAN Symp. on Principles and
  Practice of Parallel Programming}}. \bibinfo{pages}{313–328}.
\newblock
\showISBNx{9798400700156}
\urldef\tempurl%
\url{https://doi.org/10.1145/3572848.3577527}
\showDOI{\tempurl}


\bibitem[\protect\citeauthoryear{Pennington, Socher, and Manning}{Pennington
  et~al\mbox{.}}{2014}]%
        {pennington-etal-2014-glove}
\bibfield{author}{\bibinfo{person}{Jeffrey Pennington},
  \bibinfo{person}{Richard Socher}, {and} \bibinfo{person}{Christopher
  Manning}.} \bibinfo{year}{2014}\natexlab{}.
\newblock \showarticletitle{{G}lo{V}e: Global Vectors for Word Representation}.
  In \bibinfo{booktitle}{\emph{Proceedings of the 2014 Conference on Empirical
  Methods in Natural Language Processing ({EMNLP})}},
  \bibfield{editor}{\bibinfo{person}{Alessandro Moschitti},
  \bibinfo{person}{Bo~Pang}, {and} \bibinfo{person}{Walter Daelemans}} (Eds.).
  \bibinfo{publisher}{Association for Computational Linguistics},
  \bibinfo{address}{Doha, Qatar}, \bibinfo{pages}{1532--1543}.
\newblock
\urldef\tempurl%
\url{https://doi.org/10.3115/v1/D14-1162}
\showDOI{\tempurl}


\bibitem[\protect\citeauthoryear{Raasveldt and M\"{u}hleisen}{Raasveldt and
  M\"{u}hleisen}{2019}]%
        {10.1145/3299869.3320212}
\bibfield{author}{\bibinfo{person}{Mark Raasveldt} {and}
  \bibinfo{person}{Hannes M\"{u}hleisen}.} \bibinfo{year}{2019}\natexlab{}.
\newblock \showarticletitle{DuckDB: an Embeddable Analytical Database}. In
  \bibinfo{booktitle}{\emph{Proc. ACM SIGMOD Int. Conf. on Management of
  Data}}. \bibinfo{pages}{1981–1984}.
\newblock
\showISBNx{9781450356435}
\urldef\tempurl%
\url{https://doi.org/10.1145/3299869.3320212}
\showDOI{\tempurl}


\bibitem[\protect\citeauthoryear{Radford, Kim, Hallacy, Ramesh, Goh, Agarwal,
  Sastry, Askell, Mishkin, Clark, Krueger, and Sutskever}{Radford
  et~al\mbox{.}}{2021a}]%
        {Radford2021LearningTV}
\bibfield{author}{\bibinfo{person}{Alec Radford}, \bibinfo{person}{Jong~Wook
  Kim}, \bibinfo{person}{Chris Hallacy}, \bibinfo{person}{Aditya Ramesh},
  \bibinfo{person}{Gabriel Goh}, \bibinfo{person}{Sandhini Agarwal},
  \bibinfo{person}{Girish Sastry}, \bibinfo{person}{Amanda Askell},
  \bibinfo{person}{Pamela Mishkin}, \bibinfo{person}{Jack Clark},
  \bibinfo{person}{Gretchen Krueger}, {and} \bibinfo{person}{Ilya Sutskever}.}
  \bibinfo{year}{2021}\natexlab{a}.
\newblock \showarticletitle{Learning Transferable Visual Models From Natural
  Language Supervision}. In \bibinfo{booktitle}{\emph{Proc. 38th Int. Conf. on
  Machine Learning}}.
\newblock
\urldef\tempurl%
\url{https://api.semanticscholar.org/CorpusID:231591445}
\showURL{%
\tempurl}


\bibitem[\protect\citeauthoryear{Radford, Kim, Hallacy, Ramesh, Goh, Agarwal,
  Sastry, Askell, Mishkin, Clark, Krueger, and Sutskever}{Radford
  et~al\mbox{.}}{2021b}]%
        {pmlr-v139-radford21a}
\bibfield{author}{\bibinfo{person}{Alec Radford}, \bibinfo{person}{Jong~Wook
  Kim}, \bibinfo{person}{Chris Hallacy}, \bibinfo{person}{Aditya Ramesh},
  \bibinfo{person}{Gabriel Goh}, \bibinfo{person}{Sandhini Agarwal},
  \bibinfo{person}{Girish Sastry}, \bibinfo{person}{Amanda Askell},
  \bibinfo{person}{Pamela Mishkin}, \bibinfo{person}{Jack Clark},
  \bibinfo{person}{Gretchen Krueger}, {and} \bibinfo{person}{Ilya Sutskever}.}
  \bibinfo{year}{2021}\natexlab{b}.
\newblock \showarticletitle{Learning Transferable Visual Models From Natural
  Language Supervision}. In \bibinfo{booktitle}{\emph{Proc. 38th Int. Conf. on
  Machine Learning}}, Vol.~\bibinfo{volume}{139}. \bibinfo{pages}{8748--8763}.
\newblock
\urldef\tempurl%
\url{https://proceedings.mlr.press/v139/radford21a.html}
\showURL{%
\tempurl}


\bibitem[\protect\citeauthoryear{Reimers and Gurevych}{Reimers and
  Gurevych}{2019}]%
        {reimers-gurevych-2019-sentence}
\bibfield{author}{\bibinfo{person}{Nils Reimers} {and} \bibinfo{person}{Iryna
  Gurevych}.} \bibinfo{year}{2019}\natexlab{}.
\newblock \showarticletitle{Sentence-{BERT}: Sentence Embeddings using
  {S}iamese {BERT}-Networks}. In \bibinfo{booktitle}{\emph{Proceedings of the
  2019 Conference on Empirical Methods in Natural Language Processing and the
  9th International Joint Conference on Natural Language Processing
  (EMNLP-IJCNLP)}}. \bibinfo{pages}{3982--3992}.
\newblock
\urldef\tempurl%
\url{https://doi.org/10.18653/v1/D19-1410}
\showDOI{\tempurl}


\bibitem[\protect\citeauthoryear{Schuhmann, Vencu, Beaumont, Kaczmarczyk,
  Mullis, Katta, Coombes, Jitsev, and Komatsuzaki}{Schuhmann
  et~al\mbox{.}}{2021}]%
        {laion400}
\bibfield{author}{\bibinfo{person}{Christoph Schuhmann},
  \bibinfo{person}{Richard Vencu}, \bibinfo{person}{Romain Beaumont},
  \bibinfo{person}{Robert Kaczmarczyk}, \bibinfo{person}{Clayton Mullis},
  \bibinfo{person}{Aarush Katta}, \bibinfo{person}{Theo Coombes},
  \bibinfo{person}{Jenia Jitsev}, {and} \bibinfo{person}{Aran Komatsuzaki}.}
  \bibinfo{year}{2021}\natexlab{}.
\newblock \bibinfo{title}{LAION-400M: Open Dataset of CLIP-Filtered 400 Million
  Image-Text Pairs}.
\newblock
\newblock
\showeprint[arxiv]{2111.02114}~[cs.CV]
\urldef\tempurl%
\url{https://arxiv.org/abs/2111.02114}
\showURL{%
\tempurl}


\bibitem[\protect\citeauthoryear{Silpa-Anan and Hartley}{Silpa-Anan and
  Hartley}{2008}]%
        {4587638}
\bibfield{author}{\bibinfo{person}{Chanop Silpa-Anan} {and}
  \bibinfo{person}{Richard Hartley}.} \bibinfo{year}{2008}\natexlab{}.
\newblock \showarticletitle{Optimised KD-trees for fast image descriptor
  matching}. In \bibinfo{booktitle}{\emph{Proc. IEEE Int. Conf. on Computer
  Vision and Pattern Recognition}}. \bibinfo{pages}{1--8}.
\newblock
\urldef\tempurl%
\url{https://doi.org/10.1109/CVPR.2008.4587638}
\showDOI{\tempurl}


\bibitem[\protect\citeauthoryear{Teofili and Lin}{Teofili and Lin}{2025}]%
        {10.1007/978-3-031-88714-7_39}
\bibfield{author}{\bibinfo{person}{Tommaso Teofili} {and}
  \bibinfo{person}{Jimmy Lin}.} \bibinfo{year}{2025}\natexlab{}.
\newblock \showarticletitle{Patience in Proximity: A Simple Early Termination
  Strategy for HNSW Graph Traversal Approximate k-Nearest Neighbor Search}. In
  \bibinfo{booktitle}{\emph{Proc. 47th European Conf. on IR Research}}.
  \bibinfo{pages}{401–407}.
\newblock
\showISBNx{978-3-031-88713-0}
\urldef\tempurl%
\url{https://doi.org/10.1007/978-3-031-88714-7_39}
\showDOI{\tempurl}


\bibitem[\protect\citeauthoryear{Tian, Liu, Tang, Xiao, Duan, Liao, Jin, Zhang,
  Zhu, and Zhang}{Tian et~al\mbox{.}}{2025}]%
        {305214}
\bibfield{author}{\bibinfo{person}{Bing Tian}, \bibinfo{person}{Haikun Liu},
  \bibinfo{person}{Yuhang Tang}, \bibinfo{person}{Shihai Xiao},
  \bibinfo{person}{Zhuohui Duan}, \bibinfo{person}{Xiaofei Liao},
  \bibinfo{person}{Hai Jin}, \bibinfo{person}{Xuecang Zhang},
  \bibinfo{person}{Junhua Zhu}, {and} \bibinfo{person}{Yu Zhang}.}
  \bibinfo{year}{2025}\natexlab{}.
\newblock \showarticletitle{Towards High-throughput and Low-latency
  Billion-scale Vector Search via {CPU/GPU} Collaborative Filtering and
  Re-ranking}. In \bibinfo{booktitle}{\emph{Proc. 23rd USENIX Conf. on File and
  Storage Technologies}}. \bibinfo{pages}{171--185}.
\newblock
\showISBNx{978-1-939133-45-8}
\urldef\tempurl%
\url{https://www.usenix.org/conference/fast25/presentation/tian-bing}
\showURL{%
\tempurl}


\bibitem[\protect\citeauthoryear{Tibshirani}{Tibshirani}{2022}]%
        {elastic_hnsw}
\bibfield{author}{\bibinfo{person}{Julie Tibshirani}.}
  \bibinfo{year}{2022}\natexlab{}.
\newblock \bibinfo{booktitle}{\emph{Introducing approximate nearest neighbor
  search in Elasticsearch 8.0}}.
\newblock
\urldef\tempurl%
\url{https://www.elastic.co/blog/introducing-approximate-nearest-neighbor-search-in-elasticsearch-8-0}
\showURL{%
\tempurl}


\bibitem[\protect\citeauthoryear{Tyshchuk, Karpikova, Spiridonov, Prutianova,
  Razzhigaev, and Panchenko}{Tyshchuk et~al\mbox{.}}{2023}]%
        {DBLP:journals/information/TyshchukKSPRP23}
\bibfield{author}{\bibinfo{person}{Kirill Tyshchuk}, \bibinfo{person}{Polina
  Karpikova}, \bibinfo{person}{Andrew Spiridonov}, \bibinfo{person}{Anastasiia
  Prutianova}, \bibinfo{person}{Anton Razzhigaev}, {and}
  \bibinfo{person}{Alexander Panchenko}.} \bibinfo{year}{2023}\natexlab{}.
\newblock \showarticletitle{On Isotropy of Multimodal Embeddings}.
\newblock \bibinfo{journal}{\emph{Inf.}} \bibinfo{volume}{14},
  \bibinfo{number}{7} (\bibinfo{year}{2023}), \bibinfo{pages}{392}.
\newblock


\bibitem[\protect\citeauthoryear{{Vargas Muñoz}, Gonçalves, Dias, and {da S.
  Torres}}{{Vargas Muñoz} et~al\mbox{.}}{2019}]%
        {Vargas2019HCNNG}
\bibfield{author}{\bibinfo{person}{Javier {Vargas Muñoz}},
  \bibinfo{person}{Marcos~A. Gonçalves}, \bibinfo{person}{Zanoni Dias}, {and}
  \bibinfo{person}{Ricardo {da S. Torres}}.} \bibinfo{year}{2019}\natexlab{}.
\newblock \showarticletitle{Hierarchical Clustering-Based Graphs for Large
  Scale Approximate Nearest Neighbor Search}.
\newblock \bibinfo{journal}{\emph{Pattern Recognition}}  \bibinfo{volume}{96}
  (\bibinfo{year}{2019}), \bibinfo{pages}{106970}.
\newblock
\showISSN{0031-3203}
\urldef\tempurl%
\url{https://doi.org/10.1016/j.patcog.2019.106970}
\showDOI{\tempurl}


\bibitem[\protect\citeauthoryear{Vaswani, Shazeer, Parmar, Uszkoreit, Jones,
  Gomez, Kaiser, and Polosukhin}{Vaswani et~al\mbox{.}}{2017}]%
        {NIPS2017_3f5ee243}
\bibfield{author}{\bibinfo{person}{Ashish Vaswani}, \bibinfo{person}{Noam
  Shazeer}, \bibinfo{person}{Niki Parmar}, \bibinfo{person}{Jakob Uszkoreit},
  \bibinfo{person}{Llion Jones}, \bibinfo{person}{Aidan~N Gomez},
  \bibinfo{person}{\L~ukasz Kaiser}, {and} \bibinfo{person}{Illia Polosukhin}.}
  \bibinfo{year}{2017}\natexlab{}.
\newblock \showarticletitle{Attention is All you Need}. In
  \bibinfo{booktitle}{\emph{Advances in Neural Information Processing Syst.,
  Proc. 31st Annual Conf. on Neural Information Processing Syst.}},
  \bibfield{editor}{\bibinfo{person}{I.~Guyon}, \bibinfo{person}{U.~Von
  Luxburg}, \bibinfo{person}{S.~Bengio}, \bibinfo{person}{H.~Wallach},
  \bibinfo{person}{R.~Fergus}, \bibinfo{person}{S.~Vishwanathan}, {and}
  \bibinfo{person}{R.~Garnett}} (Eds.), Vol.~\bibinfo{volume}{30}.
\newblock
\urldef\tempurl%
\url{https://proceedings.neurips.cc/paper_files/paper/2017/file/3f5ee243547dee91fbd053c1c4a845aa-Paper.pdf}
\showURL{%
\tempurl}


\bibitem[\protect\citeauthoryear{Wang, Yi, Guo, Jin, Xu, Li, Wang, Guo, Li, Xu,
  Yu, Yuan, Zou, Long, Cai, Li, Zhang, Mo, Gu, Jiang, Wei, and Xie}{Wang
  et~al\mbox{.}}{2021b}]%
        {10.1145/3448016.3457550}
\bibfield{author}{\bibinfo{person}{Jianguo Wang}, \bibinfo{person}{Xiaomeng
  Yi}, \bibinfo{person}{Rentong Guo}, \bibinfo{person}{Hai Jin},
  \bibinfo{person}{Peng Xu}, \bibinfo{person}{Shengjun Li},
  \bibinfo{person}{Xiangyu Wang}, \bibinfo{person}{Xiangzhou Guo},
  \bibinfo{person}{Chengming Li}, \bibinfo{person}{Xiaohai Xu},
  \bibinfo{person}{Kun Yu}, \bibinfo{person}{Yuxing Yuan},
  \bibinfo{person}{Yinghao Zou}, \bibinfo{person}{Jiquan Long},
  \bibinfo{person}{Yudong Cai}, \bibinfo{person}{Zhenxiang Li},
  \bibinfo{person}{Zhifeng Zhang}, \bibinfo{person}{Yihua Mo},
  \bibinfo{person}{Jun Gu}, \bibinfo{person}{Ruiyi Jiang}, \bibinfo{person}{Yi
  Wei}, {and} \bibinfo{person}{Charles Xie}.} \bibinfo{year}{2021}\natexlab{b}.
\newblock \showarticletitle{Milvus: A Purpose-Built Vector Data Management
  System}. In \bibinfo{booktitle}{\emph{Proc. ACM SIGMOD Int. Conf. on
  Management of Data}}. \bibinfo{pages}{2614–2627}.
\newblock
\showISBNx{9781450383431}
\urldef\tempurl%
\url{https://doi.org/10.1145/3448016.3457550}
\showDOI{\tempurl}


\bibitem[\protect\citeauthoryear{Wang, Wu, Ke, Gao, Zhu, and Zhou}{Wang
  et~al\mbox{.}}{2025}]%
        {wang2025accelerating}
\bibfield{author}{\bibinfo{person}{Mengzhao Wang}, \bibinfo{person}{Haotian
  Wu}, \bibinfo{person}{Xiangyu Ke}, \bibinfo{person}{Yunjun Gao},
  \bibinfo{person}{Yifan Zhu}, {and} \bibinfo{person}{Wenchao Zhou}.}
  \bibinfo{year}{2025}\natexlab{}.
\newblock \showarticletitle{Accelerating Graph Indexing for ANNS on Modern
  CPUs}.
\newblock \bibinfo{journal}{\emph{arXiv preprint arXiv:2502.18113}}
  (\bibinfo{year}{2025}).
\newblock


\bibitem[\protect\citeauthoryear{Wang, Xu, Yi, Wu, Peng, Ke, Gao, Xu, Guo, and
  Xie}{Wang et~al\mbox{.}}{2024b}]%
        {10.1145/3639269}
\bibfield{author}{\bibinfo{person}{Mengzhao Wang}, \bibinfo{person}{Weizhi Xu},
  \bibinfo{person}{Xiaomeng Yi}, \bibinfo{person}{Songlin Wu},
  \bibinfo{person}{Zhangyang Peng}, \bibinfo{person}{Xiangyu Ke},
  \bibinfo{person}{Yunjun Gao}, \bibinfo{person}{Xiaoliang Xu},
  \bibinfo{person}{Rentong Guo}, {and} \bibinfo{person}{Charles Xie}.}
  \bibinfo{year}{2024}\natexlab{b}.
\newblock \showarticletitle{Starling: An I/O-Efficient Disk-Resident Graph
  Index Framework for High-Dimensional Vector Similarity Search on Data
  Segment}.
\newblock \bibinfo{journal}{\emph{Proc. ACM Manag. Data}} \bibinfo{volume}{2},
  \bibinfo{number}{1}, Article \bibinfo{articleno}{14} (\bibinfo{date}{March}
  \bibinfo{year}{2024}), \bibinfo{numpages}{27}~pages.
\newblock
\urldef\tempurl%
\url{https://doi.org/10.1145/3639269}
\showDOI{\tempurl}


\bibitem[\protect\citeauthoryear{Wang, Xu, Yue, and Wang}{Wang
  et~al\mbox{.}}{2021a}]%
        {wang2021survey}
\bibfield{author}{\bibinfo{person}{Mengzhao Wang}, \bibinfo{person}{Xiaoliang
  Xu}, \bibinfo{person}{Qiang Yue}, {and} \bibinfo{person}{Yuxiang Wang}.}
  \bibinfo{year}{2021}\natexlab{a}.
\newblock \showarticletitle{A comprehensive survey and experimental comparison
  of graph-based approximate nearest neighbor search}.
\newblock \bibinfo{journal}{\emph{Proc. VLDB Endowment}} \bibinfo{volume}{14},
  \bibinfo{number}{11} (\bibinfo{date}{July} \bibinfo{year}{2021}),
  \bibinfo{pages}{1964–1978}.
\newblock
\showISSN{2150-8097}
\urldef\tempurl%
\url{https://doi.org/10.14778/3476249.3476255}
\showDOI{\tempurl}


\bibitem[\protect\citeauthoryear{Wang, Wang, Palpanas, and Wang}{Wang
  et~al\mbox{.}}{2023}]%
        {wang2023surveygt}
\bibfield{author}{\bibinfo{person}{Zeyu Wang}, \bibinfo{person}{Peng Wang},
  \bibinfo{person}{Themis Palpanas}, {and} \bibinfo{person}{Wei Wang}.}
  \bibinfo{year}{2023}\natexlab{}.
\newblock \showarticletitle{Graph- and Tree-based Indexes for High-dimensional
  Vector Similarity Search: Analyses, Comparisons, and Future Directions}.
\newblock \bibinfo{journal}{\emph{IEEE Data Eng. Bull.}}  \bibinfo{volume}{46}
  (\bibinfo{year}{2023}), \bibinfo{pages}{3--21}.
\newblock
\urldef\tempurl%
\url{https://api.semanticscholar.org/CorpusID:265509541}
\showURL{%
\tempurl}


\bibitem[\protect\citeauthoryear{Wang, Wang, Cheng, Wang, Palpanas, and
  Wang}{Wang et~al\mbox{.}}{2024a}]%
        {10.14778/3704965.3704974}
\bibfield{author}{\bibinfo{person}{Zeyu Wang}, \bibinfo{person}{Qitong Wang},
  \bibinfo{person}{Xiaoxing Cheng}, \bibinfo{person}{Peng Wang},
  \bibinfo{person}{Themis Palpanas}, {and} \bibinfo{person}{Wei Wang}.}
  \bibinfo{year}{2024}\natexlab{a}.
\newblock \showarticletitle{Steiner-Hardness: A Query Hardness Measure for
  Graph-Based ANN Indexes}.
\newblock \bibinfo{journal}{\emph{Proc. VLDB Endowment}} \bibinfo{volume}{17},
  \bibinfo{number}{13} (\bibinfo{year}{2024}), \bibinfo{pages}{4668–4682}.
\newblock
\urldef\tempurl%
\url{https://doi.org/10.14778/3704965.3704974}
\showDOI{\tempurl}


\bibitem[\protect\citeauthoryear{Yandex and Lempitsky}{Yandex and
  Lempitsky}{2016}]%
        {7780595}
\bibfield{author}{\bibinfo{person}{Artem~Babenko Yandex} {and}
  \bibinfo{person}{Victor Lempitsky}.} \bibinfo{year}{2016}\natexlab{}.
\newblock \showarticletitle{Efficient Indexing of Billion-Scale Datasets of
  Deep Descriptors}. In \bibinfo{booktitle}{\emph{Proc. IEEE Int. Conf. on
  Computer Vision and Pattern Recognition}}. \bibinfo{pages}{2055--2063}.
\newblock
\urldef\tempurl%
\url{https://doi.org/10.1109/CVPR.2016.226}
\showDOI{\tempurl}


\bibitem[\protect\citeauthoryear{Yang, Wang, Xu, Wang, Xiao, Du, and Zhou}{Yang
  et~al\mbox{.}}{2021}]%
        {yang2021taolearningframeworkadaptive}
\bibfield{author}{\bibinfo{person}{Kaixiang Yang}, \bibinfo{person}{Hongya
  Wang}, \bibinfo{person}{Bo Xu}, \bibinfo{person}{Wei Wang},
  \bibinfo{person}{Yingyuan Xiao}, \bibinfo{person}{Ming Du}, {and}
  \bibinfo{person}{Junfeng Zhou}.} \bibinfo{year}{2021}\natexlab{}.
\newblock \bibinfo{title}{Tao: A Learning Framework for Adaptive Nearest
  Neighbor Search using Static Features Only}.
\newblock
\newblock
\showeprint[arxiv]{2110.00696}~[cs.DB]
\urldef\tempurl%
\url{https://arxiv.org/abs/2110.00696}
\showURL{%
\tempurl}


\bibitem[\protect\citeauthoryear{Yang, Fang, and Lin}{Yang
  et~al\mbox{.}}{2017}]%
        {10.1145/3077136.3080721}
\bibfield{author}{\bibinfo{person}{Peilin Yang}, \bibinfo{person}{Hui Fang},
  {and} \bibinfo{person}{Jimmy Lin}.} \bibinfo{year}{2017}\natexlab{}.
\newblock \showarticletitle{Anserini: Enabling the Use of Lucene for
  Information Retrieval Research}. In \bibinfo{booktitle}{\emph{Proc. 40th
  Annual Int. ACM SIGIR Conf. on Research and Development in Information
  Retrieval}}. \bibinfo{pages}{1253–1256}.
\newblock
\showISBNx{9781450350228}
\urldef\tempurl%
\url{https://doi.org/10.1145/3077136.3080721}
\showURL{%
\tempurl}


\bibitem[\protect\citeauthoryear{Yang, Li, Fang, and Wei}{Yang
  et~al\mbox{.}}{2020}]%
        {10.1145/3318464.3386131}
\bibfield{author}{\bibinfo{person}{Wen Yang}, \bibinfo{person}{Tao Li},
  \bibinfo{person}{Gai Fang}, {and} \bibinfo{person}{Hong Wei}.}
  \bibinfo{year}{2020}\natexlab{}.
\newblock \showarticletitle{PASE: PostgreSQL Ultra-High-Dimensional Approximate
  Nearest Neighbor Search Extension}. In \bibinfo{booktitle}{\emph{Proc. ACM
  SIGMOD Int. Conf. on Management of Data}}. \bibinfo{pages}{2241–2253}.
\newblock
\showISBNx{9781450367356}
\urldef\tempurl%
\url{https://doi.org/10.1145/3318464.3386131}
\showDOI{\tempurl}


\bibitem[\protect\citeauthoryear{Yokoi, Bao, Kurita, and Shimodaira}{Yokoi
  et~al\mbox{.}}{2024}]%
        {yokoi2024zipfian}
\bibfield{author}{\bibinfo{person}{Sho Yokoi}, \bibinfo{person}{Han Bao},
  \bibinfo{person}{Hiroto Kurita}, {and} \bibinfo{person}{Hidetoshi
  Shimodaira}.} \bibinfo{year}{2024}\natexlab{}.
\newblock \showarticletitle{Zipfian Whitening}. In
  \bibinfo{booktitle}{\emph{Advances in Neural Information Processing Syst.,
  Proc. 38th Annual Conf. on Neural Information Processing Syst.}}
\newblock


\bibitem[\protect\citeauthoryear{You}{You}{2025}]%
        {you2025semanticsanglecosinesimilarity}
\bibfield{author}{\bibinfo{person}{Kisung You}.}
  \bibinfo{year}{2025}\natexlab{}.
\newblock \bibinfo{title}{Semantics at an Angle: When Cosine Similarity Works
  Until It Doesn't}.
\newblock
\newblock
\showeprint[arxiv]{2504.16318}~[cs.LG]
\urldef\tempurl%
\url{https://arxiv.org/abs/2504.16318}
\showURL{%
\tempurl}


\bibitem[\protect\citeauthoryear{Zhang, Xu, Chen, Sui, Xie, Cai, Chen, He,
  Yang, Yang, et~al\mbox{.}}{Zhang et~al\mbox{.}}{2023}]%
        {zhang2023vbase}
\bibfield{author}{\bibinfo{person}{Qianxi Zhang}, \bibinfo{person}{Shuotao Xu},
  \bibinfo{person}{Qi Chen}, \bibinfo{person}{Guoxin Sui},
  \bibinfo{person}{Jiadong Xie}, \bibinfo{person}{Zhizhen Cai},
  \bibinfo{person}{Yaoqi Chen}, \bibinfo{person}{Yinxuan He},
  \bibinfo{person}{Yuqing Yang}, \bibinfo{person}{Fan Yang}, {et~al\mbox{.}}}
  \bibinfo{year}{2023}\natexlab{}.
\newblock \showarticletitle{$\{$VBASE$\}$: Unifying Online Vector Similarity
  Search and Relational Queries via Relaxed Monotonicity}. In
  \bibinfo{booktitle}{\emph{Proc. 17th USENIX Symp. on Operating System Design
  and Implementation}}.
\newblock


\bibitem[\protect\citeauthoryear{Zhao, Zheng, Yi, Luan, Xie, Zhou, and
  Jensen}{Zhao et~al\mbox{.}}{2023}]%
        {10.14778/3579075.3579084}
\bibfield{author}{\bibinfo{person}{Xi Zhao}, \bibinfo{person}{Bolong Zheng},
  \bibinfo{person}{Xiaomeng Yi}, \bibinfo{person}{Xiaofan Luan},
  \bibinfo{person}{Charles Xie}, \bibinfo{person}{Xiaofang Zhou}, {and}
  \bibinfo{person}{Christian~S. Jensen}.} \bibinfo{year}{2023}\natexlab{}.
\newblock \showarticletitle{FARGO: Fast Maximum Inner Product Search via Global
  Multi-Probing}.
\newblock \bibinfo{journal}{\emph{Proc. VLDB Endowment}} \bibinfo{volume}{16},
  \bibinfo{number}{5} (\bibinfo{date}{Jan.} \bibinfo{year}{2023}),
  \bibinfo{pages}{1100–1112}.
\newblock
\showISSN{2150-8097}
\urldef\tempurl%
\url{https://doi.org/10.14778/3579075.3579084}
\showDOI{\tempurl}


\end{thebibliography}

\appendix

\section{Pseudocode of Adaptive EF Search}
\begin{algorithm}[h] 
\caption{ADAPTIVE-EF-SEARCH$(\mathbf{q}, ep, r)$}\label{algo:adaptive_search}
\begin{flushleft}
\textbf{Input:} query vector $\mathbf{q}$, enter point $ep$, expected recall $r$\\
\textbf{Output:} closest neighbors to $\mathbf{q}$
\end{flushleft}
\begin{algorithmic}[1]
\State $S \gets ep$ \Comment{set of visited elements}
\State $C \gets ep$ \Comment{set of candidates}
\State $W \gets ep$ \Comment{dynamic list of found nearest neighbors}
\State $\ef \gets \infty$ \Comment{no limitation of ef at the beginning}
\State $D \gets \text{get distance between } ep \text{ and } \mathbf{q}$ \Comment{sampled distance list}
\While{$|C| > 0$}
    \State $\mathbf{c} \gets \text{extract from } C \text{ the nearest element to } \mathbf{q}$
    \State $\mathbf{f} \gets \text{get from } W \text{ the furthest element to } \mathbf{q}$
    \If{$dist(\mathbf{c}, \mathbf{q}) > dist(\mathbf{f}, \mathbf{q})$}
        \State \textbf{break} 
    \EndIf
    \For{each $\mathbf{e} \in \text{neighborhood}(\mathbf{c})$} \Comment{update $C$ and $W$}
        \If{$\mathbf{e} \notin S$}
            \State $S \gets S \cup \mathbf{e}$
            \State $\mathbf{f} \gets \text{get from } W \text{ the furthest element to } \mathbf{q}$
        
            \If{$dist(\mathbf{e},\mathbf{q}) < dist(\mathbf{f},\mathbf{q})$ or $|W| < \ef$}
                \State $C \gets C \cup \mathbf{e}$
                \State $W \gets W \cup \mathbf{e}$
                \If{$|W| > ef$}
                    \State remove from $W$ the furthest element to $\mathbf{q}$
                \EndIf        
            \EndIf
            \If{$|D|<l$} \Comment{collect distances}
                \State $D \gets dist(\mathbf{e}, \mathbf{q})$
                \If{$|D|=l$} \Comment{reach distance limit}
                    \State $\ef\gets \text{ ESTIMATE-EF}(\mathbf{q}, D, r)$
                    \While{$|W|>\ef$}
                        \State remove from $W$ the furthest element
                    \EndWhile
                \EndIf
            \EndIf
        \EndIf            
    \EndFor
\EndWhile
\State \textbf{return} $W$
\end{algorithmic}
\end{algorithm}
Algorithm~\ref{algo:adaptive_search} presents the adaptive-\ef\ search algorithm, which extends the standard HNSW search described in Section~\ref{sec:pre}, with several key distinctions elaborated below.
First, the input to Algorithm~\ref{algo:adaptive_search} does not require a predefined \ef\ value; instead, \ef\ is initially set to $\infty$, allowing the traversal to visit any nodes around the entry point $ep$. 
Additionally, the algorithm initializes a sampled distance list $D$ to record the distances between visited nodes and the query vector $\mathbf{q}$. 
The size of $D$ is bounded by $l$ that corresponds to the number of nodes reachable within 2-hop from the entry point $ep$ at the base layer. 
The primary distinction from the original HNSW search lies between lines 20 and 25 of Algorithm~\ref{algo:adaptive_search}. Once $l$ distances have been collected, the algorithm invokes the ESTIMATE-EF function, which takes the query vector $\mathbf{q}$, the list $D$, and the user-specified recall $r$ as input, and returns a dynamically estimated \ef\ value. If the estimated \ef\ exceeds $l$, the dynamic list $W$ which maintains the current nearest neighbors will be truncated to match the estimated \ef.

\section{Relaxing Identical Distribution}  
The classical CLT can be extended to sequences of non-identically distributed random variables under certain conditions—most notably, Lindeberg’s condition~\cite{Klenke2020}. 
Consider the inner product between a query vector and a data vector,  $\mathbf{q} \cdot \mathbf{v} = \sum_{i=1}^d q_i v_i$. 
For convenience, let each term $q_i v_i$ be represented by a random variable $X_i$.
Suppose each $X_i$ has mean $\mu_i$ and variance $\sigma_i^2$. Then, Lindeberg’s condition for the sequence $(X_1, \ldots, X_d)$ is satisfied if, for every $\varepsilon > 0$,
$$
\lim_{d \to \infty} \frac{1}{s_d^2} \sum_{i=1}^d \text{E}\left[(X_i - \mu_i)^2 \cdot \mathbf{1}_{\{|X_i - \mu_i| > \varepsilon s_d\}}\right] = 0
,$$ where $s_d^2 = \sum_{i=1}^d \sigma_i^2$, and $\mathbf{1}_{\{|X_i - \mu_i| > \varepsilon s_d\}}$ is the indicator function, equal to 1 when the condition inside holds, and 0 otherwise.
Lindeberg’s condition ensures that \textit{no single  $X_i$ contributes a disproportionately large amount to the sum $\sum_{i=1}^d X_i$}.
In practice, modern embedding models—especially those based on Transformers—often apply layer normalization~\cite{ba2016layer,NIPS2017_3f5ee243} during training that regulates the scale of embedding components, resulting in components with approximately balanced magnitudes. This empirical property supports the assumption that real-world embedding vectors are likely to satisfy Lindeberg’s condition.

\section{Example of Query Score}
    Consider the case $ m = 5 $ and $ \delta = 0.001 $, and using  cosine distance as the similarity metric.  
Assume that for a given query vector $ \mathbf{q} $, the estimated FDL distribution is parameterized by $ \mu = 0.936 $ and $ \sigma = 0.0739 $, as obtained from Equation~\ref{equ:cosine_distance}.  
Under this setting, the first bin $ b_1 $ contains all collected distances in $ D $ that are smaller than $ 0.936 + 0.0739 \times \Phi^{-1}(0.001) = 0.7076 $, while bin $ b_2 $ includes distances between 0.7076 and $ 0.7233 = 0.936 + 0.0739 \times \Phi^{-1}(0.002) $. The remaining three bins are defined similarly based on the corresponding quantile thresholds.
Suppose the total number of distances is $ |D| = 100 $, and the bin counts are $ c_1 = 90 $, $ c_2 = 5 $, and $ c_3 = 5 $, with $ c_4 = c_5 = 0 $. The query score for $ \mathbf{q} $ is then computed as:
$
\text{Score}(\mathbf{q}) = 0.9 \times 100.0 + 0.05 \times 36.79 + 0.05 \times 13.53 = 92.516.
$
This high score suggests that $ \mathbf{q} $ could be an “easy” query, requiring a small \ef\ to achieve target recall. Indeed, 90\% of the first 100 visited nodes have distances smaller than the 0.001 quantile of the FDL distribution, indicating a high concentration of strong candidates early in the search.

\end{document}